\documentclass[footinbib,a4paper,aps,superscriptaddress,reprint,twocolumn,preprintnumbers,amsmath,amssymb,nobalancelastpage,10pt]{revtex4-1}
\usepackage{amsmath}
\usepackage{verbatim}
\usepackage{graphicx}
\usepackage{color}
\usepackage{tikz}
\usepackage{siunitx}
\usepackage{subfigure}
\usepackage{float}
\usepackage{slashed}
\usepackage{mathptmx}
\usepackage{amssymb}
\usepackage{amsmath}
\usepackage{amsfonts}
\usepackage{array}

\usepackage{bm}
\usepackage{xcolor}
\graphicspath{{figures/}}

\usepackage{braket}

\definecolor{mygrey}{gray}{0.35}
\definecolor{myblue}{rgb}{0.2,0.2,0.8}
\definecolor{mygreen}{rgb}{0.2,0.8,0.5}
\definecolor{myzard}{cmyk}{0,0,0.05,0}
\definecolor{mywhite}{rgb}{1,1,1}
\definecolor{myred}{rgb}{1,0.,0.3}

\usepackage[colorlinks=true,citecolor=myblue,linkcolor=myred]{hyperref}

 \def\ee{\mathord{\rm e}}
 
 \def\ii{\mathord{\rm i}}

\def\half{\textstyle\frac{1}{2}}

\renewcommand{\ii}{{\rm i}}
\renewcommand{\ee}{{\rm e}}
\def\beq{\begin{equation}}
\def\eeq{\end{equation}}
\def\barray{\begin{eqnarray}}
\def\earray{\end{eqnarray}}

\begin{document}

\title{Synthetic $\mathbb{Z}_2$ gauge theories based on parametric excitations of trapped ions}



\author{O. Băzăvan}
\affiliation{Department of Physics, University of Oxford, Clarendon Laboratory, Parks Road, Oxford OX1 3PU, United Kingdom}

\author{S. Saner} 
\affiliation{Department of Physics, University of Oxford, Clarendon Laboratory, Parks Road, Oxford OX1 3PU, United Kingdom}

\author{E. Tirrito}
\affiliation{The Abdus Salam International Centre for Theoretical Physics (ICTP), Strada Costiera 11, 34151 Trieste, Italy}
\affiliation{Pitaevskii BEC Center, CNR-INO and Dipartimento di Fisica, Università di Trento, Via Sommarive 14, Trento, I-38123, Italy}

\author{G. Araneda}
\affiliation{Department of Physics, University of Oxford, Clarendon Laboratory, Parks Road, Oxford OX1 3PU, United Kingdom}

\author{ R. Srinivas}
\affiliation{Department of Physics, University of Oxford, Clarendon Laboratory, Parks Road, Oxford OX1 3PU, United Kingdom}

\author{A. Bermudez}
\altaffiliation[On  sabbatical at ]{Department of Physics, University of Oxford, Clarendon Laboratory, Parks Road, Oxford, UK.}
\affiliation{Instituto de F\'isica Teorica, UAM-CSIC, Universidad Autónoma de Madrid, Cantoblanco, 28049 Madrid, Spain. }

\begin{abstract}

We present a detailed scheme for the analog quantum simulation of $\mathbb{Z}_2$ gauge theories  in  crystals of trapped ions, which exploits a more efficient hybrid encoding of the gauge and matter fields using the native  internal and motional degrees of freedom. We introduce a versatile toolbox based on  parametric excitations corresponding to different spin-motion-coupling schemes that induce a tunneling of the ions vibrational excitations conditioned to their internal qubit state. 
This building block, when implemented with a single trapped ion, corresponds to a minimal $\mathbb{Z}_2$ gauge theory, where the qubit plays the role of the gauge field on a synthetic link, and the vibrational excitations along different trap axes mimic the dynamical matter fields two synthetic sites, each carrying a $\mathbb{Z}_2$ charge. To evaluate their feasibility, we perform numerical simulations  of the state-dependent tunneling using realistic  parameters, and identify the leading sources of error in future experiments. We discuss how to generalise this minimal case to more complex settings by increasing the number of ions, moving  from a single link to a $\mathbb{Z}_2$ plaquette, and to an entire $\mathbb{Z}_2$  chain. We present analytical expressions for the gauge-invariant dynamics and the corresponding confinement, which are benchmarked using matrix product state simulations.

\end{abstract}

\maketitle

\setcounter{tocdepth}{2}
\begingroup
\hypersetup{linkcolor=black}
\tableofcontents
\endgroup

\section{\bf Introduction}

{Understanding the role of symmetry and its breakdown in the emergence of  various forms of order has played a key historical role in  many-body physics~\cite{Anderson393}. In this context, the  spontaneous breakdown of a global symmetry can unveil a local order parameter~\cite{landau_37}, which is crucial to understand phase transitions between different phases of matter, as well as  scaling and universality in the vicinity of certain types of critical points separating these phases~\cite{fradkin_2013}. Ultimately, it is this scaling and universality that underly our understanding of renormalization-group fixed points and the very definition of a quantum field theory (QFT)~\cite{WILSON197475,hollowood_2013}. It is within the realm of a particular type of QFTs, relativistic ones, where  symmetry has also played a pivotal historic role~\cite{doi:10.1073/pnas.93.25.14256}. In addition to symmetry breaking, and the consequences it brings when the global symmetry being broken is continuous~\cite{PhysRev.117.648,PhysRev.127.965}, the gauging of some of these global symmetries has also been  paramount of importance~\cite{PhysRev.96.191}. By  the introduction of  additional gauge fields,  these symmetries are converted into  local ones, and determine the way in which particles interact with each other.  

In particular, gauging of SU(2)$_{\rm L}\times$U(1) and SU(3)  symmetries underlies our understanding of nature at its most fundamental level, leading to our models of the electroweak~\cite{GLASHOW1959107,Salam1959,PhysRevLett.19.1264,HOOFT1971167,THOOFT1972189} and  strong~\cite{PhysRev.125.1067,GELLMANN1964214,Fritzsch:1973pi,PhysRevLett.30.1343,PhysRevLett.30.1346}  interactions, as well as their interplay with  the breakdown of continuous symmetries~\cite{PhysRev.130.439,PhysRevLett.13.508}. Altogether, the interplay of global and local symmetries, and how gauge fields get intertwined with matter fields, culminates in the standard model of particle physics~\cite{Peskin:1995ev}, our most fundamental theory of nature that has been tested with unprecedented precision. In spite of all the progress and detailed understanding of many facets of the standard model of physics, there  still remain open questions that do not require looking for other theories beyond.  These open problems arise in non-perturbative phenomena that are somehow linked  to  the (de)confinement of  particles~\cite{greensite_2020}, and the dynamical and static phenomena that occur at high densities~\cite{kogut_stephanov_2003}, or in non-equilibrium heavy ion collisions~\cite{Brambilla:2014jmp}. These questions   must be addressed non-perturbatively, e.g. on a lattice~\cite{PhysRevD.10.2445}, where great progress has taken place over the years leading, for instance, to the precise ab-initio determination  of the hadron masses in agreement with the quark-model predictions~\cite{durr2009abinitio}. Yet, the  so-called sign problem~\cite{PhysRevLett.94.170201,NAGATA2022103991} has partially hindered further progress for finite-density and real-time problems  using numerical Monte Carlo path-integral techniques. }

To  make further progress,  important insights can be gained by looking at lower dimensions and simpler gauge groups. For example, the study of   gauge theories in one spatial dimension has improved our understanding of confinement in high-energy physics~\cite{PhysRev.128.2425,PhysRevLett.31.792,COLEMAN1975267,THOOFT1974461, PhysRevD.13.1649}. Focusing also on  discrete  groups, such as the  $\mathbb{Z}_2$ gauge theory in two spatial dimensions~\cite{doi:10.1063/1.1665530}, one can  unveil the role of  non-local order parameters in  a confinement-deconfinement phase transition~\cite{PhysRevD.10.2445,doi:10.1063/1.1665530}. The deconfined phase of this model displays an exotic collective ordering, so-called topological order~\cite{Xiao:803748,RevModPhys.89.041004}, with a ground state with two characteristic features. First, it has a degeneracy that depends on topological invariants related to the homology of the low-energy excitations. Secondly, it displays long-range entanglement in spite of  a non-zero energy gap. 

It is worth mentioning that discrete-group  gauge theories  also appear as effective descriptions in condensed matter~\cite{fradkin_2013}, e.g. high-temperature superconductivity and  magnetism ~\cite{PhysRevB.37.580,PhysRevLett.66.1773,KITAEV20062}. Here, one typically deals  with  Hamiltonian gauge theories~\cite{PhysRevD.11.395} in which  Gauss' law restricts the system to a specific super-selection sector characterised by a background charge density. Whereas the vacuum is the privileged  super-selection  sector  in high-energy physics, any other sector  is in principle equally valid in condensed matter~\cite{PhysRevB.65.024504}. Moreover,  Gauss' law need not be  a strict constraint, but can instead arise  as soft constraint due to an energetic penalty  in the Hamiltonian~\cite{KITAEV20032}. For a  $\mathbb{Z}_2$ gauge theory, this approach allows to unveil  another characteristic property of topological order, namely the mutual anyonic statistics of the excitations, which becomes relevant for   fault-tolerant quantum computation~\cite{doi:10.1063/1.1499754,RevModPhys.87.307}.

Due to all of these cross-disciplinary connections between high-energy physics, condensed matter, and quantum computation, the study of $\mathbb{Z}_2$ lattice gauge theories~\cite{PhysRevX.6.041049,Gazit2017,doi:10.1073/pnas.1806338115, PhysRevB.97.245137, PhysRevX.10.041057, PhysRevB.102.155143,PhysRevD.102.074501,PhysRevLett.126.050401,PhysRevB.105.075132}, and also the larger $\mathbb{Z}_d$  groups~\cite{PhysRevD.19.3715}, has shown  a remarkable progress in recent years~\cite{PhysRevD.98.074503,PhysRevD.99.014503,PhysRevB.100.115152,PhysRevLett.124.120503,PhysRevLett.127.167203,Magnifico2020realtimedynamics,Surace_2021,PhysRevB.106.L041101,https://doi.org/10.48550/arxiv.2111.13205, delpino2023dynamical, https://doi.org/10.48550/arxiv.2208.07099,domanti2023coherence,https://doi.org/10.48550/arxiv.hep-lat/0509045,PhysRevX.10.041007,PhysRevResearch.3.013133,PhysRevB.105.245105,https://doi.org/10.48550/arxiv.2208.04182} {\cite{florio2023mass}}. 
 This interest has been further encouraged by key advances in the field of quantum simulators (QSs)~\cite{Feynman_1982,Cirac2012,Bloch2012,Blatt2012,PRXQuantum.2.017003}: experimental systems that can be controlled to realise a target Hamiltonian gauge theory in the laboratory. This approach   exploits the discrete nature of the gauge groups to find  efficient experimental encodings of the gauge  fields.  Starting with the pioneering cold-atom proposals for the digital~\cite{Weimer2010,TAGLIACOZZO2013160,Tagliacozzo2013,PhysRevX.4.041037} and analog~\cite{PhysRevLett.95.040402,PhysRevLett.107.275301,PhysRevLett.109.125302,PhysRevLett.109.175302,PhysRevLett.110.055302,PhysRevLett.110.125303,PhysRevLett.110.125304}  QSs for lattice gauge theories, a considerable effort has been devoted to push these QSs along various novel research directions (see reviews ~\cite{https://doi.org/10.1002/andp.201300104,Zohar,doi:10.1080/00107514.2016.1151199,Banuls2020,Carmen_Banuls_2020,doi:10.1098/rsta.2021.0064,Klco_2022, https://doi.org/10.48550/arxiv.2204.03381} {~\cite{Bauer:2023qgm,halimeh2023coldatom}}). In this regard, an alternative to  discrete gauge groups is the so-called ``quantum-link'' approach~\cite{HORN1981149,ORLAND1990647,CHANDRASEKHARAN1997455},  both  for Abelian and non-Abelian gauge groups~\cite{Banerjee_2013,PhysRevLett.112.201601,PhysRevA.90.042305, PhysRevX.6.011023, PhysRevLett.119.180402, PhysRevLett.122.250401, PhysRevD.100.074512,PhysRevLett.124.123601, PhysRevLett.126.220601,PhysRevX.10.041040,PhysRevB.102.041118,PhysRevB.102.165132,Magnifico2021,https://doi.org/10.48550/arxiv.2104.00025,https://doi.org/10.48550/arxiv.2112.04501,10.21468/SciPostPhys.13.2.017,https://doi.org/10.48550/arxiv.2201.07171,https://doi.org/10.48550/arxiv.2206.11273}. These advances have stimulated   the experimental efforts to build the first prototype QSs for lattice gauge theories~\cite{Martinez2016,Dai2017,PhysRevA.98.032331,Schweizer2019,Kokail2019,PhysRevX.10.021041,PhysRevD.101.074512,Mil1128,Yang2020,doi:10.1126/science.abl6277,Atas2021, PhysRevLett.127.212001,PRXQuantum.3.020324,PhysRevD.103.094501,PhysRevD.104.034501,PhysRevD.105.074504,PhysRevResearch.4.L022060,https://doi.org/10.48550/arxiv.2207.03473,https://doi.org/10.48550/arxiv.2207.01731,https://doi.org/10.48550/arxiv.2209.10781,Frolian2022,charles2023simulating}.
 
Gauge-theory QSs do not suffer from the sign problem of Monte Carlo methods   with fermionic matter at finite densities and real-time dynamics~\cite{PhysRevLett.94.170201,NAGATA2022103991}. Therefore, they have the potential of addressing  questions that have remained elusive for decades. Since solving the sign problem lies in the class of $NP$ (nondeterministic polynomial time)-hard problems~\cite{PhysRevLett.94.170201}, such that no polynomial-time classical algorithm is likely to be found, large-scale QSs of gauge theories are good candidates to demonstrate  practical quantum advantage. 
In fact,  QSs of the  real-time dynamics of even simpler  field theories have been proven to be  $BQP$ (bounded-error quantum polynomial time)-hard problems~\cite{doi:10.1126/science.1217069,https://doi.org/10.48550/arxiv.1112.4833,Jordan_2018} and, thus,  among the hardest problems that can be solved  with a quantum computer. Unless  a collapse of the complexity classes occurs, large-scale QSs of gauge theories should lead to stronger instances of quantum advantage that go beyond  the superiority with respect to the sign problem of a certain type of classical algorithms, and are instead ultimately supported by the  complexity of problems that can be solved by  quantum computers. 

In light of this promising future, an outstanding question  for  gauge-theory QSs  is to find viable schemes that allow one to move from the initial prototypes towards  the large-scale regime, both in terms of lattice sizes (i.e. qubit numbers) and simulation times (i.e. circuit depths). Experiments based on schemes that use specific  concatenation of gates 
have already allowed for   small-scale QSs of certain gauge theories~\cite{Martinez2016,PhysRevA.98.032331,Kokail2019,PhysRevD.101.074512,Atas2021, PhysRevLett.127.212001,PRXQuantum.3.020324,PhysRevD.103.094501,PhysRevD.104.034501,PhysRevD.105.074504,https://doi.org/10.48550/arxiv.2207.03473,https://doi.org/10.48550/arxiv.2207.01731,https://doi.org/10.48550/arxiv.2209.10781,charles2023simulating}. However, the existing levels of  noise and  errors, which accumulate along the circuits, will most likely require the use of future quantum-errror-corrected devices in order to reach very large scales . Although less flexible, the experiments on  analog QSs for gauge fields~\cite{Dai2017,PhysRevX.10.021041,Schweizer2019,Mil1128,Yang2020,doi:10.1126/science.abl6277,PhysRevResearch.4.L022060,Frolian2022} are, in principle, more amenable for scaling, even in the presence of noise~\cite{Flannigan_2022}. 
In this work, we thus focus on analog QSs, and choose the $\mathbb{Z}_2$ lattice gauge theory with dynamical matter as our target. In spite of its apparent simplicity, analog QSs of this gauge theory have only been recently realised for  two matter sites coupled by an intermediate gauge link in  recent  cold-atom experiments~\cite{Schweizer2019}. Other experiments targeting this model in  superconducting-qubit arrays~\cite{PhysRevResearch.4.L022060} are limited by the appearance of terms that explicitly break the gauge symmetry. Therefore, it would be desirable to find  alternatives that allow one to reach the desired large-scale QSs.
In this manuscript, we present a detailed toolbox for the QS of $\mathbb{Z}_2$ lattice gauge theories coupled to dynamical matter using trapped-ion systems that can overcome  these limitations. 

\subsection{Summary of the results}

Our toolbox exploits the versatility of trapped-ion platforms, which  not only contain qubits/spins that can be used to encode directly the $\mathbb{Z}_2$  fields, which can be understood as a binary truncation of an electric-field line,  but also have quantised vibrational degrees of freedom, i.e. phonons,  that can play the role of matter fields. 
As discussed in detail below, this hybrid spin-motional toolbox includes simple building blocks that can be realised already with state-of-the-art  technologies, benefiting from the current techniques used by ion trappers   for quantum computation in the context of high-precision phonon-mediated  gates. The key novel aspect is that these phonons are no longer used as auxiliary  buses to mediate the entangling gates between the qubit degrees of freedom, but instead become dynamical degrees of freedom themselves mimicking discretised matter fields in the QS. In order to make the connection to $\mathbb{Z}_2$ gauge theories, we show how  a particular state-dependent parametric excitation  can lead to a frequency conversion between a pair of excitations of different phonon modes mediated by the ion qubit, which can be interpreted as a gauge-invariant tunneling along a single link. Building on this result, we present more scalable schemes, going from a single plaquette to a full chain, which could be implemented upon realistic technological developments. We believe that the results  presented open an interesting direction to extend the QSs of $\mathbb{Z}_2$ lattice gauge theories  to more   challenging scenarios, and set the stage for future trapped-ion studies that explore gauge-theory QSs in higher spatial dimensions.

Our results are presented as follows. In Sec.~\ref{sec:state_dep_param_tunn}, building on the application of parametric driving reviewed in Appendix~\ref{sec_bg:gauge},  we show how one can generate a gauge-invariant Hamiltonian where a $\mathbb{Z}_2$ gauge field mediates the tunnelling between a pair of bosonic modes. In the single-boson sector, where the tunnelling of the boson is correlated to the stretching/compressing of the electric field at the link, the dynamics corresponds to detuned Rabi oscillations. In the two-boson sector, we show that the system gives rise to bright and dark states, and that the gauge-invariant dynamics leas to mode entanglement in the matter sector. In Sec.~\ref{sec:trapped-ion_toolbox}, we describe various realistic schemes for the implementation of this $\mathbb{Z}_2$ gauge theory on a link using a single trapped ion. We discuss  a light-shift- or a M{\o}lmer-S{\o}rensen-type scheme to implement a state-dependent parametric tunnelling between the phonons along two different trap axes, presenting a thorough numerical comparison of the ideal and realistic dynamics  with current trapped-ion parameters. We also discuss a different  possibility that combines two orthogonal state-dependent forces, and can be combined in a pulsed Trotterization to yield the desired gauge-invariant model with realistic trapped-ion parameters. Several technical details and extended discussion are presented in Appendix~\ref{sec:trapped_ion_dg}. In Sec.~\ref{sec:dim_reduction}, we discuss how one can also realise other $\mathbb{Z}_2$ models by including more ions. In particular, we show that the center-of-mass modes of a two-ion system can be used to simulate  a gauge-invariant model  with two links forming a plaquette. We show that the superposition principle of the possible encircling paths is that a single boson can lead to an entangled state for the $\mathbb{Z}_2$ gauge fields. In this section, we also show how a string of ions with dimerised center-of-mass modes can also be used to implement a $\mathbb{Z}_2$ gauge theory on an entire chain. We present analytical solutions for the confinement dynamics on the single- and two-boson sector, and also explore the phenomenon of string breaking in the half-filled sector using Matrix-Product-State simulations. We present our conclusions in Sec.~\ref{sec:conclusion_outlook}.

\section{\bf  Dynamical gauge fields:  the {$\mathbb{Z}_2$}  theory on a   link}
\label{sec:state_dep_param_tunn}

\subsection{State-dependent parametric tunnelling}

The use of periodic resonant modulations allows to  design quantum simulators  on a lattice that has a connectivity   different from the original one. This leads to the concept of synthetic dimensions ~\cite{PhysRevLett.108.133001,PhysRevLett.112.043001}, as recently reviewed in~\cite{Ozawa2019}, which have been realised  in various experimental platforms~\cite{doi:10.1126/science.aaa8736,doi:10.1126/science.aaa8515,PhysRevLett.117.220401,PhysRevLett.122.065303,Bernier2017,doi:10.1126/sciadv.1602685,Fang2017,Zilberberg2018,lustig2019photonic,Dutt2019,doi:10.1126/science.aaz3071,Chalopin2020}. A particular form of these resonant modulations can be achieved by a parametric excitation/driving, which will be the main tool exploited in this work. In its original context, a parametric excitation induces couplings between different modes of the electromagnetic field,  leading to  a well-known technique for frequency conversion and linear amplification of photons~\cite{PhysRev.124.1646,PhysRev.128.2407,PhysRev.129.481,PhysRev.160.1076,PhysRevD.26.1817}. For instance, as originally discussed in~\cite{PhysRev.124.1646}, a small periodic modulation of the dielectric constant of a cavity can lead to different couplings between the cavity modes that can be controlled by tuning the modulation frequency to certain resonances. In the particular case of frequency conversion, this scheme can be understood in terms of  a parametric tunnelling term between two synthetic lattice sites labelled by the frequencies of the two modes, which is the essence of the schemes for synthetic dimensions discussed in~\cite{PhysRevLett.108.133001,PhysRevLett.112.043001}. As  emphasised in~\cite{PhysRevLett.108.153901}, this simple parametric tunnelling already inherits the phase of the drive~\cite{PhysRev.124.1646}, such that one could design and implement~\cite{PhysRevB.87.060301} non-trivial schemes where this phase mimics the Aharonov-Bohm effect of charged particles moving under a static background magnetic field~\cite{PhysRev.115.485}. These ideas can also be exploited when the modes belong to distant resonators  coupled via intermediate components, such as mixers~\cite{Fang2012} or tunable inductors~\cite{Chen_2014}. This  leads to parametric tunnelling terms where the phase can be tuned locally, leading to quantum simulators of quantum Hall-type physics~\cite{Roushan2017}. Furthermore, periodic modulations of the mode frequencies with a relative phase difference can also lead to these synthetic background gauge fields~\cite {PhysRevLett.107.150501,PhysRevLett.109.145301}, as demonstrated in experiments that exploit Floquet engineering in optical lattices~\cite{PhysRevLett.111.185301,PhysRevLett.111.185302,Atala2014,Aidelsburger2015}, symmetrically-coupled resonators~\cite{Estep2014}, and  trapped-ion crystals~\cite{PhysRevLett.123.213605}. We should  mention that there are other schemes for   static background gauge fields  {that} do not exploit periodic modulations, but instead mediate the tunnelling by an intermediate quantum system~\cite{PhysRevA.82.043811,Hafezi2011,Hafezi2013,PhysRevLett.113.087403,Mittal2018,PRXQuantum.1.020303,https://doi.org/10.48550/arxiv.2205.11178}.

  In Appendix~\ref{sec_bg:gauge}, we present a detailed description of the use of parametric excitations in this context, and the possible implementation of   quantum Hall-type physics in crystals of trapped ions. We emphasize that the parametric schemes discussed in the Appendix lead to a background static gauge field, the dynamics of which does not follow from a gauge theory.
In this section,  we start by discussing  how to generalise the scheme towards the quantum simulation of the simplest discrete gauge theory: a $\mathbb{Z}_2$ gauge link.  We  consider    two  modes $d\in\mathcal{D}=\{1,2\}$ of  energies $\omega_d$ ( $\hbar=1$ henceforth) and an additional spin-1/2 system/qubit ~\cite{nielsen00}, which is initially decoupled from the modes. The bare Hamiltonian is
\beq
\label{eq:H_0_st}
\tilde{H}_0=\omega_1 a^{{\dagger}}_{1}a^{\phantom{\dagger}}_{1}+\omega_2 a^{{\dagger}}_{2}a^{\phantom{\dagger}}_{2}+\frac{\omega_{0}}{2}\sigma^z_{1,{\bf e}_1},
\eeq
where we have  introduced the qubit transition frequency $\omega_0$,  the  Pauli matrix $\sigma^z_{1,{\bf e}_1}=\ket{\uparrow_{1,{\bf e}_1}}\!\bra{\uparrow_{1,{\bf e}_1}}-\ket{\downarrow_{1,{\bf e}_1}}\!\bra{\downarrow_{1,{\bf e}_1}}$, and the creation annihilation operators $a^{{\dagger}}_{d}, a^{\phantom{\dagger}}_{d}$ for each mode of frequency $\omega_d$. The choice of the convoluted notation for the index of the qubit will be justified below once we interpret the effective model in the light of a synthetic lattice gauge theory. 

The idea now is to consider a   generalisation of a frequency-conversion parametric driving that includes two tones $\tilde{V}(t)=\tilde{V}_1+\tilde{V}_2$. One of them yields  a  parametric drive 
\beq
\label{eq:parametric_tunneling_state_dep}
\tilde{V}_1=\Omega_{\rm d} \sigma^z_{1,{\bf e}_1}a^{{\dagger}}_{2}a^{\phantom{\dagger}}_{1}\cos(\phi_{\rm d}-\omega_{\rm d}t)+{\rm H.c.},
\eeq
which has an amplitude that depends on the state of the qubit. 
 Additionally, the other  tone drives transitions on the qubit 
\beq
\label{eq:second_tone}
\tilde{V}_2=\tilde{\Omega}_{\rm d}\sigma^x_{1,{\bf e}_1}\cos(\tilde{\phi}_{\rm d}-\tilde{\omega}_{{\rm d}}t)+{\rm H.c.},
\eeq
where we have introduced an additional Pauli matrix $\sigma^x_{1,{\bf e}_1}=\ket{\uparrow_{1,{\bf e}_1}}\!\bra{\downarrow_{1,{\bf e}_1}}+\ket{\downarrow_{1,{\bf e}_1}}\!\bra{\uparrow_{1,{\bf e}_1}}$. Considering that  the frequency and strength of this additional  driving are constrained by
\beq
\label{eq:constraints_carrier}
\tilde{\omega}_{\rm d}=\omega_0, \hspace{2ex} |\tilde{\Omega}_{\rm d}|\ll 4\omega_0,
\eeq
we can  follow the exact same steps as those discussed in the Appendix to show that, after  setting $\phi_{\rm d}=\tilde{\phi}_{\rm d}=0$, the two-tone drive  leads to a time-independent effective Hamiltonian $\tilde{V}_I(t)\approx \tilde{H}_{\rm eff}$ that supersedes Eq.~\eqref{eq:tunneling}, and reads
\beq
\label{eq:tunneling_gauge}
 H_{\rm eff}=\left(t^{\phantom{\dagger}}_{1,\textbf{e}_1}a^{{\dagger}}_{2}\sigma^z_{1,{\bf e}_1}a^{\phantom{\dagger}}_{1}+{\rm H.c.}\right)+h\sigma^x_{1,{\bf e}_1}
\eeq
where we have introduced the effective couplings
\beq
\label{eq:link_couplings}
t^{\phantom{\dagger}}_{1,\textbf{e}_1}=\frac{\Omega_{\rm d}}{2}, \hspace{1ex}\text{and}\hspace{1ex}h=\frac{\tilde{\Omega}_{\rm d}}{2}.
\eeq
\begin{figure}[t]
	\centering
	\includegraphics[width=1\columnwidth]{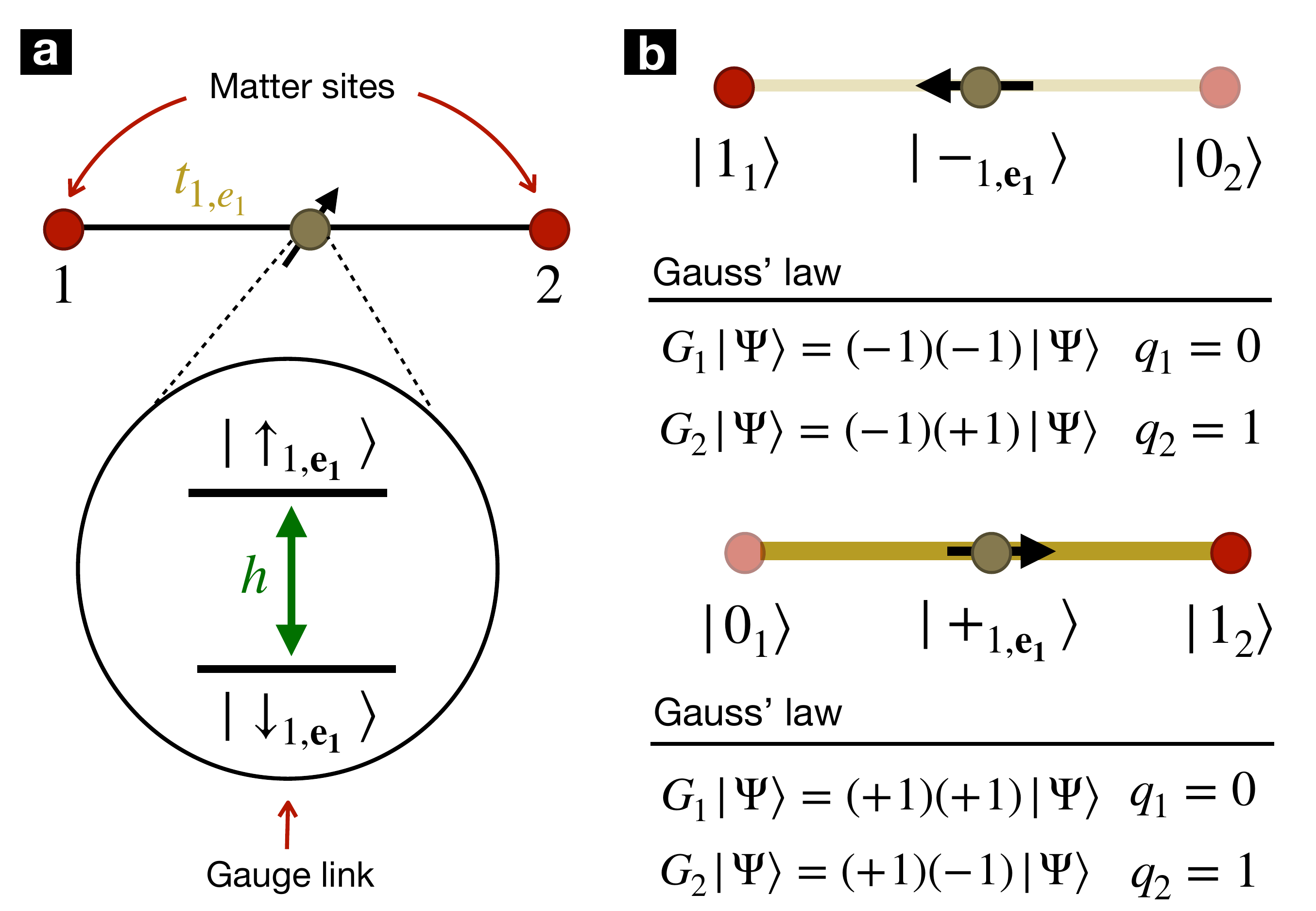}
	\caption{{\bf Synthetic $\mathbb{Z}_2$ gauge links and Gauss' law:}  {\bf (a)}     Schematic representation of the effective Hamiltonian in Eq.~\eqref{eq:tunneling_gauge}. The two modes labelled by 1,2, which play the role of matter fields,  are  coupled by a synthetic tunnelling of strength  $t_{1,{\bf e}_1}$ that is mediated by a qubit that  plays the role of the gauge field, and effectively sits on the synthetic link. In addition to the tunnelling, the electric-field term of strength $h$ drives transitions in the qubit  (inset). {\bf (b)} For a single particle, Gauss' law~\eqref{eq:gauss_law_link} for a distribution of background charges $q_1=0$, $q_2=1$, is fulfilled by the two states $\ket{1_1,-_{1,{\bf e}_1},0_2},\ket{0_1,+_{1,{\bf e}_1},1_2}$,  characterised by the absence or presence of an electric field attached to the matter particle sitting on the leftmost or rightmost site. These electric-field states are represented by arrows  parallel (anti-parallel) to the external field $h$, and the presence (absence) of the corresponding electric-field line is represented by a thicker (shaded) golden link.         }
	\label{fig:gauge_invariant_states_link}
\end{figure}

There are two important aspects to highlight. First, we have again  managed to engineer  a synthetic link  connecting the two modes via  the tunneling of particles. However, in contrast to the previous case~\eqref{eq:tunneling},  the qubit enters in this process and mediates the tunnelling. In the spirit of synthetic dimensions, we can say  that the qubit effectively sits on the synthetic link (see  Fig.~\ref{fig:gauge_invariant_states_link}{\bf (a)}). It is for this reason that the label used for the qubit $(1,{\bf e}_1)$ refers to the link that connects the synthetic site $1$ to its nearest neighbour $2$ via the direction specified by the vector ${\bf e}_1$. This notation has a clear generalisation to larger lattices and different geometries, and is common in the context of lattice gauge theories~\cite{gattringer_lang_2010}.  The second important aspect to remark  is that the dynamics dictated by the Hamiltonian of Eq.~\eqref{eq:tunneling_gauge}, considering also the term proportional to $h$, has a local/gauge $\mathbb{Z}_2$  symmetry. This gauge symmetry is related to the previous $U(1)$ phase rotation of the modes discussed below Eq.~\eqref{eq:trap_freq} when restricting to  a $\pi$ phase. More importantly, this $\pi$ phase can be chosen  locally.  We can  transform either $a^{\phantom{\dagger}}_1,a_1^\dagger\mapsto-a^{\phantom{\dagger}}_1,-a_1^\dagger$, or $a^{\phantom{\dagger}}_2,a_2^\dagger\mapsto-a^{\phantom{\dagger}}_2,-a_2^\dagger$ by a local $\pi$ phase, and retain gauge invariance in the Hamiltonian by simultaneously inverting the link qubit $\sigma^z_{1,{\bf e}_1}\mapsto-\sigma^z_{1,{\bf e}_1}, \sigma^x_{1,{\bf e}_1}\mapsto\sigma^x_{1,{\bf e}_1}$. Accordingly, the qubit can be interpreted as a $\mathbb{Z}_2$  gauge field introduced to gauge the global $\mathbb{Z}_2$ inversion symmetry of the Hamiltonian~\eqref{eq:tunneling}, paralleling the situation with other groups where gauge fields are introduced in the links of the lattice and mediate the tunnelling of matter particles~\cite{gattringer_lang_2010}. Accordingly, we see that the analogy with gauge theories is not a mere notational choice due to our qubit labelling, but  rests on the effective  gauging of the symmetry: the engineered tunnelling leads to  a discretised version of the covariant derivative that is required to upgrade  a global symmetry into a local one. 

In Hamiltonian approaches to lattice gauge theories~\cite{PhysRevD.11.395}, one   {can} work in  Weyl's temporal gauge, such that there is  a residual  redundancy that is dealt with by imposing  Gauss' law. As usually in the literature~\cite{PhysRevX.6.041049,Gazit2017,doi:10.1073/pnas.1806338115, PhysRevB.97.245137, PhysRevX.10.041057, PhysRevB.102.155143,PhysRevD.102.074501,PhysRevLett.126.050401,PhysRevB.105.075132,PhysRevD.98.074503,PhysRevD.99.014503,PhysRevB.100.115152,PhysRevLett.124.120503,PhysRevLett.127.167203,Magnifico2020realtimedynamics,Surace_2021,PhysRevB.106.L041101,https://doi.org/10.48550/arxiv.2111.13205, https://doi.org/10.48550/arxiv.2208.07099,https://doi.org/10.48550/arxiv.hep-lat/0509045,PhysRevX.10.041007,PhysRevResearch.3.013133,PhysRevB.105.245105,https://doi.org/10.48550/arxiv.2208.04182}, one refers to the external field $h$ as the electric field term, and talks about the Hadamard states $\ket{\pm_{1,{\bf e}_1}}=(\ket{\uparrow_{1,{\bf e}_1}}\pm\ket{\downarrow_{1,{\bf e}_1}})/\sqrt{2}$ as the electric-field basis, such that the $\ket{+_{1,{\bf e}_1}}$ state describes an electric field line connecting two neighbouring sites. The generators of the aforementioned local symmetries $[H_{\rm eff},G_1]=[H_{\rm eff},G_2]=[G_1,G_2]=0$ are thus 
\beq
\label{eq:generators_link}
G_1=\ee^{\ii\pi a_1^\dagger a_1^{\phantom{\dagger}}}\sigma_{1,{\bf e}_1}^x, \hspace{2ex}  G_2=\sigma_{1,{\bf e}_1}^x\ee^{\ii\pi a_2^\dagger a_2^{\phantom{\dagger}}},
\eeq
and Gauss' law is imposed by fixing a specific  super-selection sector  that fulfills 
\beq
\label{eq:gauss_law_link}
G_1\ket{\Psi_{\rm phys}}=\ee^{\ii\pi q_1} \ket{\Psi_{\rm phys}},\hspace{1ex} G_2\ket{\Psi_{\rm phys}}= \ee^{\ii\pi q_2}\ket{\Psi_{\rm phys}}.
\eeq
Here, $q_i\in\{0,1\}$ are the so-called background charges, and $\ket{\Psi_{\rm phys}}$ is a generic state of the physical system that is a  common eigenstate of the local-symmetry generators. These generators, which fulfill $G_i^2=\mathbb{I}$, and thus only have the two discrete eigenvalues $\pm 1$, can be used to define orthogonal projectors  $P_{\{q_i\}}=\prod_j\frac{1}{{2}}(\mathbb{I}+\ee^{\ii\pi q_j}G_j)$. Due to the gauge symmetry, the Hamiltonian is block-diagonal in subspaces with a fixed arrangement of background charges $H_{\rm eff}=\sum_{\{q_i\}}P_{\{q_i\}}H_{\rm eff}P_{\{q_i\}}$, and the different super-selection sectors  cannot be  connected by the gauge-invariant dynamics.

In Fig.~\ref{fig:gauge_invariant_states_link}{\bf (b)}, we depict  the possible  states for this physical subspace in the case of a single  particle, where one can see how the link qubit must lie in a specific electric-field state such that Gauss' law is fulfilled for the choice $q_1=0, q_2=1$. In particular, when a single particle sits in the rightmost site 2, an electric field line is created by flipping the qubit in the  link that connects it to the leftmost site 1. This is a clear  analogy to  Maxwell electrodynamics, where the regions with a net charge act as sinks/sources of the electric field.  In fact, for (1+1)-dimensional quantum electrodynamics, i.e. the Schwinger model~\cite{PhysRev.128.2425,PhysRevLett.31.792,COLEMAN1975267}, a $U(1)$ positive charge acts as a source of  an electric field that  remains constant in space  until a negative charge  is found, which acts a sink. This  leads to  an electric field string connecting the electron-positron pair~\cite{COLEMAN1975267}. As we will discuss in more detail below, a similar effect occurs  when the $\mathbb{Z}_2$ gauge theory~\eqref{eq:tunneling_gauge} is extended to larger chains. 
Here, the particle moves by stretching or compressing the electric field string, which resembles the Dirac string construction that attaches an infinitely-thin solenoid carrying magnetic flux to a magnetic charge, i.e. a magnetic monopole~\cite{doi:10.1098/rspa.1931.0130}. The electric analog is, nonetheless, a 1D effect and, furthermore, it is gauge-invariant and observable as   emphasised  below.

Before going to these generalisations, let us  discuss simple dynamical manifestations of this electric field string that will be a guide for the trapped-ion implementation discussed in the following section. Although the quantum simulation scheme leading to Eq.~\eqref{eq:tunneling_gauge} works for both fermionic and bosonic matter, we will thus focus on the later that will be mapped onto trapped-ion phonons. For two bosonic modes and a single gauge qubit, the Hilbert space is infinite dimensional $\mathcal{H}=\mathcal{F}\otimes\mathbb{C}^2$, where $\mathcal{F}=\oplus_{n=0}^\infty\mathcal{F}_{n}$, and  each subspace $ \mathcal{F}_{n}={\rm span}\{\ket{n_1}\otimes\ket{n_2}:n_1+n_2=n\}$ contains $n$ bosonic particles in total. These subspaces can be spanned by the corresponding Fock states $\ket{n_i}=(a_{i}^\dagger)^{n_i}\ket{0_i}/{\sqrt{n_i!}}$, where  $\ket{0_i}$ is the vacuum of the $i$-th mode. As a consequence of the global $U(1)
$ symmetry and   Gauss' law~\eqref{eq:gauss_law_link}, we can reduce the size of the subspace where the dynamics takes place, and find some neat manifestations of the correlations between the charge and electric field degrees of freedom. In particular, we will show how the dynamics of  the $\mathbb{Z}_2$ link can be understood in terms of typical phenomena in quantum optics, such as Rabi oscillations, dark states, and mode entanglement.

\subsection{One-boson sector: Rabi oscillations and matter-gauge-field correlated dynamics}

\begin{figure}[t]
	\centering
	\includegraphics[width=0.95\columnwidth]{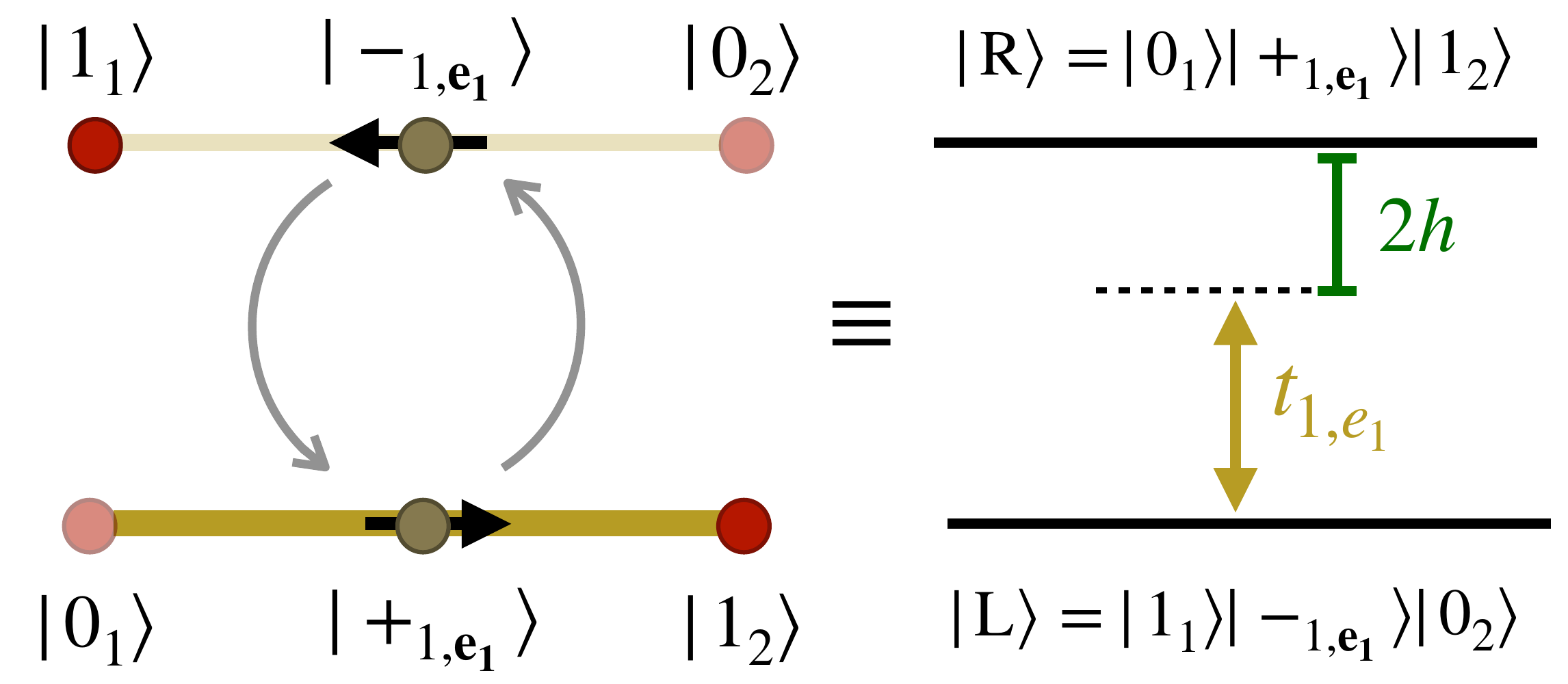}
	\caption{{\bf Effective two-level system for $\mathbb{Z}_2$-link tunnelling:}  In the super-selection sector~\eqref{eq:gauss_law_link} with background charges $q_1=0,q_2=1$, the physical subspace for a single particle is composed of two states~\eqref{eq:basis} depicted in Fig.~\ref{fig:gauge_invariant_states_link} {\bf (b)}. The gauge-invariant Hamiltonian~\eqref{eq:tunneling_gauge} can then be mapped onto the problem of detuned Rabi oscillations of a two-level atom in the rotating frame, where the tunnelling plays the role of the Rabi frequency, and the electric-field term is proportional to the detuning of the Rabi drive. }
	\label{fig:z2_link_tunneling_rabi_oscillations}
\end{figure}

When the initial state contains a single particle, we can restrict the Hilbert space to $\mathcal{F}_1\otimes\mathbb{C}^2$ using the global $U(1)$ symmetry. Moreover, Gauss' law~\eqref{eq:gauss_law_link} for $q_1=0,q_2=1$ allows us to restrict the physical states further to a two-dimensional subspace $\mathcal{V}_{\rm phys}	\subset\mathcal{H}$, such that the  states have the form $\ket{\Psi_{\rm phys}(t)}=c_{r}(t)\ket{\rm R}+c_l(t)\ket{\rm L}$, where we have introduced 
\beq
\begin{split}
\label{eq:basis}
\ket{\rm L}=\ket{1_1}\otimes\ket{-_{1,{\bf e}_1}}\otimes\ket{0_2}, \hspace{2ex}
\ket{\rm R}=\ket{0_1}\otimes\ket{+_{1,{\bf e}_1}}\otimes \ket{1_2}.
\end{split}
\eeq 
According to the effective Hamiltonian of Eq.~\eqref{eq:tunneling_gauge}, these levels are split in energy by $2h$, and transitions between them are induced by the gauge-invariant tunnelling of strength $t_{1,\textbf{e}_1}$~\eqref{eq:link_couplings}. The problem thus reduces to that of Rabi oscillations~\cite{PhysRev.51.652} of a driven two-level atom~\cite{osti_7365050} (see Fig.~\ref{fig:z2_link_tunneling_rabi_oscillations}), and has an exact solution 
$\boldsymbol{c}(t)=\ee^{-\ii\Omega_0t\boldsymbol{n}\cdot\boldsymbol{\sigma}}\boldsymbol{c}(0)$, where $\boldsymbol{c}(t) = (c_r(t),c_l(t))^{\rm t}$, and  we have introduced the vector of Pauli matrices $\boldsymbol{\sigma}=(\sigma^x,\sigma^y,\sigma^z)$, and the following quantities
\beq
\label{eq_eff_rabi_flop}
\Omega_0=\sqrt{t_{1,\textbf{e}_1}^2+h^2},\hspace{2ex} \boldsymbol{n}=\frac{1}{\Omega_0}(t_{1,\textbf{e}_1},0,h).
\eeq
Assuming that the particle occupies initially  the leftmost site $\ket{\Psi_{\rm phys}(0)}=\ket{\rm L}$, we see that the tunnelling to the right is accompanied by the build-up of an electric field line across the gauge link, which is thus attached  to the dynamical $\mathbb{Z}_2$ charge carried by the particle.   This correlated dynamics can be observed by measuring the periodic oscillations of the following gauge-invariant observables 
\beq
\label{eq_Rabi flops_link}
\begin{split}
\overline{n}_2(t):=\langle a_2^\dagger a_2^{\phantom{\dagger}}(t)\rangle&=\frac{t_{1,\textbf{e}_1}^2}{\Omega_0^2}\sin^2(\Omega_0t),\\
\overline{n}_1(t):=\langle a_1^\dagger a_1^{\phantom{\dagger}}(t)\rangle&=1-\frac{t_{1,\textbf{e}_1}^2}{\Omega_0^2}\sin^2(\Omega_0t),\\
\overline{s}_x(t):=\langle \sigma^x_{1,{\bf e}_1}(t)\rangle&=\frac{2t_{1,\textbf{e}_1}^2}{\Omega_0^2}\sin^2(\Omega_0t)-1.\\
\end{split}
\eeq

\begin{figure}[t]
	\centering
	\includegraphics[width=0.75\columnwidth]{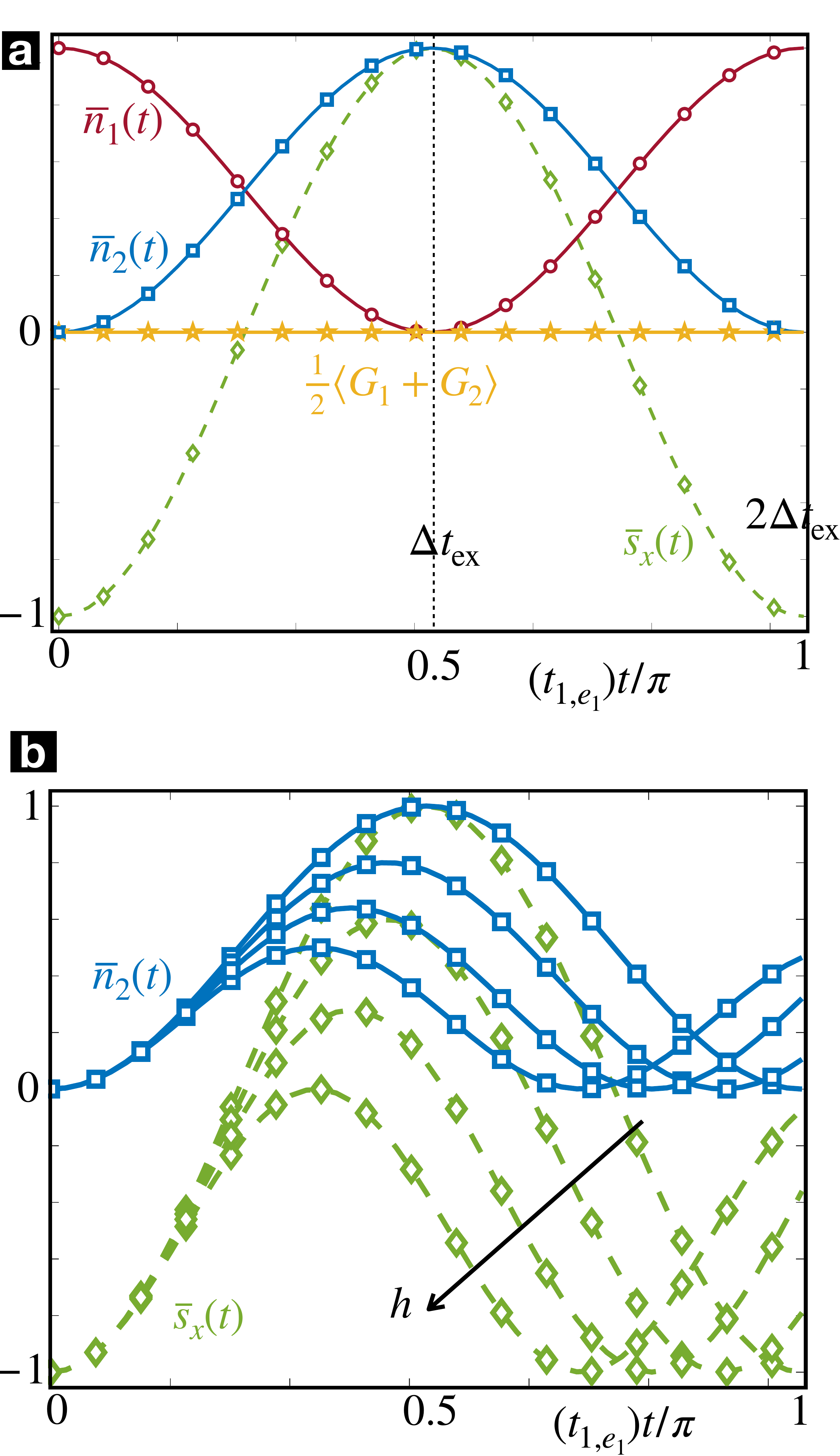}
	\caption{{\bf $\mathbb{Z}_2$-invariant tunnelling and correlated Rabi flopping:} {\bf (a)} Dynamics of an initial state $\ket{\Psi(0)}=\ket{\rm L}$ characterised by  the gauge invariant observables $\overline{n}_1(t)=\langle a_1^\dagger a_1^{\phantom{\dagger}}(t)\rangle$, $ \overline{n}_2(t)=\langle a_2^\dagger a_2^{\phantom{\dagger}}(t)\rangle$ and $\overline{s}_x(t)=\langle \sigma^x_{1,\textbf{e}_1}(t)\rangle$, as well as the averaged expectation value of the local-symmetry  generators  $\langle G_1(t)+G_2(t)\rangle/2$. The symbols correspond to the numerical evaluation in the full Hilbert space, whereas the lines display the analytical predictions for $h=0$~\eqref{eq_Rabi flops_link}.  {\bf (b)} Dynamics for the $\mathbb{Z}_2$ gauge link when the electric field $h$ is increased. The symbols correspond to the numerical simulations, and the lines to the corresponding analytical expressions~\eqref{eq_Rabi flops_link}. 
	}
	\label{fig:rabi_flopping_dynamics}
\end{figure}

In Fig.~\ref{fig:rabi_flopping_dynamics} {\bf (a)}, we compare these analytical predictions~\eqref{eq_Rabi flops_link} for $h=0$ to the numerical simulation for an initial state $\ket{\Psi(0)} = \ket{1_1}\ket{-_{1,\textbf{e}_1}}\ket{0_2}$. Note that, for  the  numerical simulation,  we do not restrict the Hilbert space to the single-particle subspace, nor to the gauge-invariant basis of Eq.~\eqref{eq:basis}. We truncate the maximal number of Fock states in each site to  $n_i\leq n_{\rm max}$, and compute the exact dynamics of the $\mathbb{Z}_2$-link Hamiltonian~\eqref{eq:tunneling_gauge} after this truncation, checking that no appreciable changes appear when increasing $n_{\rm max}$. The lines depicted in this figure represent the numerical results for matter and gauge observables $\overline{n}_1(t)=\langle a_1^\dagger a_1^{\phantom{\dagger}}(t)\rangle$, $ \overline{n}_2(t)=\langle a_2^\dagger a_2^{\phantom{\dagger}}(t)\rangle$ and $\overline{s}_x(t)=\langle \sigma^x_{1,\textbf{e}_1}(t)\rangle$. Fig.~\ref{fig:rabi_flopping_dynamics} {\bf (a)} also shows the expectation value of the sum of Gauss' generators~\eqref{eq:generators_link}  which, according to  the specific distribution of external background charges $q_1=0,q_2=1$, should vanish exactly at all times, i.e.,  $\langle G_1(t)+G_2(t)\rangle/2=(\ee^{\ii\pi q_1}+\ee^{\ii\pi q_2})/2=0$. The  symbols represent the respective analytical expressions in Eq.~\eqref{eq_Rabi flops_link}. The picture  shows a clear agreement of the numerical and  exact  solutions, confirming  the validity of  the picture of the correlated  Rabi flopping in the matter and gauge  sectors. As the boson tunnels to the right $\ket{1_1,0_2}\to\ket{0_1,1_2}$, the electric field line stretches to comply with Gauss' law until, right at the exchange duration, the  qubit gets flipped  $\ket{-_{1,\textbf{e}_1}}\to\ket{+_{1,\textbf{e}_1}}$. This behaviour is repeated periodically as the boson tunnels back and forth, and is a direct manifestation of gauge invariance.

From the two-level  scheme in the right panel of Fig.~\ref{fig:z2_link_tunneling_rabi_oscillations}, we see that a non-zero electric field  $h>0$ plays the role of a detuning in the Rabi problem~\cite{osti_7365050}. Accordingly, as the electric field gets stronger, i.e.  $h\gg |t_{1,\textbf{e}_1}|$, it costs more  energy to create an electric field line, and the particle ceases to tunnel, i.e. the contrast of the Rabi oscillations between the L/R levels diminishes (see Fig.~\ref{fig:rabi_flopping_dynamics} {\bf (b)}). It is worth comparing to the case of Peierls' phases and static/background gauge fields of Sec.~\ref{sec:Peierls:ions}. There, four modes were required  to define a plaquette and get an effective  flux that can lead to  Aharonov-Bohm destructive interference, which inhibits  the tunnelling of a single boson between the corners of the synthetic plaquette (see Fig.~\ref{fig:transverse_phonons_syntehtic_ladder} {\bf (d)}. In the case of the $\mathbb{Z}_2$ gauge model on a link, only two modes and a gauge qubit are required. The tunnelling of the boson is inhibited by increasing the energy cost of stretching/compressing the accompanying electric field line.  As discussed in more detail below, for larger lattices, this electric-field energy penalty  is responsible for the confinement of matter particles in this $\mathbb{Z}_2$ gauge theory, a characteristic feature of this type of discrete gauge theories~\cite{PhysRevX.6.041049,Gazit2017,doi:10.1073/pnas.1806338115, PhysRevB.97.245137, PhysRevX.10.041057, PhysRevB.102.155143,PhysRevD.102.074501,PhysRevLett.126.050401, PhysRevB.105.075132,PhysRevD.98.074503,PhysRevD.99.014503,PhysRevB.100.115152,PhysRevLett.124.120503,PhysRevLett.127.167203,Magnifico2020realtimedynamics,Surace_2021,PhysRevB.106.L041101,https://doi.org/10.48550/arxiv.2111.13205, https://doi.org/10.48550/arxiv.2208.07099,https://doi.org/10.48550/arxiv.hep-lat/0509045,PhysRevX.10.041007,PhysRevResearch.3.013133,PhysRevB.105.245105,https://doi.org/10.48550/arxiv.2208.04182}.

\subsection{Two-boson sector: Dark states and    entanglement between   modes of the matter fields }

\begin{figure}[t]
	\centering
	\includegraphics[width=0.95\columnwidth]{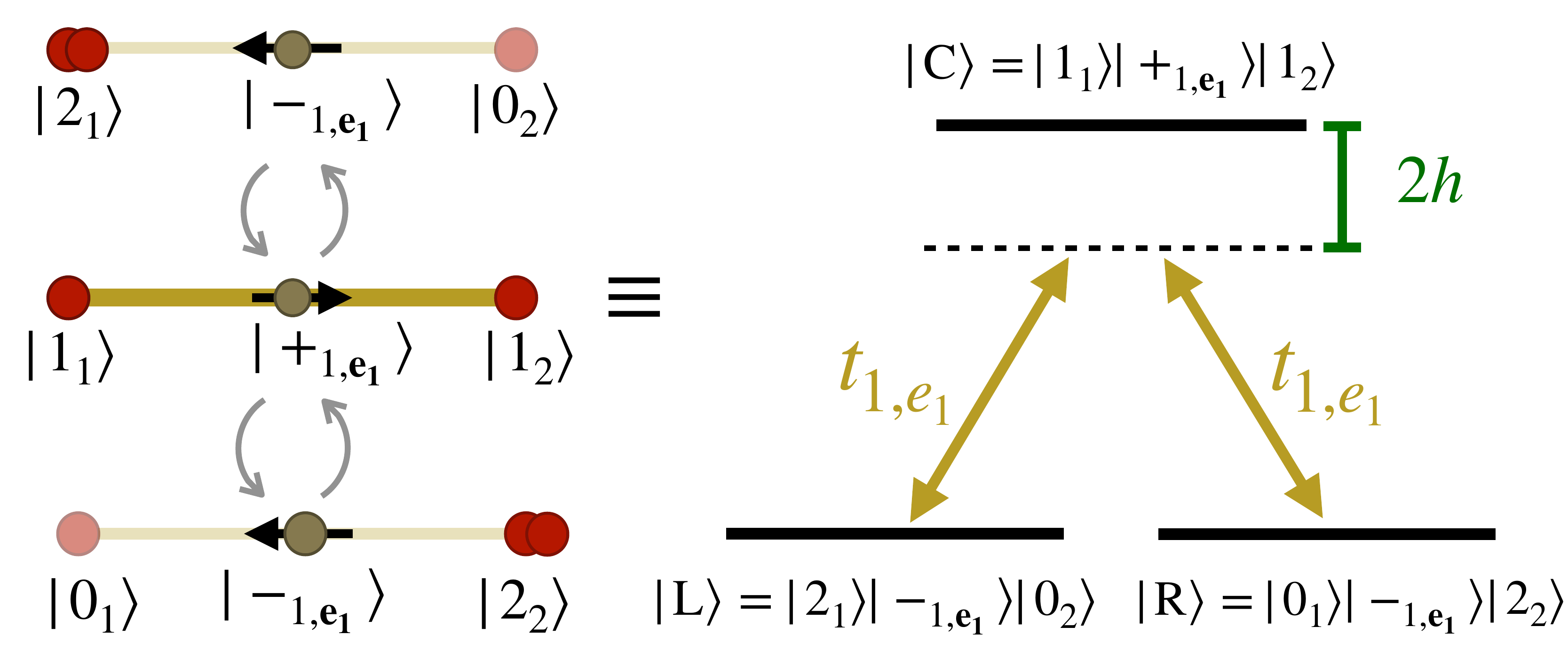}
	\caption{{\bf $\Lambda$-scheme for 2-boson  $\mathbb{Z}_2$-invariant tunnelling:}  In the left panel, we depict the three possible states in Eq.~\eqref{eq:2_boson_basis} for the distributions of the $\mathbb{Z}_2$ charges and electric field. In the right panel, we depict the quantum-optical level scheme, in which the gauge-invariant tunneling couples the $\ket{\rm L}$ and $\ket{\rm R}$ states to the state $\ket{\rm C}$ with one boson at each site, and a electric-field string in the link. The electric field $h$  {acts} as a detuning of these transitions, leading to a $\Lambda$-scheme. 
	}
	\label{fig:z2_link_tunneling_rabi_oscillations_2bosons}
\end{figure}

Let us now move to the two-particle case, and describe how the connection to well-known effects in quantum optics can be pushed further depending on the exchange statistics. 
A pair of  fermions  can only occupy the state  $\ket{1_1}\otimes\ket{+_{1,{\bf e}_1}}\otimes\ket{1_2}$, and do not display any dynamics due to the Pauli exclusion principle. On the other hand, if the particles are bosonic, the dynamics can be non-trivial and lead to interesting effects such as mode entanglement. Due to the $U(1)$ symmetry and Gauss' law~\eqref{eq:gauss_law_link} for $q_1=q_2=0$, the physical subspace can now be spanned by three different states
\beq
\label{eq:2_boson_basis}
\begin{split}
\ket{\rm L}&=\ket{2_1}\otimes\ket{-_{1,{\bf e}_1}}\otimes\ket{0_2}, \\
\ket{\rm C}&=\ket{1_1}\otimes\ket{+_{1,{\bf e}_1}}\otimes\ket{1_2}, \\
\ket{\rm R}&=\ket{0_1}\otimes\ket{-_{1,{\bf e}_1}}\otimes \ket{2_2}.
\end{split}
\eeq
A pair of charges sitting on the same site have a vanishing net $\mathbb{Z}_2$ charge $1\oplus1=(1+1){\rm mod}\, 2=0$, and  cannot act as a source/sink of electric field. Therefore, the  L and R states in Eq.~\eqref{eq:2_boson_basis} do not sustain any  electric field. On the other hand, when the pair of $\mathbb{Z}_2$ charges occupy the two different sites, Gauss' law imposes that an electric field line must be established at the link. Since creating this electric field costs energy, these three levels are then separated in energy by $2h$, and the gauge-invariant tunnelling  of the Hamiltonian in Eq.~\eqref{eq:tunneling_gauge} leads to a $\Lambda$-scheme in quantum optics (see Fig.~\ref{fig:z2_link_tunneling_rabi_oscillations_2bosons}).

As it is known to occur for  three-level atoms~\cite{Arimondo1976,ARIMONDO1996257}, one can find the so-called  bright $\ket{\rm B}=(\ket{\rm L}+\ket{\rm R})/\sqrt{2}$ and dark  $\ket{\rm D}=(\ket{\rm L}-\ket{\rm R})/\sqrt{2}$ states, which here correspond to the symmetric and anti-symmetric super-positions of the doubly-occupied sites at the left and right sites. In general, the state of the system can be expressed as a superposition of $\ket{\rm B}$, $\ket{\rm D}$ and $\ket{\rm C}$, namely $\ket{\Psi_{\rm phys}(t)}=d(t)\ket{\rm D}+c_{b}(t)\ket{\rm B}+ c_c(t)\ket{\rm C}$. However, as the dark state  decouples completely from the dynamics, its amplitude evolves by acquiring a simple phase  $d(t)=\ee^{\ii ht}d(0)$. Conversely,  the amplitudes of the remaining states mix and display periodic Rabi oscillations  $
\boldsymbol{c}(t)=\ee^{-\ii\tilde{\Omega}_0t\tilde{\boldsymbol{n}}\cdot\boldsymbol{\sigma}}\boldsymbol{c}(0)$ where,  in this case $\boldsymbol{c}(t) = (c_b(t),c_c(t))^{\rm t}$, and 
\beq
\tilde{\Omega}_0=\sqrt{4t_{1,\textbf{e}_1}^2+h^2},\hspace{2ex} \tilde{\boldsymbol{n}}=\frac{1}{\tilde{\Omega}_0}(2t_{1,\textbf{e}_1},0,h).
\eeq

We can now discuss a different manifestation of the gauge-invariant dynamics with respect to the single-particle case~\eqref{eq_Rabi flops_link}. Let us consider the initial state to be $\ket{\Psi_{\rm phys}(0)}=\ket{\rm C}$ with one boson at each site, and an electric-field line at the link in between. If we look at the local number of bosons, we do not observe any apparent dynamics  $\overline{n}_1(t):=\langle a_1^\dagger a_1^{\phantom{\dagger}}(t)\rangle=1=\langle a_2^\dagger a_2^{\phantom{\dagger}}(t)\rangle=:\overline{n}_2(t)$. However, looking into the electric field at the link, we find periodic Rabi flopping  again, i.e.
\beq
\label{eq_Rabi flops_link_2_exc}
\overline{s}_x(t):=\langle \sigma^x_{1,{\bf e}_1}(t)\rangle=1-\frac{8t_{1,\textbf{e}_1}^2}{\tilde{\Omega}_0^2}\sin^2(\tilde{\Omega}_0t).
\eeq
Since the gauge field cannot have independent oscillations with respect to the matter particles, there must be a non-trivial dynamics within the matter sector which, nonetheless, cannot be inferred by looking at the local number of particles. In this context, it is  the  interplay of the superposition principle of quantum mechanics and gauge symmetry, which underlies a neat dynamical effect. This effect becomes manifest by inspecting the state after a single exchange period $ \Delta t_{\rm ex}=\pi/2\tilde{\Omega}_0$ for $h=0$.
After this time, a boson can either tunnel to the left or to the right. In both cases, the electric field string compresses, since a doubly-occupied site amounts to a vanishing net $\mathbb{Z}_2$  charge, and  there is thus no sink/source of electric field. Accordingly, when the bosons tunnel along either path respecting gauge invariance, the state ends up with the same link configuration, namely $\ket{-}_{1,\textbf{e}_1}$. Then, according to the superposition principle, both paths must be added, and  the state of the system at time $t_e$ is given by
\beq
\label{eq:noon}
\ket{\Psi_{\rm phys}(\Delta t_{\rm ex})}=\frac{1}{\sqrt{2}}\left(\ket{2_1,0_2}+\ket{0_1,2_2}\right)\otimes\ket{-_{1,{\bf e}_1}}.
\eeq
We see that, as a consequence of the dynamics,   mode entanglement~\cite{BENATTI20201} has been generated in the matter sector since the state  cannot be written as a separable state $\ket{\Psi_{\rm phys}(\Delta t_{\rm ex})}\neq P(a_1^\dagger)Q(a_{2}^\dagger)\ket{0_1,0_2}\otimes\ket{-_{1,{\bf e}_1}}$ for any polynomials $P, Q$. The specific state (Eq.~\eqref{eq:noon}) in the matter sector is a particular type of NOON states, which have been studied in the context of metrology~\cite{PhysRevA.40.2417,PhysRevA.54.R4649,PhysRevLett.85.2733,doi:10.1080/0950034021000011536,Giovannetti2011}. Note that this state cannot be distinguished from the initial state if one only looks at  $\overline{n}_1(t)=\overline{n}_2(t)=1$. The non-trivial dynamics becomes  manifest via the link  and the quantum mode-mode correlations.

\begin{figure}[t]
	\centering
	\includegraphics[width=0.75\columnwidth]{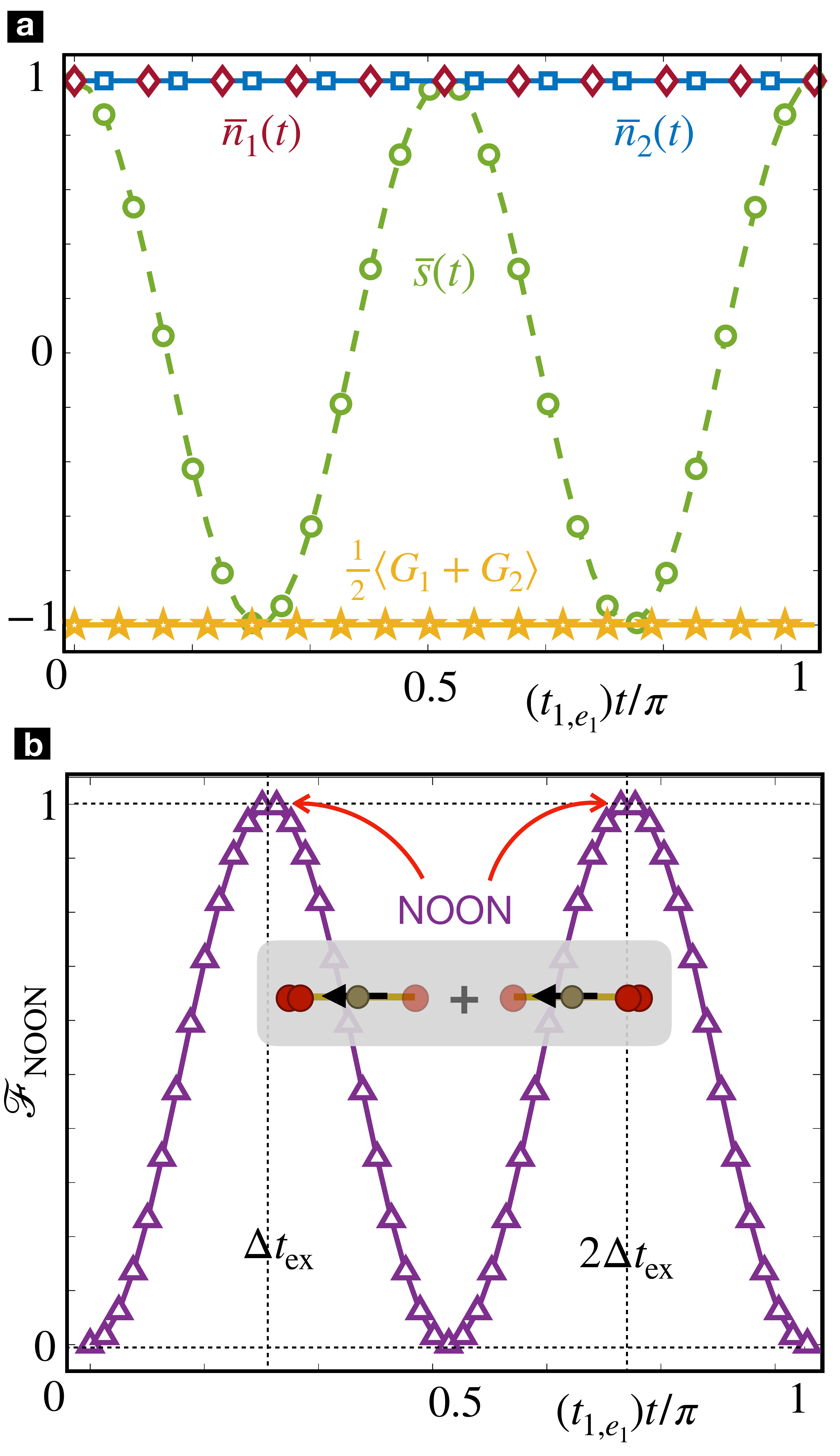}
	\caption{{\bf Two-boson $\mathbb{Z}_2$-invariant tunneling and  entanglement:} {\bf (a)} Dynamics of an initial state $\ket{\Psi(0)}=\ket{\rm L}$ characterised by  the gauge invariant observables $\overline{n}_1(t)=\langle a_1^\dagger a_1^{\phantom{\dagger}}(t)\rangle$, $ \overline{n}_2(t)=\langle a_2^\dagger a_2^{\phantom{\dagger}}(t)\rangle$ and $\overline{s}_x(t)=\langle \sigma^x_{1,\textbf{e}_1}(t)\rangle$, as well as the averaged expectation value of the local-symmetry  generators  $\langle G_1(t)+G_2(t)\rangle/2$.   The symbols correspond
    to the numerical evaluation in the full Hilbert space, whereas the lines display the analytical predictions for $h = 0$. {\bf (b)} State fidelity with respect to a mode-entangled 2-boson NOON state~\eqref{eq:noon}, which tends to unity at the integer exchange periods $\Delta t_{\rm ex}$, and $2\Delta t_{\rm ex}$. 
	\label{fig:noon_flopping_dynamics}}
\end{figure}

In Fig.~\ref{fig:noon_flopping_dynamics}, we present a comparison of the analytical predictions with the corresponding  numerical results where, once more, we do not restrict to the basis in Eq.~\eqref{eq:2_boson_basis}, nor to the 2-boson subspace. We initialise the system in $\ket{\Psi(0)}=\ket{1_1}\ket{+_{1,\textbf{e}_1}}\ket{1_2}$, and numerically truncate the Hilbert space such, that $n_i\leq n_{\rm max}$,  computing numerically  the Schr\"{o}dinger dynamics for  the $\mathbb{Z}_2$-link Hamiltonian~\eqref{eq:tunneling_gauge}. In Fig.~\ref{fig:noon_flopping_dynamics}{\bf (a)}, we represent these numerical results with  lines for the observables $\overline{n}_1(t),\overline{n}_2(t),\overline{s}_x(t)$, as well as the average of the local-symmetry  generators $\langle G_1(t)+G_2(t)\rangle/2$. Once again, these numerical results agree perfectly with the corresponding analytical expressions~\eqref{eq_Rabi flops_link_2_exc}, which are represented by the symbols. In Fig.~\ref{fig:noon_flopping_dynamics}{\bf (b)}, we show the fidelity of the system state with respect the NOON state of Eq.~\eqref{eq:noon}, namely $\mathcal{F}_{\rm NOON}(t)=|\bra{\Psi_{\rm phys}(t_{e})}\ee^{\-\ii tH_{\rm eff}}\ket{\Psi(0)}|^2$ at different evolution times. We see that this fidelity tends to unity at the periodic exchange periods $t=m t_{e},\,\,m\in\mathbb{Z}^+$. Note that the timescale in the horizontal axis is the same as the one for the one-particle case in Fig.~\ref{fig:rabi_flopping_dynamics}, but the periodic oscillations are twice as fast. This two-fold speed-up is caused by the bosonic enhancement due to the presence of two particles in the initial state, providing a $\sqrt{2}$ factor, and the enhancement due to the bright state, which brings the additional $\sqrt{2}$ factor. This total speed-up is the only difference if one compares the dynamics of the bosonic sector with that of a standard beam splitter leading to the Hong-Ou-Mandel interference~\cite{PhysRevLett.59.2044}. In fact, in the trapped-ion literature, the bare tunneling terms between a pair of vibrational modes are commonly referred to as a beam splitter due to the formal analogy with the optical device that splits an incoming light mode  into the transmitted and  reflected modes~\cite{PhysRevLett.89.247901}.


After the results presented in this section, we can more to the discussion of  two possible  schemes for implementing the state-dependent parametric excitation~\eqref{eq:parametric_tunneling_state_dep} in trapped-ion experiments.
The first scheme (I) is based on trapped-ion analog quantum simulators that generalise  straightforwardly from Eq.~\eqref{eq:ac_beams}. The second scheme (II) exploits recent ideas developed for continuous-variable quantum computing~\cite{sutherland2021universal}.
\section{\bf Trapped-ion toolbox:   phonons and  qubits}
\label{sec:trapped-ion_toolbox}

Before delving into the details of the trapped-ion schemes, let us first review the progress of trapped-ion-based QSs for lattice gauge theories. As discussed in~\cite{Martinez2016, Muschik_2017, PhysRevResearch.2.023015}, certain gauge theories can be mapped exactly onto spin models that represent the fermionic matter with effective long-range interactions mediated by the gauge fields. Following these ideas, the $U(1)$ Schwinger model of quantum electrodynamics in 1+1 dimensions~\cite{Martinez2016,PRXQuantum.3.020324} and variational quantum eigensolvers~\cite{Kokail2019} have been simulated digitally in recent trapped-ion experiments. As discussed in~\cite{ PhysRevResearch.2.023015,PRXQuantum.2.030334}, there are theoretical proposals to generalise this approach to gauge theories in 2+1 dimensions. Although not considered in the specific context of trapped ions, digital quantum simulators, and variational eigensolvers have also been recently considered for  $\mathbb{Z}_2$ gauge theories~\cite{PhysRevD.103.054507,PRXQuantum.3.020320,PhysRevLett.129.051601,https://doi.org/10.48550/arxiv.2203.08905}. 
Rather than eliminating the gauge fields as in the cases above~\cite{Martinez2016, Muschik_2017, PhysRevResearch.2.023015}  one could consider the opposite, and obtain effective models for the gauge fields after eliminating the matter content~\cite{https://doi.org/10.48550/arxiv.2206.00685,https://doi.org/10.48550/arxiv.2206.08909}.

In order to move beyond those specific models, it would be desirable to simulate matter and gauge fields on the same footing. Trapped-ion schemes for the quantum-link approach to the Schwinger model have been proposed in~\cite{PhysRevX.3.041018,Andrade_2022}. In particular, for the specific spin-$1/2$ representation of the link operators, the gauge-invariant tunneling becomes a three-spin interaction. This could be implemented using only the native two-spin interactions in trapped-ion experiments and by imposing an additional energetic Gauss penalty~\cite{PhysRevX.3.041018}. Alternatively, one may also generate three-spin couplings~\cite{Andrade_2022} directly by exploiting second-order sidebands that use the phonons as carriers of these interactions~\cite{PhysRevA.79.060303,PhysRevLett.129.063603,https://doi.org/10.48550/arxiv.2209.05691}. We note that there have also been other proposals~\cite{PhysRevA.94.052321,PhysRevResearch.3.043072} to use the motional modes to encode the $U(1)$ gauge field, whereas the fermionic matter is represented by spin-$1/2$ operators. In this case, the gauge-invariant tunneling can be achieved via other second-sideband motional couplings~\cite{PhysRevA.94.052321}, or by combining  digital and analog ingredients in a ``hybrid'' approach~\cite{PhysRevResearch.3.043072}. A different  possibility is to use the collective motional modes to simulate bosonic matter and reserve the spins to represent the quantum link operators for the gauge fields~\cite{PhysRevA.94.052321}. In this way, one can simulate a quantum link model provided that all the collective vibrational modes can be individually addressed in frequency space~\cite{PhysRevA.94.052321}, which can be complicated by frequency crowding as the number of ions increases. We note that engineering the collective-motional-mode couplings has also been recently considered in the context of continuous-variable quantum computing, boson sampling, and quantum simulation of condensed-matter models~\cite{https://doi.org/10.48550/arxiv.2205.14841,PhysRevA.104.032609,https://doi.org/10.48550/arxiv.2207.13653,Chen2023}. In the following, we present a trapped-ion scheme for the quantum simulation of $\mathbb{Z}_2$ gauge theories based on our previous idea of a state-dependent parametric tunneling~\eqref{eq:parametric_tunneling_state_dep}, and using motional states along two different transverse directions, and a pair of electronic states to encode the particles and the gauge field.

\subsection{  Analog scheme for the  
$\mathbb{Z}_2$ gauge link}
\label{sec:scheme_I}

\subsubsection{Light-shift-type parametric tunneling}
\label{subsec:ls_impl}
In Appendix \ref{sec:Peierls:ions}, we discuss how a parametric excitation between the two local transverse vibrations in an ion chain could be synthesised by exploiting an optical potential~\eqref{eq:ac_beams} created by a far-detuned two-beam laser field. In order to achieve this, we considered that all of the ions were initialised in the same ground-state level. However, depending on the nuclear spin of the ions, the ground-state manifold can contain a variety of levels $\{\ket{s}$\} that can be used to obtain the state-dependent parametric tunneling of Eq.~\eqref{eq:parametric_tunneling_state_dep}. In general, when the field is far-detuned from any  atomic transition,  
the  light-shift potential  becomes state-dependent~\cite{3981}, namely 
\beq
\label{eq:ac_beams_st_dep}
V_1(t)=\sum_{n,n'=1,2}\sum_{i,s\phantom{l}}\Omega_{n,n'}^{(s)}\ket{s_i}\bra{s_i}\ee^{\ii(\boldsymbol{k}_{{\rm L},n}-\boldsymbol{k}_{{\rm L},n'})\cdot\boldsymbol{r}_i-(\omega_{{\rm L},n}-\omega_{{\rm L},n'})t}+{\rm H.c.}.
\eeq
Here, $\Omega_{n,n'}^{(s)}$ is the amplitude of the light-shift terms discussed after Eq.~\eqref{eq:ac_beams}, in which the corresponding Rabi frequencies now refer to the particular ground-state level $\ket{s_i}$ of the $i$-th ion involved in the two virtual transitions. 
These light shifts then depend on the specific state and the intensity and polarisation of the laser fields. As discussed in the context of state-dependent dipole forces~\cite{3981,leibfried2003experimental,doi:10.1098/rsta.2003.1205}, one can focus on a particular pair of states $s_1,s_2$ and tune the polarisation, detuning and intensity of the light, such that the corresponding amplitudes for the crossed beat note terms  attain a  differential value~\cite{leibfried2003experimental,doi:10.1098/rsta.2003.1205}. 
To obtain the local gauge symmetry, it is important that this amplitude is equal in absolute value but opposite in sign for each of the two electronic states 
\beq
\label{eq:oposs_shifts} 
\Omega_{1,2}^{(0)}=-\Omega_{1,2}^{(1)}.
\eeq
If this condition is not satisfied, one can still obtain a state-dependent tunneling, but this would not have the desired local gauge invariance under the above $\mathbb{Z}_2$  group~\eqref{eq:gauss_law_link}. Nonetheless, such state-dependent tunneling can be interesting for other purposes in the context of hybrid discrete-continuous variable quantum information processing, as realised in~\cite{gan_2020}.

One can now follow  the same steps  as in Sec.~\ref{sec:state_dep_param_tunn}, introducing the local transverse phonons via the position operators, 
and performing a Lamb-Dicke expansion discussed in the Appendix. Using the same set of constraints as for the  {standard} parametric drive, i.e. Eq.~\eqref{eq:constraints_param_tunneling_ions}, we find that 
\beq
\label{eq:state_dep_parametric_ions}
V_1(t)\approx\sum_i\Delta E_{\rm ac}\sigma_{i}^z+\sum_{i}\Omega_{\rm d}\sigma_{i}^z\cos(\phi_i-\omega_{\rm d}t)a^\dagger_{i,y}a^{\phantom{\dagger}}_{i,x}+{\rm H.c.},
\eeq
where we have introduced $\omega_{\rm d}=\omega_{{\rm L},1}-\omega_{{\rm L},2}$, together with 
\beq
\label{eq:parameters_parametric_ions_main_text}
\Omega_{\rm d}=|\Omega_{1,2}|\eta_{x}\eta_y, \hspace{2ex}\phi_i=\boldsymbol{k}_{\rm d}\cdot\boldsymbol{r}_i^0+{\arg}(-\Omega_{1,2}).
\eeq
where we have introduced
\beq
\label{eq:lamb_dicke_regime_main}
\boldsymbol{k}_{\rm d}=\boldsymbol{k}_{{\rm L},1}-\boldsymbol{k}_{{\rm L},2},\hspace{2ex} \eta_\alpha={\boldsymbol{k}_{\rm d}\cdot\textbf{e}_\alpha}/{\sqrt{2m\omega_{\alpha}}}.
\eeq

\begin{figure}[t]
	\centering
	\includegraphics[width=0.85\columnwidth]{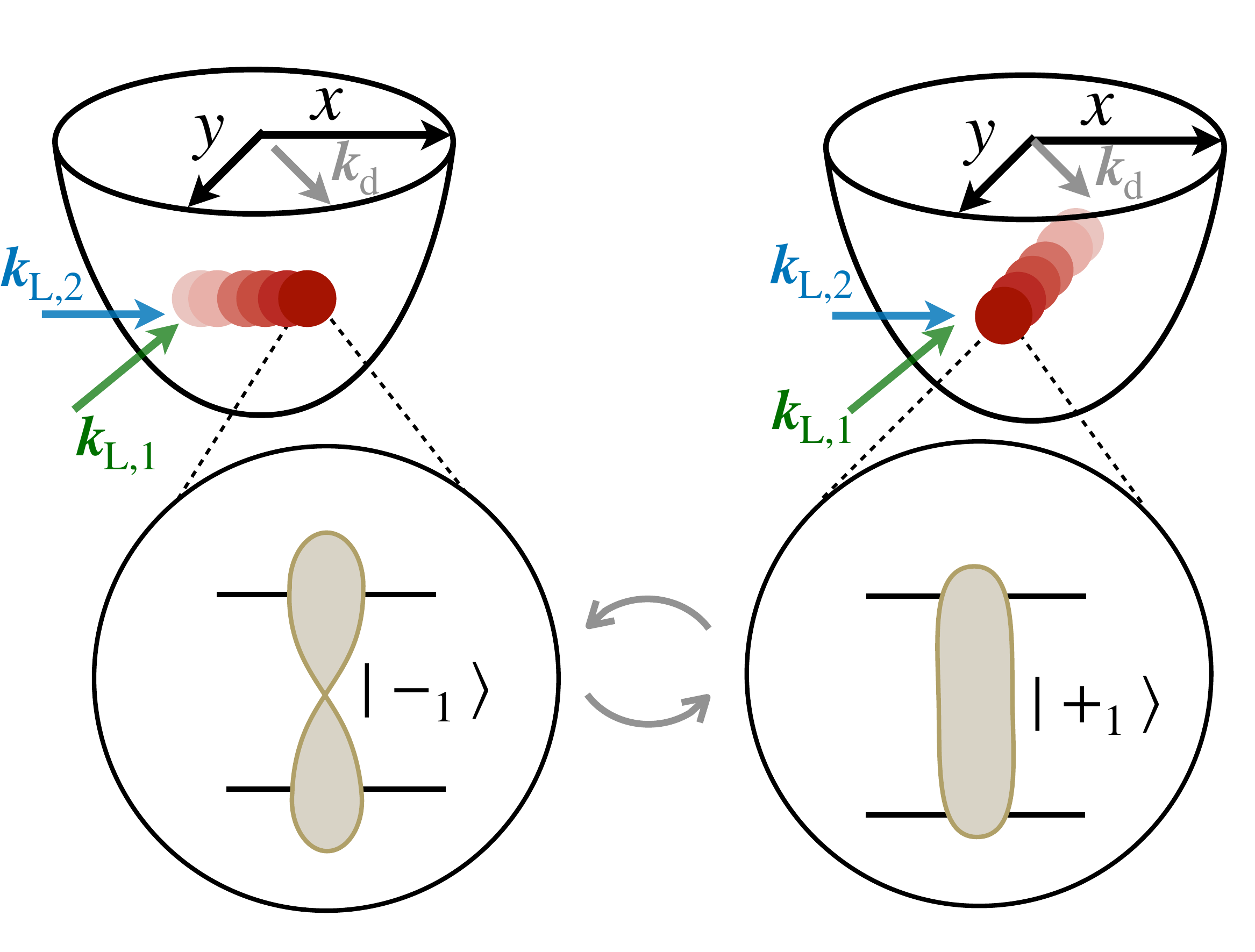}
	\caption{{\bf Trapped-ion synthetic $\mathbb{Z}_2$ gauge theory on a link:} Schematic representation of the  single-ion system that can realise the $\mathbb{Z}_2$ gauge theory on a synthetic link~\eqref{eq:tunneling_gauge}. On the left, we depict an ion vibrating in the transverse $x$ direction, and the inset represents the state of the corresponding qubit in $\ket{-_1}=(\ket{\uparrow}_1-\ket{\downarrow_1})/\sqrt{2}$. On the right, we can see how,  as a consequence of the trapped-ion effective Hamiltonian~\eqref{eq:link_ions}, the vibrational excitation along $x$ is transferred into a vibrational excitation along $y$, while simultaneously flipping the qubit into $\ket{+_1}=(\ket{\downarrow}_1+\ket{\uparrow_1})/\sqrt{2}$. This dynamics, which is fully consistent with the local gauge symmetry, can be engineered by shining a far-detuned two-beam laser field with wave-vectors associated to each frequency represented by green and blue arrows, leading to a beat note along the grey arrow that yields the desired term~\eqref{eq:link_ions}. }
	\label{fig:z2_link_ions}
\end{figure}
We are already close to the idealised situation in which the effective model would lead to a gauge-invariant tunneling to a full ion chain. 
However,  a simple counting argument shows that the effective model cannot achieve $\mathbb{Z}_2$ gauge invariance. For a string of $N$ ions, we have $2N$ local motional modes along the transverse directions, which lead to the synthetic two-leg ladder with  $3N-2$ links,  each of which would requires a gauge qubit to achieve a local $\mathbb{Z}_2$ symmetry. Since we only have $N$ trapped-ion qubits at our disposal, i.e. one qubit per ion, it is not possible to build a gauge-invariant model for the synthetic ladder in a straightforward manner. In  Sec.~\ref{sec:dim_reduction}, we present  a solution to this problem by introducing a mechanism that we call synthetic dimensional reduction. 
 
Prior to that, we can discuss the minimal case in which gauge invariance can be directly satisfied in the trapped-ion experiment: a single $\mathbb{Z}_2$ link. This link requires a single ion: one gauge qubit for the link, and two motional modes for the matter particles, which can  be the vibrations along any of the axes. In  Fig.~\ref{fig:z2_link_ions}, we consider the two transverse modes, and thus restrict  Eq.~\eqref{eq:state_dep_parametric_ions} to a single ion. Following the same steps as in the derivation of Eq.~\eqref{eq:tunneling_gauge}, we move to an interaction picture and neglect rapidly rotating terms under the conditions of Eq.~\eqref{eq:constraints_param_tunneling_ions}. We obtain a time-independent term that corresponds to a $\mathbb{Z}_2$ gauge-invariant tunneling 
\beq
\label{eq:link_ions}
V_{1}(t)\approx 
\frac{\Omega_{\rm d}}{2}\ee^{\ii\phi_{\rm 1}}a^{{\dagger}}_{{1},y}\sigma^{z\phantom{\dagger}\!\!\!\!}_{1}a^{\phantom{\dagger}}_{1,x}+{\rm H.c.}, 
\eeq
At this point, the driving phase $\phi_{\rm d}=\phi_1$ is  irrelevant and can be set to zero without loss of generality. Identifying  the trapped-ion operators with those of the lattice gauge theory
\beq
\label{eq:guage_matter_mapping}
\begin{split}
a^{\phantom{\dagger}}_{1,x},a^\dagger_{1,x} &\mapsto  a^{\phantom{\dagger}}_1,a_1^\dagger,\\ a^{\phantom{\dagger}}_{1,y},a^\dagger_{1,y} &\mapsto  a^{\phantom{\dagger}}_2,a_2^\dagger,\\ \sigma_{1}^x,\sigma_{1}^z \hspace{2ex}&\mapsto \sigma_{1,{\bf e}_1}^x,\sigma_{1,{\bf e}_1}^z,
\end{split}
\eeq
we obtain a realisation of the  $\mathbb{Z}_2$ gauge-invariant tunneling  on a link~\eqref{eq:tunneling_gauge} using a single trapped ion, such that
\beq
\label{eq:eff_tunneling_light_shift}
t_{1,{\bf e}_1}=\frac{\Omega_{\rm d}}{2}=\frac{|\Omega_{1,2}|}{2}\eta_x\eta_y.
\eeq
As explained above, this exploits the qubit as the gauge field, and two vibrational modes to host the $\mathbb{Z}_2$-charged  matter.  In the next subsection, we  present alternative schemes that do not depend on this condition for the differential light shift.

Let us now test the validity of this scheme for realistic trapped-ion parameters, considering a $^{88}\mathrm{Sr}^+$ ion confined in the setup of Refs.~\cite{schafer2018a,thirumalai2019high}, and a ground state-qubit encoding with  experimentally-realistic parameters  described  in Appendix~\ref{app:dp_ls}. Note that for a single $\mathbb{Z}_2$ link, any two of the three motional modes can be used. In the following simulations, with the trapped-ion parameters of the considered setup, we encode the matter particles in an axial ($z$) and a transverse ($x$) mode, such that we can benefit from the larger Lamb-Dicke parameter of the axial mode, as well as the higher frequency separation of the two motional modes. We model numerically the possible deviations of a realistic trapped-ion implementation from the above-idealised expressions used in Fig.~\ref{fig:rabi_flopping_dynamics}. 
For the simulations presented below, we perform exact numerical integration of the Schr\"{o}dinger equation under the time-dependent trapped-ion Hamiltonians  using the QuantumOptics.jl package in Julia~\cite{kramer2018quantumoptics}. In particular, we consider  the full Hamiltonian \eqref{eq:ac_beams_st_dep}, using experimentally feasible parameters, and do not assume the Lamb-Dicke expansion, and thus include possible off-resonant carrier excitations as well as other non-linear terms neglected in Eq.~\eqref{eq:state_dep_parametric_ions}. We use a single ion and two of its motional modes, truncating their individual Hilbert spaces at phonon number $n_{\rm max}=7$.

\begin{figure}[t]
	\centering
	\includegraphics[width=0.9\columnwidth]{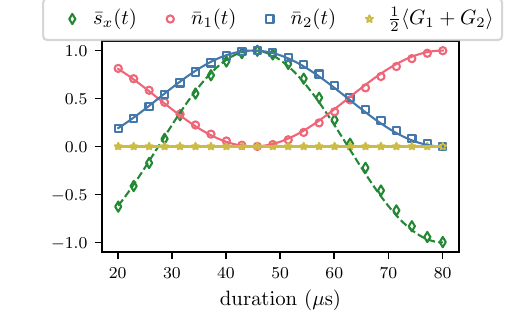}	\caption{\textbf{$\mathbb{Z}_2$ gauge  dynamics  with analog trapped-ion scheme}: We simulate the $\mathbb{Z}_2$ dynamics using the Raman-based light-shift parametric tunneling and try to replicate Fig.~\ref{fig:rabi_flopping_dynamics}~{\bf (a)}. The markers are full numerical simulations including non-linear terms~\eqref{eq:ac_beams_st_dep} beyond the desired Lamb-Dicke expansion, while the continuous lines are analytical predictions in Eq.~(\ref{eq_Rabi flops_link}) using the effective tunneling strength~\eqref{eq:eff_tunneling_light_shift} and $h=0$. The adiabatic pulse shaping sets the minimum pulse duration to the rising and falling edge ($\SI{10}{\micro\second}$ each).}
	\label{fig:scheme_1_BS_time_scan}
\end{figure}

 The results are  shown in Fig.~\ref{fig:scheme_1_BS_time_scan}, where the coloured lines  represent the analytical predictions for the various observables in Eq.~\eqref{eq_Rabi flops_link} using the effective tunneling strength~\eqref{eq:eff_tunneling_light_shift}. The coloured symbols stand for  the full
numerical simulations including non-linear terms in the trapped-ion case, leading thus to a trapped-ion counterpart  of Fig.~\ref{fig:rabi_flopping_dynamics}~{\bf (a)} following Scheme~I. We note that, in order to find a better agreement with the idealised evolution~(\ref{eq_Rabi flops_link}), we have incorporated an adiabatic pulse shaping of the light-matter coupling that restricts the minimal duration of the real-time dynamics as discussed in the caption of Fig.~\ref{fig:scheme_1_BS_time_scan}. For the specific choice of parameters detailed in Appendix~\ref{app:dp_ls}, we see that the exchange duration of the phonon tunneling  and the stretching of the electric-field line is about several tens of $\mu$s, which is sufficiently fast compared to other possible sources of noise such as heating and dephasing, as discussed in more detail below. In  Appendix~\ref{app:dp_ls}, we  make a more detailed error analysis, distinguishing  those that arise from non-linearities or from non-resonant corrections. As discussed in Appendix~\ref{app:qdp_ls},  the analog scheme can also be accomplished with optical qubits. 

\subsubsection{M\o lmer-S\o rensen-type parametric tunneling}\label{sec:MS_impl}

In this section, we present an alternative scheme for synthesising the gauge-invariant tunneling that doe not require a fine tuning of the differential ac-Stark shifts~\eqref{eq:oposs_shifts} to  {achieve} the desired gauge invariance and, moreover,  leads to a technically simpler method to induce the electric-field term. In this case, the parametric tunneling  arises from a ``bichromatic'' field that is no longer far-detuned from the qubit transition but, instead, has two components symmetrically detuned from the qubit frequency, which connects to the M\o lmer-S\o rensen(MS) scheme used for high-fidelity trapped-ion gates~\cite{PhysRevLett.82.1971,PhysRevLett.82.1835,Soerensen2000}. The main difference of our scheme is that the bichromatic field is not tuned to first sidebands, but to the frequency conversion between the two motional modes.

As discussed in more detail in Appendix~\ref{app:MS_scheme}, either for ground state or optical qubits, the MS-type scheme leads to the following term instead of  Eq.~\eqref{eq:ac_beams_st_dep}, 
\begin{equation}
\label{eq:int_term}
   \tilde{V}_1(t) = \Omega \cos(\delta t)\sum_i\ket{\uparrow_i}\bra{\downarrow_i} \ee^{\ii\boldsymbol{k}_{\rm d}\cdot\boldsymbol{r}_i} + {\rm H.c.}.
\end{equation}
 As in our previous derivation, we expand 
  in the Lamb-Dicke parameters assuming Eq.~\eqref{eq:lamb_dicke_regime}. By focusing again on  a single trapped ion, and choosing the detuning  to be resonant with  $\delta= \omega_x -\omega_z$, we reach the frequency conversion. 
We  refer to this scheme as a  M\o lmer-S\o rensen(MS) parametric tunneling
\begin{equation}
	\tilde{V}_1(t) \approx \frac{\Omega_{\rm d}}{2} {a}_{1,z}^\dagger {\sigma}_1^{x\phantom{\dagger}\!\!\!\!} {a}_{1,x}^{\phantom{\dagger}}+{\rm H.c.}, \hspace{1ex} \Omega_{\rm d} = \Omega\eta_x \eta_z.
 \label{eq:MS_second_order_only}
\end{equation}
The  gauge-invariant tunneling rate~\eqref{eq:tunneling_gauge} then reads
\beq
\label{eq:eff_tunneling_MS}
t_{1,{\bf e}_1}=\frac{\Omega_{\rm d}}{2}=\frac{|\Omega|}{2}\eta_x\eta_z.
\eeq
which is analogous to the light-shift case of Eq.~\eqref{eq:eff_tunneling_light_shift}.
We thus obtain a rotated version  of the  gauge-invariant tunneling in Eq.~\eqref{eq:link_ions}, the only difference being that the operators need to be transformed as $\sigma_1^z\mapsto\sigma_1^x$, and $a_{1,y}^{\phantom{\dagger}}, a_{1,y}^{\dagger}\mapsto a_{1,z}^{\phantom{\dagger}}, a_{1,z}^{\dagger}$, which must also be considered in the mapping to the operators of the lattice gauge theory in Eq.~\eqref{eq:guage_matter_mapping}.  Accordingly, the generators of the local symmetries now read
\beq
\label{eq:generators_link_1}
G_1=\ee^{\ii\pi a_1^\dagger a_1^{\phantom{\dagger}}}\sigma_{1,{\bf e}_1}^z, \hspace{2ex}  G_2=\sigma_{1,{\bf e}_1}^z\ee^{\ii\pi a_2^\dagger a_2^{\phantom{\dagger}}}.
\eeq
In  Appendix~\ref{app:MS_scheme}, we  provide a more detailed error analysis of the validity of this scheme using  {realistic} experimental parameters for $^{88}\mathrm{Sr}^+$ ions~\cite{schafer2018a,thirumalai2019high}, and include figures that support  the validity of the MS scheme at the same level as the previous dipole light-shift one. The advantage will become  {apparent} in the following section.

\subsubsection{Electric field and experimental considerations }

So far, we have  {restructured} to the gauge-invariant tunneling in Eq.~\eqref{eq:tunneling_gauge}, but we also need a term that drives the qubit transition~\eqref{eq:second_tone} with Rabi frequency $\tilde{\Omega}_{\rm d}$, which corresponds to the electric field $h=\tilde{\Omega}_{\rm d}/2$ in Eq.~\eqref{eq:link_couplings}. The technique to induce this term  depends on the specific scheme. For the scheme based on the light-shift potential (Sec.~\ref{subsec:ls_impl}), one needs to add a field driving the qubit transition resonantly. For an optical qubit, this term would arise from a resonant laser driving the quadrupole-allowed transition. On the other hand, if the qubit is encoded in the ground state, this term can be induced by either a resonant microwave field or a pair of Raman laser beams. In both cases, trapped-ion experiments routinely work in the regime of Eq.~\eqref{eq:constraints_carrier}, where the value of $\tilde{\Omega}_{\rm d}$ can be controlled very precisely by tuning the amplitude of the laser or microwave field~\cite{doi:10.1063/1.5088164}. Note that the resonance condition~\eqref{eq:constraints_carrier} must  account for the ac-Stark shifts shown in Eq.~\eqref{eq:state_dep_parametric_ions}, namely
\beq
\label{eq:constraints_carrier_ac_stark}
\tilde{\omega}_{\rm d}=\omega_0+2\Delta E_{\rm ac}, \hspace{2ex} |\tilde{\Omega}_{\rm d}|\ll 4(\omega_0+2E_{\rm ac}).
\eeq
This leads to the desired Hamiltonian 
\beq
\label{eq:link_ions_h}
V_{1}(t)\approx \left(
\frac{\Omega_{\rm d}}{2}\ee^{\ii\phi_{\rm d}}a^{{\dagger}}_{{1},z}\sigma^{z\phantom{\dagger}\!\!\!\!}_{1}a^{\phantom{\dagger}}_{1,x}+{\rm H.c.}\right)+\frac{\tilde{\Omega}_{\rm d}}{2}\sigma_1^x,
\eeq
which maps directly onto the  $\mathbb{Z}_2$ gauge link in Eqs.~\eqref{eq:tunneling_gauge}-\eqref{eq:link_couplings} with the new term playing the role of the electric field 
\beq
\label{eq:el_field_LS}
h=\frac{\tilde{\Omega}_{\rm d}}{2}.
\eeq

For the M{\o}lmer-S{\o}rensen-type scheme discussed in Sec.~\ref{sec:MS_impl}, the spin conditioning of the tunneling occurs in the Hadamard basis $\ket{\pm_1}$, such that the effective electric field must also be rotated with respect to Eq.~\eqref{eq:tunneling_gauge}. We can introduce this term by simply shifting the centre frequency of the bichromatic laser field~\eqref{eq:int_term} relative to the qubit resonance by a detuning $\delta_{\rm s}$. In a rotating frame, this modifies Eq.~\eqref{eq:int_term} by introducing an additional term, namely
\begin{equation}
	\tilde{V}_1(t) \approx \left(\frac{\Omega_{\rm d}}{2} {a}_{1,z}^\dagger {\sigma}_1^{x\phantom{\dagger}\!\!\!\!} {a}_{1,x}^{\phantom{\dagger}}+{\rm H.c.}\right)+\frac{\delta_{\rm s}}{2} {\sigma}_1^{z\phantom{\dagger}\!\!\!\!},
 \label{eq:ms-z2}
\end{equation}
which leads to the effective electric-field term 
\beq
\label{eq:el_field_MS}
h=\frac{\delta_{\rm s}}{2}.
\eeq
This is a considerable advantage with respect to the light-shift scheme, as no additional tones are required to implement the electric field term. A detailed analysis of the errors for current trapped-ion parameters is presented in Appendix~\ref{app:MS_scheme}.

\subsection{ Pulsed scheme for the $\mathbb{Z}_2$ gauge link}\label{subsec:pulsed_scheme}

\subsubsection{Orthogonal-force parametric tunneling}

\begin{figure}[t]
	\centering
	\includegraphics[width=0.9\columnwidth]{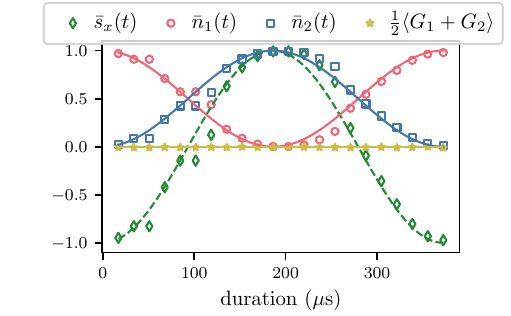}
	\caption{\textbf{ $\mathbb{Z}_2$ gauge  dynamics with pulsed trapped-ion scheme}: We simulate the $\mathbb{Z}_2$ dynamics using two orthogonal spin-dependent forces and try to replicate Fig.~\ref{fig:rabi_flopping_dynamics}~(a). The markers are the numerical simulation data considering the evolution under the two state-dependent forces in Eq.~\eqref{eqn:scheme_2_H_sdf}, but also including the additional off-resonant carriers that stem from the Lamb-Dicke expansion of Eq.~\eqref{eq:int_term_2}. The coloured lines are the analytical predictions in Eq.~(\ref{eq_Rabi flops_link}), considering the effective tunneling strength in Eq.~\eqref{eq:eff_tunneling_trotter} and $h=0$.}
	\label{fig:scheme_2_BS_time_scan}
\end{figure}

We now discuss an alternative strategy to realise the $\mathbb{Z}_2$ gauge link based on digital quantum simulation and the concatenation of gates.
First, we focus on a new way of engineering the gauge-invariant tunneling term using two orthogonal state-dependent forces and, then, explain how it can be used to experimentally implement the $\mathbb{Z}_2$ gauge model in Eq.~\eqref{eq:tunneling_gauge}. 
As before, we consider the case of a single $\mathbb{Z}_2$ link, i.e. one single ion and two vibrational modes. Following the scheme proposed in Ref.~\cite{sutherland2021universal} for hybrid discrete-continuous variable approaches in trapped-ion quantum information processing, we consider two orthogonal state-dependent forces acting on the two vibrational modes with  Lamb-Dicke parameters $\eta_z$ and $\eta_x$, respectively. We thus start from two terms like Eq.~\eqref{eq:int_term}, each of which will be tuned to yield a different state-dependent force, i.e. $\boldsymbol{k}_{{\rm d},1}||{\bf e}_x$, and $\boldsymbol{k}_{{\rm d},2}||{\bf e}_z$, 
\begin{equation}
\label{eq:int_term_2}
   \tilde{V}_1(t) = \Omega \cos(\delta t)\sum_i\sum_{n=1,2}\ket{\uparrow_i}\bra{\downarrow_i} \ee^{\ii\boldsymbol{k}_{{\rm d},n}\cdot\boldsymbol{r}_i+\phi_{{\rm d},n}} + {\rm H.c.}.
\end{equation}
In the interaction picture with respect to the qubit frequency, $\omega_0$, motional frequencies, $\omega_z$ and $\omega_x$, the interaction reads
\begin{equation}\label{eqn:scheme_2_H_sdf}
{V}_1(t) \approx \eta_x\Omega{\sigma}_1^x{a}_{1,x}\ee^{-\ii\delta t} + \eta_z\Omega{\sigma}_1^y{a}_{1,z}\ee^{-\ii\delta t} + \text{H.c.},
\end{equation}
where $\eta_\alpha\Omega$ is the coupling strength to the respective vibrational mode, and $\delta$ is the detuning away from the sidebands $\omega_0 \pm \omega_{x/z}$. These two terms can be derived using similar steps as before, which only differ on the specific selection of the leading contribution by the appropriate choice of the laser frequencies. The interference of these two forces can lead, in the second order, to an effective state-dependent tunneling. After using a Magnus expansion for the time-ordered evolution operator $U(t)=\mathcal{T}\{{\rm exp}\{-\ii\int_0^t{\rm d}s V_1(s)\}\}$~\cite{magnus1954exponential}, the second-order term $U(t)\approx{\rm exp}\{-\ii H_{\rm eff} t\}$ yields the following interaction
\begin{equation}\label{eqn:scheme_2_H_eff}
{H}_\textrm{eff} = \frac{\Omega_{\rm d}}{2}{a}_{1,z}^{\dagger}{\sigma}_1^z{a}_{1,x}^{\phantom{\dagger}} + {\rm H.c.}, \hspace{2ex} \Omega_{\rm d} = \ii\frac{\Omega^2}{\delta}\eta_x\eta_z,
\end{equation}
which maps directly onto the desired gauge-invariant tunneling of  Eq.~\eqref{eq:tunneling_gauge} with a tunneling strength of 
\beq
\label{eq:eff_tunneling_trotter}
t_{1,{\bf e}_1}=\frac{\Omega_{\rm d}}{2}=\ii\frac{\Omega^2}{2\delta}\eta_x\eta_z.
\eeq
In this derivation,  we have neglected higher-order contributions in the Magnus expansion that would lead to  errors $\epsilon = O([\eta\Omega/\delta]^3)$ that must be kept small (we assumed that $\eta_x$ and $\eta_z$ are the same order of magnitude $\eta$). If a fixed error $\epsilon$ in these higher-order terms is considered, $\delta$ is then linear in $\eta$. Consequently, the tunneling coupling rate is also linear in $\eta$. This linear dependence is in contrast to the analog scheme, where the coupling is quadratic in $\eta$.
Additionally, the first-order term in the Magnus expansion must be accounted for, 
which leads to additional state-dependent displacements in the joint phase space of both vibrational modes. As in the case of trapped-ion entangling gates, these displacements vanish for specific evolution times corresponding to integer multiples of $2\pi/\delta$. Hence, the tunneling term~\eqref{eq:tunneling_gauge} can be achieved by applying the interaction for a duration that is  multiple of $2\pi/\delta$.

In Appendix~\ref{app:orthogonal_forces}, we present the specific experimental parameters for this pulsed scheme, which are then used to numerically validate the above derivations.  A characteristic  numerical simulation is shown in Fig.~\ref{fig:scheme_2_BS_time_scan}, which shows that one can also recover the desired gauge-invariant dynamics using this pulsed scheme, provided on considers the pulse switching discussed in the Appendix. In this Appendix, we also provide a more detailed discussion about the errors.

\subsubsection{Electric field and experimental considerations}
For this scheme, the electric field term $h\sigma_1^x$ in Eq.~\eqref{eq:tunneling_gauge} can be introduced through Trotterization. For this, we split the interaction time into segments with durations that are integer multiples of $2\pi/\delta$, and we alternate between applying the tunneling term and the external field term $h\sigma_1^x$, which can be achieved by a  carrier driving (see Fig~\ref{fig:scheme_2_pulses}). In this way, the electric field term can be easily introduced by interleaving short carrier pulses  with periods of small evolution under the combination of the two orthogonal state-dependent forces.

\subsection{ Comparison of $\mathbb{Z}_2$ gauge link schemes }
\label{sec:exp challenges trapped ions}

Let us start by  discussing how an experiment would proceed. One of the advantages of trapped ions is that they  offer a variety of high-precision techniques for state preparation and readout~\cite{doi:10.1063/1.5088164}. For a single trapped ion, it is customary to perform optical pumping to the desired qubit state, say $\ket{\uparrow}$. One can then use laser cooling in the resolved-sideband limit for both vibrational modes, and prepare them very close to the vibrational ground state. Using a blue sideband  directed along a particular axis, say the $x$-axis, one can flip the state of the qubit and, simultaneously, create a Fock state with a single vibrational excitation in the mode. We note that the initial state of the $\mathbb{Z}_2 $ gauge theory would correspond to $\ket{\rm L}=\ket{1_1}\otimes\ket{\downarrow_{1,{\bf e}_1}}\otimes\ket{0_2}$, which is the rotated version of the one discussed  previously and is thus directly valid for the M{\o}lmer-S{\o}rensen-type scheme. For the light-shift scheme, one must apply a Hadamard gate to initialise the system in  $\ket{\rm L}=\ket{1_1}\otimes\ket{-_{1,{\bf e}_1}}\otimes\ket{0_2}$, which can be accomplished by driving a specific single-qubit rotation.

One can then let the system evolve for a fixed amount of time under the effective Hamiltonian in either Eq.~\eqref{eq:link_ions_h} for the light-shift scheme, or Eq.~\eqref{eq:ms-z2} for the M{\o}lmer-S{\o}rensen-type scheme. We have shown that the effective Hamiltonians approximate the ideal Hamiltonian accurately. After this real-time evolution, the laser fields are switched off. Then, the measurement stage starts, where one tries to infer the matter-gauge field correlated dynamics in Eq.~\eqref{eq_Rabi flops_link}. In order to do that, one would take advantage of the readout techniques developed in trapped ions~\cite{doi:10.1063/1.5088164}, which typically map the information of the desired observable onto the qubit. After this mapping, the qubit can be projectively measured in the $z$-basis via state-dependent resonance fluorescence. In order to measure the electric field operator of the M{\o}lmer-S{\o}rensen-type scheme $\overline{s}_z(t)=\langle \sigma^z_{1,{\bf e}_1}(t)\rangle$, one can collect the state-dependent fluorescence. For the light-shift scheme, one needs to measure $\overline{s}_x(t)=\langle \sigma^x_{1,{\bf e}_1}(t)\rangle$, which requires an additional single-qubit rotation prior to the fluorescence measurement. On the other hand, in order to infer the phonon population $\langle a_2^\dagger a_2^{\phantom{\dagger}}(t)\rangle$=$\langle a_{1,z}^\dagger a_{1,z}^{\phantom{\dagger}}(t)\rangle$, and observe the gauge-invariant tunneling, one would need to map the vibrational information onto the qubit first prior to the fluorescence measurement~\cite{RevModPhys.75.281}. These techniques are well developed, e.g. \cite{Wineland1998experimental}.

At this point, it is worth commenting on the relative strengths of the different schemes. The analog and pulsed schemes presented above outline different viable strategies to implement the gauge-invariant model~\eqref{eq:tunneling_gauge} using current trapped-ion hardware. There are, however, several experimental challenges worth highlighting. It is crucial that the tunneling rates are large compared to any noise process present in the physical system. The dominating sources are the qubit and motional decoherence. In the considered experimental setup \cite{schafer2018a,thirumalai2019high}, the decoherence time of the qubit is the most stringent one with $T_2\approx2.4\,$ms for the ground state qubit and $T_2\approx5\,$ms for the optical qubit. These numbers could be improved by several orders of magnitude by using a clock-qubit encoding. However, in this encoding, the differential dipole light-shift that leads to Eq.~\eqref{eq:link_ions} would vanish and, hence, the method proposed in Sec.~\ref{subsec:ls_impl} would not work~\footnote{Light shifts can be created using quadrupole-Raman transitions~\cite{baldwin2021high} but extra care needs to be taken to symmetrise the shift induced on each qubit state.}. An alternative way of increasing the qubit coherence would be to use magnetic shielding or active magnetic-field stabilisation. 

In terms of motional coherence, the heating rate is limiting the coherence time for the longitudinal motional mode to be ca.~14$\,$ms. The coherence time for the transverse modes has not been properly characterised in the current system and it might be further limited by noise in the trap rf drive. However, actively stabilising the amplitude of the rf drive has shown improvement in the coherence time~\cite{johnson2016active}. We note that a similar, yet non-gauge-invariant tunneling, has been implemented in a trapped-ion experiment~\cite{gan_2020} in the context of continuous-variable quantum computing. This experiment successfully used the transverse modes and measured coherence times of 5.0(7) and 7(1) ms for $^{171}\textrm{Yb}^{+}$ ions.

Therefore, we can conclude that the exchange timescales of the $\mathbb{Z}_2$ tunneling implementations investigated in this paper using the analog scheme in Secs.~\ref{subsec:ls_impl} ($\approx \SI{40}{\micro\second}$, using Raman beams) and~\ref{sec:MS_impl} ($\approx\SI{100}{\micro\second}$, using Raman beams), as well as  the pulsed scheme in Sec.~\ref{subsec:pulsed_scheme} ($\approx\SI{200}{\micro\second}$, using quadrupole beams) are an order of magnitude faster than the qubit or motional decoherence, and hence experimentally feasible. We note that the analog schemes are operationally simpler and do not suffer from Trotterization and loop closure errors. However, the pulsed scheme can obtain substantially higher tunneling rates for fixed intensity and Lamb Dicke factors as the tunneling rate scales linearly in $\eta$ instead of quadratically. For the considered parameters of the quadrupole transition, the exchange duration in the pulsed scheme is  four times faster than  the analog one (\ref{sec:MS_impl}). Hence, this method would be amenable to the quadrupole transition or if the laser intensities are limited.


Before closing this section, it is also worth commenting on the realization of the $\mathbb{Z}_2$ gauge link in other experimental platforms. In fact, there has been a  pioneering experiment with cold atoms~\cite{Schweizer2019,Barbieroeaav7444}, which  relies on a different scheme designed for  double-well optical lattices that exploit Floquet engineering to activate a density-dependent tunneling in the presence of strong Hubbard interactions~\cite{Bermudez_2015,PhysRevA.96.053602,Gorg2019}. By playing with the ratio of the modulation and interaction strengths, the tunneling of the atoms in one electronic state (i.e. matter field)  can depend on  the density distribution of  atoms in a different internal state (i.e. gauge field). On the contrary, the atoms playing the role of the gauge field can  tunnel freely between the  minima of the double well. As realised in~\cite{Schweizer2019,Barbieroeaav7444}, using a single atom of the gauge species per double well, one can codify the  $\mathbb{Z}_2$ gauge qubit, those being the qubit states where the atom resides in either the left or the right well. Then,  its bare tunneling realises directly the electric-field term, whereas the density-dependent tunneling of the matter atoms can be designed to simulate the gauge-invariant tunneling of the $\mathbb{Z}_2$ gauge theory~\eqref{eq:tunneling_gauge}. Remarkably, the dynamics of the matter atom observed experimentally is consistent with Eq.~\eqref{eq_Rabi flops_link},  displaying periodic Rabi oscillations that get damped due to several noise sources~\cite{Schweizer2019}. Although the cold-atom experiments can also infer $\langle \sigma^z_{1,{\bf e}_1}(t)\rangle$ via the measure of the atomic density of the gauge species, measuring $\langle \sigma^x_{1,{\bf e}_1}(t)\rangle$ amounts to a bond density that would require  measuring the equal-time Green's function. This would require inferring  correlation functions between both sites, which is more challenging and  was not   measured in the experiment~\cite{Schweizer2019}. Although the reported measurements show $\langle \sigma^z_{1,{\bf e}_1}(t)\rangle\approx 0$, which is consistent with the link field being always in the electric-field basis; it would be desirable to measure the correlated Rabi flopping of the gauge link~\eqref{eq_Rabi flops_link}, which directly accounts for how the electric-field line stretches/compresses synchronous with the tunneling of the matter particle according to Gauss' law. Since in the trapped-ion case $\langle \sigma^x_{1,{\bf e}_1}(t)\rangle$ can be inferred  by applying a single-qubit gate and collecting the resonance fluorescence, the  present scheme proposed in this work  could thus go beyond these limitations, and directly observe the consequences of gauge invariance in the correlated oscillations~\eqref{eq_Rabi flops_link}. Moreover, as discussed in the following section, there are also promising pathways in the trapped-ion case that could allow extending the quantum simulation beyond the single link case.

\section{\bf Minimal plaquettes and synthetic dimensional reduction for a $\mathbb{Z}_2$  gauge chain}
\label{sec:dim_reduction}

There are various directions to increase the complexity of the trapped-ion quantum simulator of $\mathbb{Z}_2$ gauge fields. The first non-trivial extension is to consider two matter sites joined by two gauge links, which form the smallest-possible plaquette that is consistent with  $\mathbb{Z}_2$ gauge symmetry. We discuss, in this section, how to achieve this $\mathbb{Z}_2$ plaquette simulator by using a pair of ions and exploiting their collective vibrational modes~\cite{Steane1997,James1998,Marquet2003}. We then present a scheme that effectively reduces the  dimension of the synthetic ladder in Fig.~\ref{fig:transverse_phonons_syntehtic_ladder} {\bf (a)}, and allows us  to scale the gauge-invariant model of Eq.~\eqref{eq:tunneling_gauge} to a full lattice, in this case, a one-dimensional chain. We note that these extensions require additional experimental tools, and longer timescales,  making the quantum simulation  more challenging. 
Nonetheless, the proposed schemes set a clear  road map that emphasises the potential of trapped ions for the simulation of  real-time dynamics in lattice gauge theories.   

\subsection{ $\mathbb{Z}_2$ plaquette: Wegner-Wilson and 't Hooft loops for gauge-field  entanglement}

\begin{figure}[t]
	\centering
	\includegraphics[width=0.95\columnwidth]{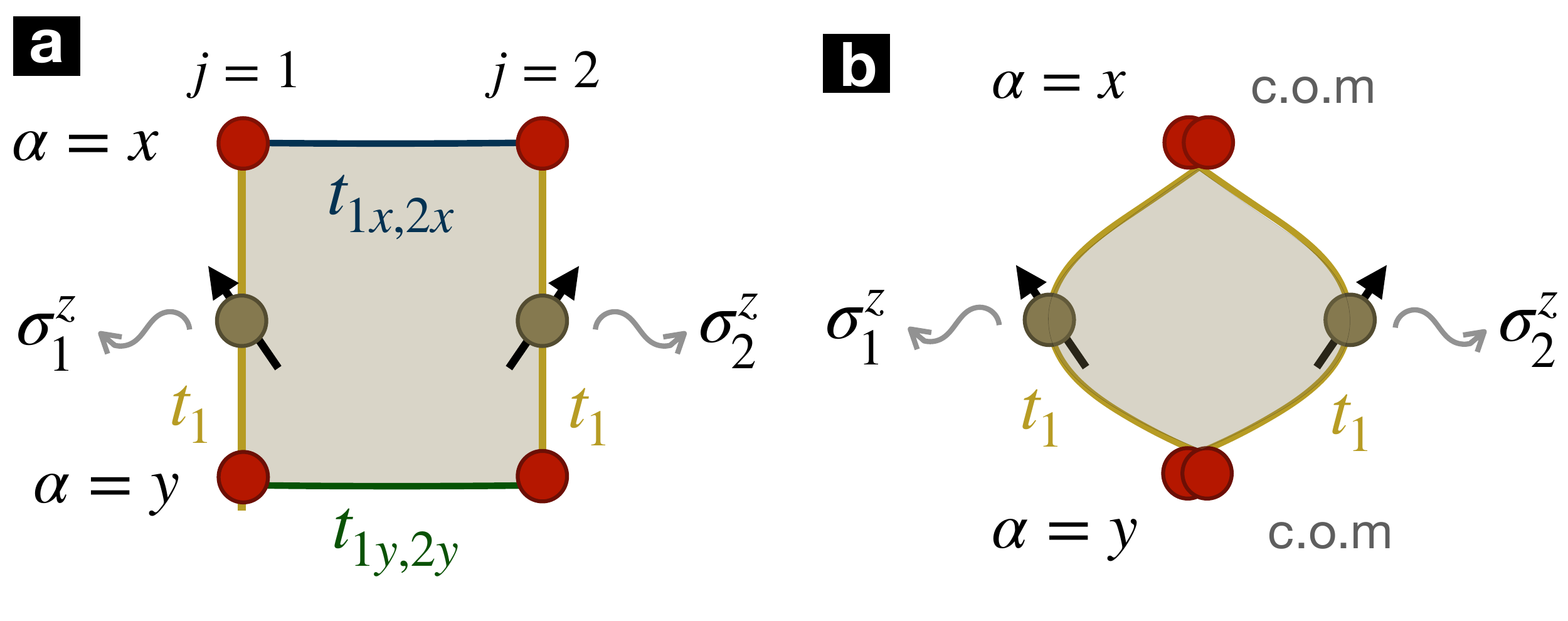}
	\caption{{\bf Scheme for synthetic $\mathbb{Z}_2$ plaquette: } {\bf (a)} The application of Eq.~\eqref{eq:state_dep_parametric_ions} to $N=2$ ions would result in a synthetic rectangular plaquette, where the four sites correspond to the local transverse modes of the two ions. The vertical links are induced by the state-dependent parametric tunneling of Eq.~\eqref{eq:link_ions}, and thus incorporate a gauge qubit (note that  in the M{\o}lmer-S{\o}rensen-type scheme, the Pauli matrices in the links need to be rotated). The horizontal links describe the bare tunneling caused by dipole-dipole interactions in Eq.~\eqref{eq:tunneling_ions}, such that no gauge qubit mediates the tunneling, and gauge invariance is explicitly broken. {\bf (b)} By modifying the set of constraints on the optical potential according to Eq.~\eqref{eq:constraints_param_tunneling_ions_plaquette}, one can resolve the collective vibrational modes, and reduce the quadrangular synthetic plaquette  of {\bf (a)} with 4 sites and 4 links, into a rhomboidal one composed of two sites and two links, both of which contain now a gauge qubit such that the effective tunneling respects a local $\mathbb{Z}_2$ symmetry.}
	\label{fig:z2_plaquette_ions_scheme}
\end{figure}

We  consider a couple of ions  leading to a pair of qubits and four vibrational modes, two per transverse direction. In principle, we could apply the previous scheme based on a global state-dependent parametric drive of Eq.~\eqref{eq:state_dep_parametric_ions}. However, this would lead to a synthetic plaquette where two of the links have a tunneling that does not depend on any  gauge qubit (see Fig.~\ref{fig:z2_plaquette_ions_scheme} {\bf (a)}), failing in this way to meet the requirements for local gauge invariance. In fact, this goes back to the simple counting of synthetic lattice sites and effective gauge fields we mentioned below Eq.~\eqref{eq:state_dep_parametric_ions}. To remedy this problem, the idea is to modify the constraints on the strength of the light-shift  optical potential of Eq.~\eqref{eq:constraints_param_tunneling_ions}, such that it becomes possible to address certain common vibrational modes instead of the local ones. Although we will focus on the light-shift scheme for now on, we note that similar ideas would apply to the M{\o}lmer-S{\o}rensen-type and orthogonal-force schemes. We will show in this section that, by addressing the collective modes, we can effectively deform the   plaquette (see Fig.~\ref{fig:z2_plaquette_ions_scheme} {\bf (b)}) such that
the   model is consistent with the local $\mathbb{Z}_2$ symmetry.

The transverse collective modes of the two-ion crystal are the symmetric and anti-symmetric superpositions of the local vibrations, and are referred to as the center-or-mass (c) and zigzag (z) modes within the trapped-ion community. The creation operators for these modes are then defined by
\beq
\label{eq:com_zz_modes}
\begin{split}
a_{{ \rm c},\alpha}=\frac{1}{\sqrt{2}}(a_{1,\alpha}+a_{2,\alpha}), \hspace{2ex}\omega_{{\rm c},\alpha}=\omega_\alpha,\\
a_{{\rm z},\alpha}=\frac{1}{\sqrt{2}}(a_{1,\alpha}-a_{2,\alpha}), \hspace{2ex}\omega_{{\rm z},\alpha}<\omega_\alpha.\\
\end{split}
\eeq
We now  substitute these equations in the expressions for the  light-shift optical potential of Eq.~\eqref{eq:ac_beams_st_dep}, and  proceed by performing the subsequent Lamb-Dicke expansion that leads to a sum of terms containing all possible powers of the creation-annihilation operators. We can now select the desired tunneling term between a single mode, say the center of mass, along the two transverse directions (see Fig.~\ref{fig:z2_plaquette_ions_scheme} {\bf (b)}). Since we can also get terms that couple the center of mass and the zigzag modes, we need to modify the  constraints in Eq.~\eqref{eq:constraints_param_tunneling_ions} to be
\beq
\label{eq:constraints_param_tunneling_ions_plaquette}
\omega_{\rm d}=\omega_{{\rm  c},y}-\omega_{{\rm c},x},\hspace{2ex}|\Omega_{\rm d}|\ll|\omega_{{\rm c}, y}-\omega_{{\rm c}, x}|,\frac{|\omega_{{\rm c},\alpha}-\omega_{\rm z,\beta}|}{\eta_\alpha\eta_\beta},
\eeq
such that those terms become off-resonant and can be neglected.
Note that these new constraints will make the gauge-invariant tunneling weaker, and the targeted dynamics   slower, making the  experimental realisations more challenging.

\begin{figure}[t]
	\centering
	\includegraphics[width=1\columnwidth]{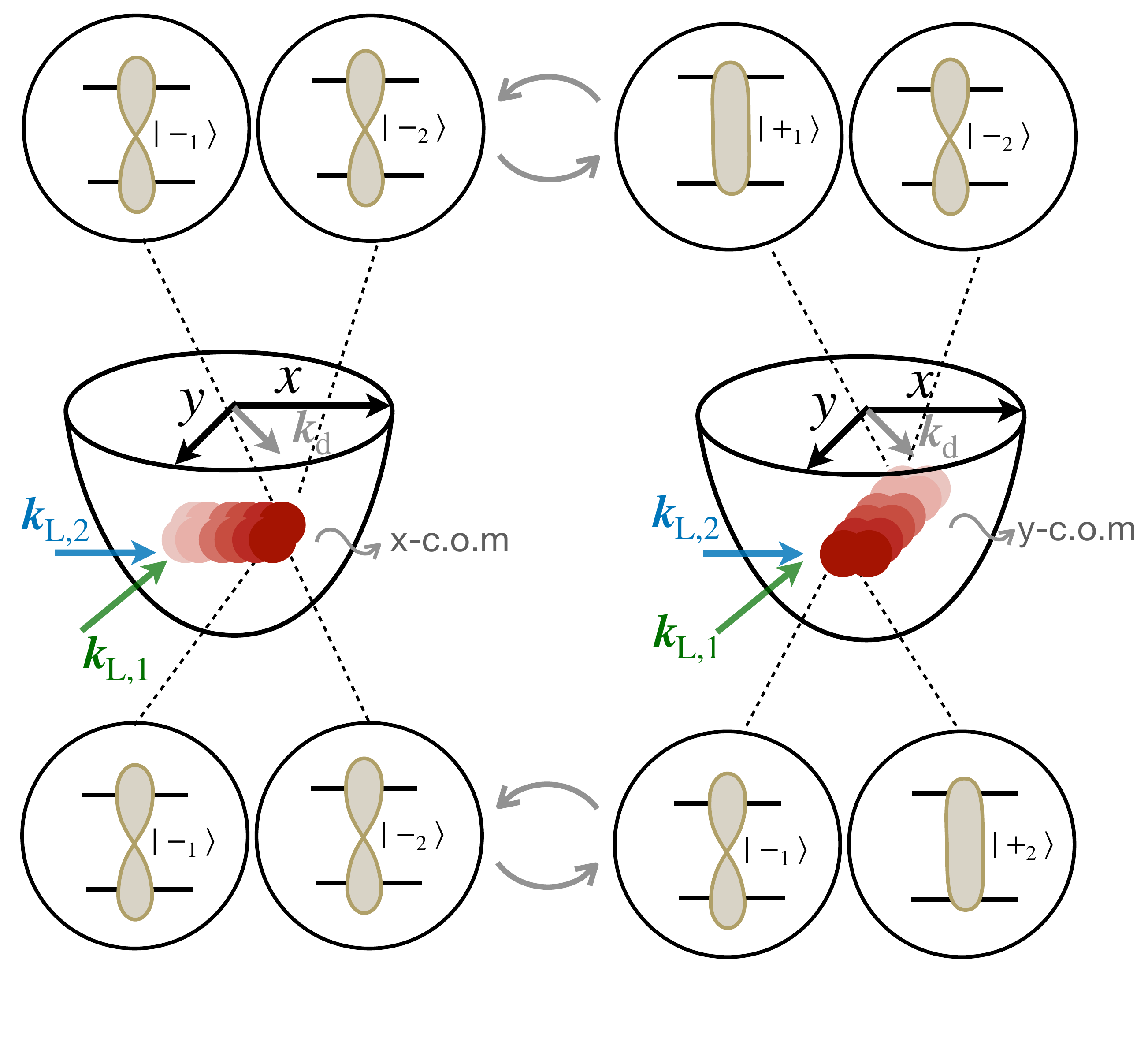}
	\caption{{\bf Trapped-ion  $\mathbb{Z}_2$ tunneling in a rhomboidal plaquette: } Schematic representation of the the gauge-invariant tunneling of a vibrational excitation, which is initially  in the center of mass (c.o.m) mode along the $x$ axis, and ``tunnels'' into the c.o.m mode along the $y$ axis. In the upper insets, this tunneling is mediated by a spin flip in the Hadamard basis of the first ion qubit $\ket{-}_1\mapsto\ket{+}_1$, whereas in the lower inset it involves the second ion qubit $\ket{-}_2\mapsto\ket{+}_2$. These two paths can be interpreted as the two effective  links of the synthetic rhomboidal plaquette displayed in Fig.~\ref{fig:z2_plaquette_ions_scheme}.   }
	\label{fig:z2_plaquette_ions}
\end{figure}

In the following, we will present the expressions for the light-shift schemes, although any of the other possibilities should be analogous. By moving to the interaction picture with respect to the full vibrational Hamiltonian~\eqref{eq:trapped_ion_harmonic}, and neglecting the off-resonant terms by a rotating-wave approximation that rests upon Eq.~\eqref{eq:constraints_param_tunneling_ions_plaquette}, the leading term stemming from the aforementioned light-shift optical potential is
\beq
\label{eq:tunneling_ions_plaquette}
V_{I}(t)\approx 
\frac{\Omega_{\rm d}}{4}a^{{\dagger}}_{{\rm c},y}\sigma^z_{1}a^{\phantom{\dagger}}_{{\rm c},x}+\frac{\Omega_{\rm d}}{4}a^{{\dagger}}_{{\rm c},y}\sigma^z_{2}a^{\phantom{\dagger}}_{{\rm c},x}+{\rm H.c.}.
\eeq
Here, we recall that the drive strength  is $\Omega_{\rm d}=\eta_{x}\eta_y\Omega_{1,2}$, as defined in Eq.~\eqref{eq:parameters_parametric_ions}, and we have neglected the irrelevant  phase that can be gauged away in this simple two-mode setting.  As depicted in Fig.~\ref{fig:z2_plaquette_ions}, there are now two different gauge-invariant processes in which a center-of-mass phonon along the $x$-axis can tunnel into a center-of-mass phonon along the $y$-axis. Each of these processes flips the Hadamard state of one, and only one, of the trapped-ion qubits (see  Fig.~\ref{fig:z2_plaquette_ions}).

We can now modify the interpretation in terms of synthetic matter sites and $\mathbb{Z}_2$ gauge qubits~\eqref{eq:guage_matter_mapping}, which must now include a pair of $\mathbb{Z}_2$ links, as we have two qubits  dressing the tunneling
\beq
a^{\phantom{\dagger}}_{{\rm c},x},a^{{\dagger}}_{{\rm c},x}\mapsto a^{\phantom{\dagger}}_1,a^{{\dagger}}_1,\hspace{2ex} a^{\phantom{\dagger}}_{{\rm c},y},a^{\dagger}_{{\rm c},y}\mapsto a^{\phantom{\dagger}}_2, a^{{\dagger}}_2 \hspace{2ex} \sigma_{i}^x,\sigma_{i}^z\mapsto \sigma_{1,{\bf e}_i}^x,\sigma_{1,{\bf e}_i}^z.
\eeq
As depicted in Fig.~\ref{fig:z2_plaquette_ions_scheme} {\bf (b)}, we need to introduce two links that connect the synthetic site 1 to 2, requiring two synthetic directions specified by the vectors  ${\bf e}_1, {\bf e}_2$, and allowing us to interpret the model in terms of a rhomboidal plaquette. In addition  to the gauge-invariant tunneling, we  also apply the additional tone of Eq.~\eqref{eq:second_tone}, which   drives the carrier transition on both qubits, and leads to the electric-field term. Altogether, the $\mathbb{Z}_2$ gauge theory on this plaquette  is 
\beq
\label{eq:gauge_plaquette}
H_{\rm eff}=\sum_{n=1,2}\left(t^{\phantom{\dagger}}_{1,\textbf{e}_n}a^{{\dagger}}_{2}\sigma^z_{1,{\bf e}_n}a^{\phantom{\dagger}}_{1}+{\rm H.c.}\right)+\sum_{n=1,2}h\sigma^x_{1,{\bf e}_n},
\eeq
where the microscopic parameters for the tunneling strength and the electric field are the same as in Eq.~\eqref{eq:tunneling_gauge}, except for  the tunneling strengths. These get halved with respect to the previous ones by working with the center-of-mass mode instead of the local vibrations, namely 
\beq
t_{1,\textbf{e}_1}=t_{1,\textbf{e}_2}=\frac{\Omega_{1,2}}{4}{\eta_{x}\eta_y},\hspace{2ex} h=\frac{\tilde{\Omega}_{\rm d}}{2}.
\eeq


Let us also note that, since we now have an increased connectivity, the generators of the  $\mathbb{Z}_2$ gauge symmetry which, in the single-link case were defined in Eq.~\eqref{eq:generators_link}, now read
\beq
\label{eq:generators_plaquette}
G_1=\ee^{\ii\pi a_1^\dagger a_1^{\phantom{\dagger}}}\sigma_{1,{\bf e}_1}^x\!\sigma_{1,{\bf e}_2}^x, \hspace{2ex}  G_2=\sigma_{1,{\bf e}_1}^x\!\sigma_{1,{\bf e}_2}^x\ee^{\ii\pi a_2^\dagger a_2^{\phantom{\dagger}}}.
\eeq
As we have a pair of synthetic $\mathbb{Z}_2$ links emanating from each of the two matter sites, the generators include products of the corresponding Pauli matrices. Note that these generators fulfill the same algebra as before, and  define projectors onto super-selection sectors, such that the effective  Hamiltonian gauge theory~\eqref{eq:gauge_plaquette} can be block decomposed into the different sectors~\eqref{eq:gauss_law_link} characterised by two static charges $q_1,q_2\in\{0,1\}$. In addition to the previous effective Hamiltonian, one could also include other gauge-invariant terms, such as 
\beq
\label{eq:gauge_plaquette_total2}
\tilde{H}_{\rm eff}=H_{\rm eff}+\Delta_1 a_1^\dagger a_1^{\phantom{\dagger}}+\Delta_2 a_2^\dagger a_2^{\phantom{\dagger}},
\eeq
where $\Delta_i$ can be controlled by a  small detuning  of the state-dependent parametric drive.

\begin{figure}[t]
	\centering
	\includegraphics[width=0.95\columnwidth]{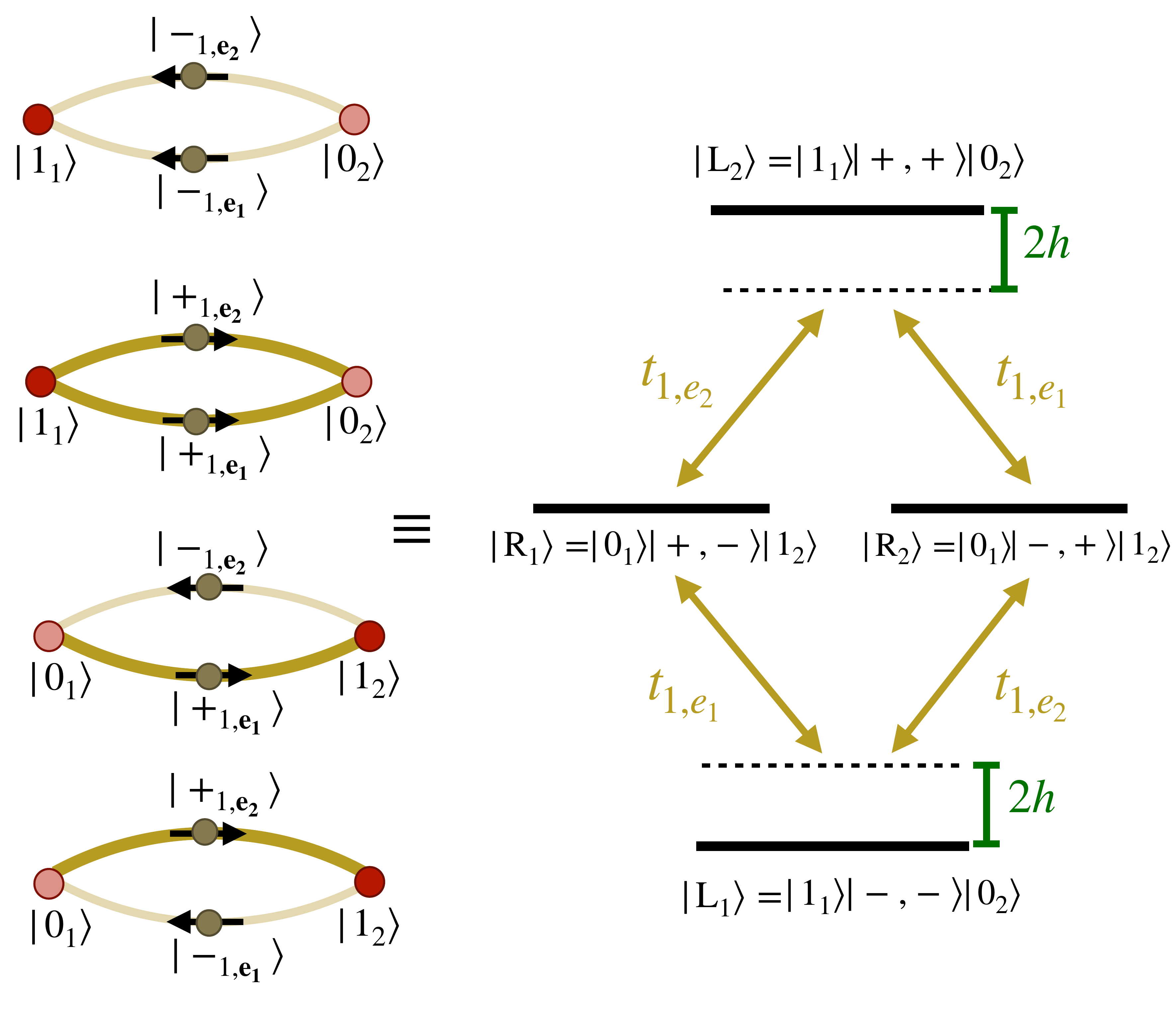}
	\caption{{\bf $\Diamond $-scheme for  $\mathbb{Z}_2$ tunneling in a plaquette: } On the left, we represent schematically the four possible gauge-invariant states in the sector with background charges $q_1=1,q_2=0$, which correspond to Eq.~\eqref{eq:gauge_inv_basis_plaquette}. We see that, when the matter particle resides on the left or right site, a 't Hooft electric field line that winds around the plaquette can be present or absent, doubling the number of possible states. On the right, we depict an effective  $\Diamond $-scheme in quantum optics, in which the gauge-invariant tunneling induces two copies of the  $\Lambda$-scheme of Fig.~\ref{fig:z2_link_tunneling_rabi_oscillations_2bosons}, which appeared for a single link and two bosons that lead to  bright and dark states, and mode entanglement.  }
	\label{fig:z2_plaquette_tunneling}
\end{figure}

Once the scheme for the  quantum simulation of the single $\mathbb{Z}_2$ plaquette has been discussed, let us describe some interesting dynamical effects that arise when considering, as in the single-link case,  the one-particle sector. Following our previous approach, one can exploit the global $U(1)$ symmetry and Gauss' law to reduce the dimensionality of the subspace where the dynamics takes place. If we consider a single bosonic particle, this subspace is spanned by four states 
\beq
\label{eq:gauge_inv_basis_plaquette}
\begin{split}
\ket{\rm L_1}&=\ket{1_1}\otimes\ket{-_{1,{\bf e}_1}}\otimes\ket{-_{1,{\bf e}_2}}\otimes\ket{0_2}, \\
\ket{\rm L_2}&=\ket{1_1}\otimes\ket{+_{1,{\bf e}_1}}\otimes\ket{+_{1,{\bf e}_2}}\otimes\ket{0_2}, \\
\ket{\rm R_1}&=\ket{0_1}\otimes\ket{+_{1,{\bf e}_1}}\otimes\ket{-_{1,{\bf e}_2}}\otimes\ket{1_2},\\
\ket{\rm R_2}&=\ket{0_1}\otimes\ket{-_{1,{\bf e}_1}}\otimes\ket{+_{1,{\bf e}_2}}\otimes\ket{1_2},
\end{split}
\eeq
where the corresponding background charges are $q_1=1, q_2=0$. In comparison to the single-link case,  the plaquette gives us  further possibilities for the stretching and compressing of the electric-field line when the matter boson tunnels back and forth (see Fig.~\ref{fig:rabi_flopping_dynamics}). On the one hand,  an electric-field loop around the plaquette, a so-called 't Hooft loop, does not require further sinks/sources since the electric field line enters and exists all sites in the plaquette. In addition, the stretched electric field can now wind along the two possible paths of the loop. This leads to the  doubling of the gauge arrangements for a fixed layout of the matter boson in Eq.~\eqref{eq:gauge_inv_basis_plaquette} and Fig.~\ref{fig:z2_plaquette_tunneling}.

\begin{figure}[t]
	\centering
	\includegraphics[width=0.95\columnwidth]{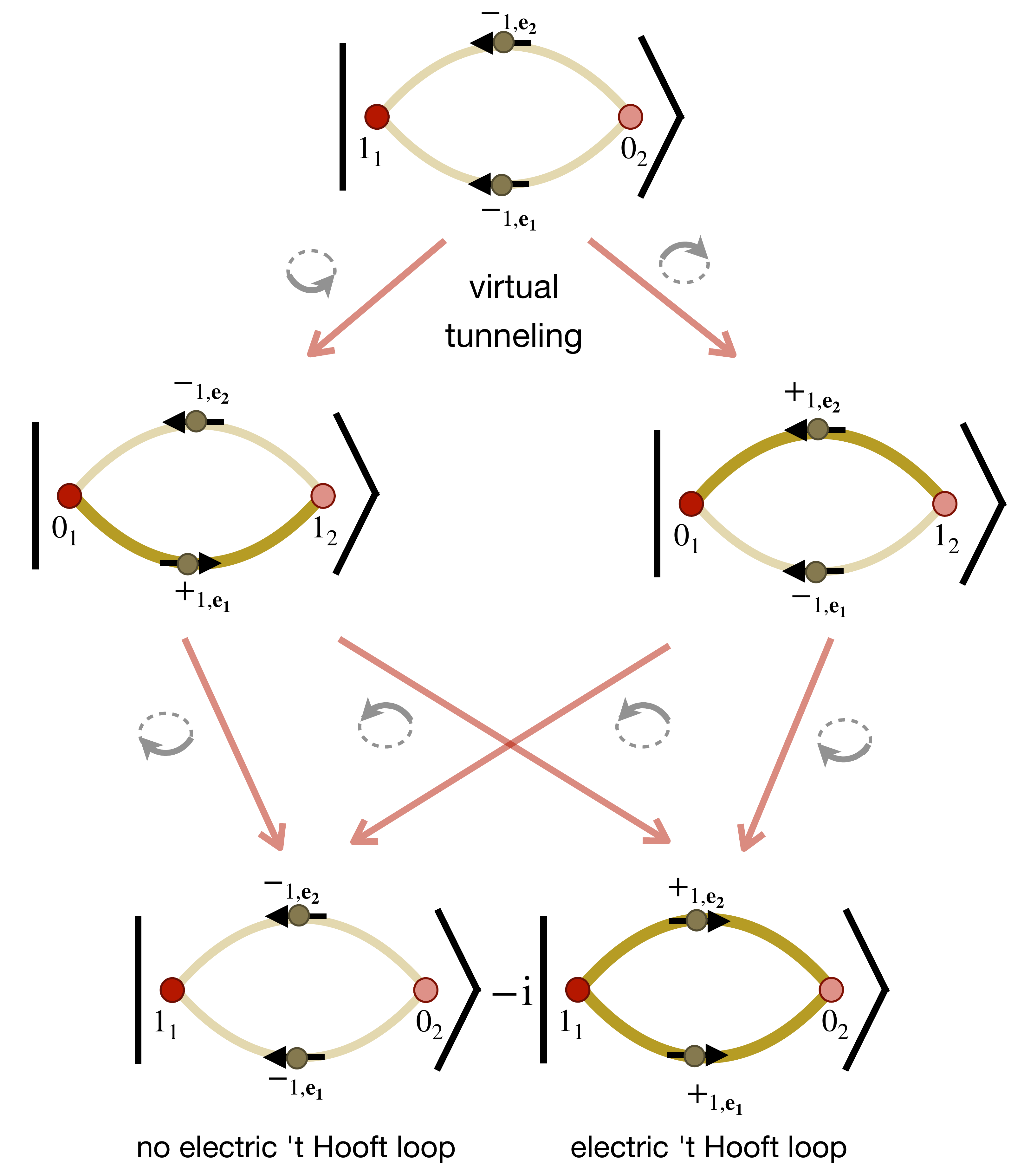}
	\caption{{\bf 't Hooft loop entanglement in the gauge qubits: } We represent a scheme for the second-order virtual tunneling of a boson, which initially occupies the left site (upper state) and aims to tunnel to right, which is penalised energetically by the  {on-site} energy $\Delta_2\gg t_{1,{\bf e}_n}$. After the virtual tunneling and the accompanying stretching of the electric-field line (intermediate states), the second tunneling process back to the left site can either compress the string, or stretch it further until it winds around the plaquette. The superposition of these possible transitions (lower state) leads to an entangled state.   }
	\label{fig:z2_plaquette_entanglement_scheme}
\end{figure}

We are interested in exploring new dynamical effects, in particular the possibility of creating entanglement between the $\mathbb{Z}_2$ gauge fields by the tunneling of the bosonic $\mathbb{Z}_2$ charge. 
The dynamics due to this effective Hamiltonian can now be depicted as a four-level system in a $\Diamond $-scheme. Setting $\Delta_1=\Delta_2=0$ in Eq.~\eqref{eq:gauge_plaquette_total2}, the states with the particle on the right site have zero energy, whereas those with the particle on the left have energies $\pm 2h$. Moreover, they are coupled by the gauge-invariant tunneling with strengths $t_{1,\textbf{e}_1},t_{1,\textbf{e}_2}$ according to the  $\Diamond $ scheme on the right panel of   Fig.~\ref{fig:z2_plaquette_tunneling}. As apparent from  this figure, we can define bright $\ket{\rm B}=(\ket{\rm R_1}+\ket{\rm R_2})/\sqrt{2}$ and dark  $\ket{\rm D}=(\ket{\rm R_1}-\ket{\rm R_2})/\sqrt{2}$ states once more, such that the effective dynamics corresponds to that of a 3-level atom. This case also has an exact solution in terms of an effective spin-1 particle that precesses   under an effective  magnetic field. Defining $\ket{\Psi_{\rm phys}(t)}=d(t)\ket{\rm D}+ c_{l,1}(t)\ket{\rm L_1}+c_{b}(t)\ket{\rm B}+c_{l,2}(t)\ket{\rm L_2}$,  the amplitude of the dark state remains constant  $d(t)=d(0)$, whereas the amplitudes of the remaining states evolve as  $
\boldsymbol{c}(t)=\ee^{-\ii t\boldsymbol{B}_0\cdot\boldsymbol{S}}\boldsymbol{c}(0)$. Here, $
\boldsymbol{c}(t)=(c_{l,1}(t),c_b(t),c_{l,2}(t))^{\rm t}$,
\beq
\boldsymbol{B}_0=(2t_{1,\textbf{e}_1},0,2h),
\eeq
and the spin-1 operators are defined as
\beq
{S}_z=\ket{\rm L_2}\bra{\rm L_2}-\ket{\rm L_1}\bra{\rm L_1},{S}_x=\textstyle{\frac{1}{\sqrt{2}}}(\ket{\rm L_2}\bra{\rm B}+\ket{\rm B}\bra{\rm L_1})+{\rm H.c.}. 
\eeq

\begin{figure}[t]
	\centering
	\includegraphics[width=0.75\columnwidth]{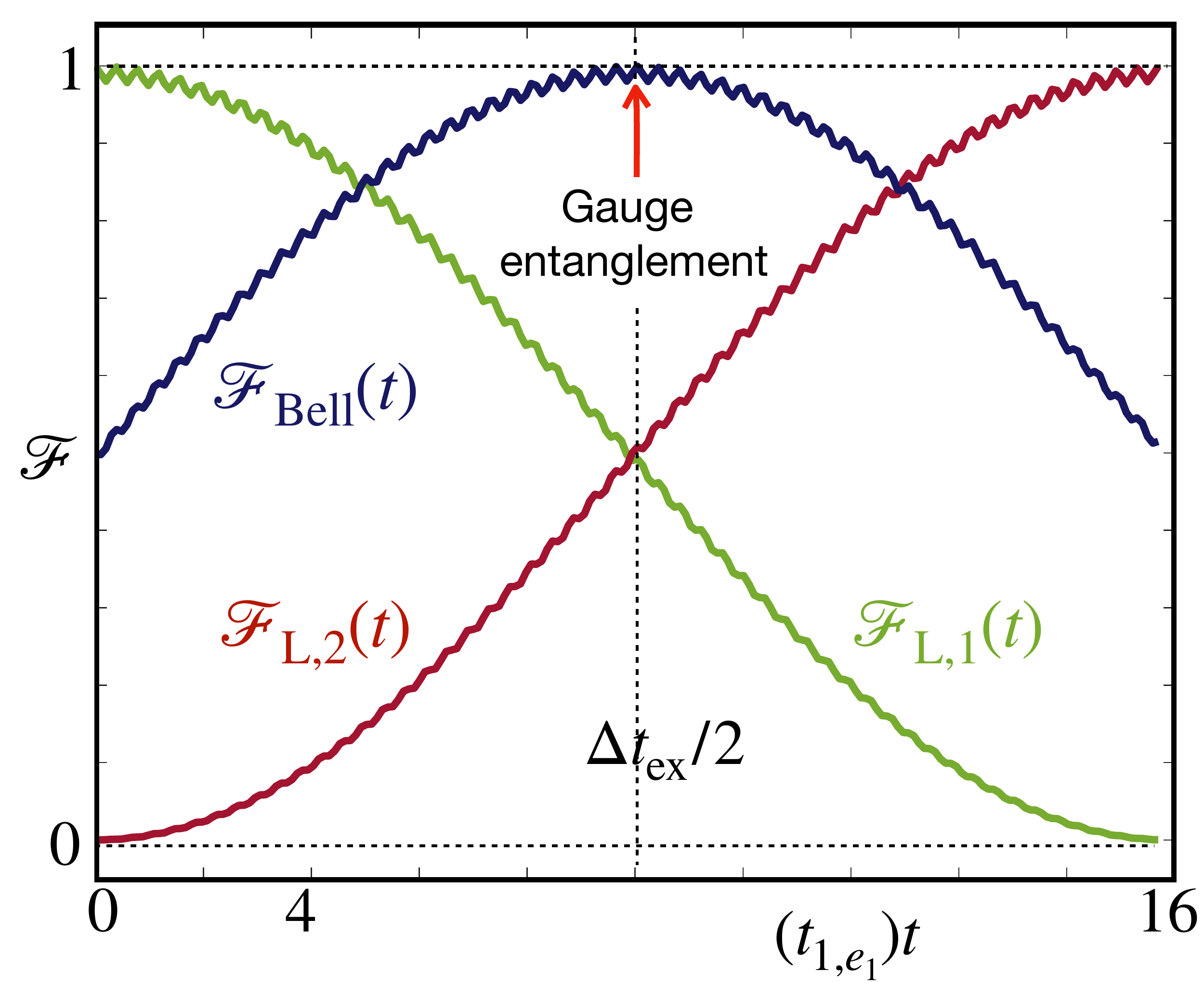}
	\caption{{\bf Gauge entanglement in a minimal plaquette: } We represent the state fidelities $\mathcal{F}_i=|\langle\Psi_i\ket{\Psi(t)}|^2$ for the target state $\ket{\Psi_i}$ that can either be $\ket{{\rm L}_1},\ket{{\rm L}_2}$ in Eq.~\eqref{eq:gauge_inv_basis_plaquette}, or the Bell state $\ket{\Psi_{\rm Bell}^-}$ in Eq.~\eqref{eq:Bell_state}. The time evolved state $\ket{\Psi(t)}$ is obtained by solving numerically the dynamics for $\Delta_2=10t_{1,{\bf e}_1}$, and $h=\Delta_1=0$.}
	\label{fig:z2_plaquette_entanglement_fidelities}
\end{figure}
If one now switches on the detuning $\Delta_2>0=h=\Delta_1$, the intermediate bright state gets shifted in energies such that, for $|t_{1,\textbf{e}_n}|\ll \Delta_2$, the bright state will only be virtually populated when one starts from the configuration $ \ket{\Psi_{\rm phys}(0)}=\ket{\rm L_1}$. In this initial state,   the matter particle is in the left site, and no electric-field line winds around the plaquette (see Fig.~\ref{fig:z2_plaquette_entanglement_scheme}). The dynamics can then be understood from a second-order process in which the particle tunnels to the right, while the electric field line stretches, followed by a second tunneling event to the left. Due to the high energy offset $\Delta_2\gg t_{1,{\bf e}_1}$, the right site is only virtually populated in the intermediate states of Fig.~\ref{fig:z2_plaquette_entanglement_scheme}.   Note  that, during the subsequent tunneling from these virtual states, the particle can either follow the same link/path of the first tunneling event, flipping  the corresponding link qubit back to the original one $\ket{\rm L_1}$ or, alternatively,    it can  choose to follow the other link/path, going around the plaquette and ending in state $\ket{\rm L_2}$. Since one must superpose  the possible histories in quantum mechanics, the  state after half the exchange  time $\Delta t_{\rm ex}/2=\pi\Delta_2/4t_{1,t_{1,\textbf{e}_1}}^2$ becomes 
\beq
\label{eq:Bell_state}
\ket{\Psi_{\rm phys}(t_e)}=\textstyle{\frac{1}{\sqrt{2}}}(\ket{\rm L_1}-\ii\ket{\rm L_2})=\ket{1_1}\otimes\ket{\Psi_{\rm Bell}^{-}}\otimes\ket{0_2}.
\eeq
Here, the boson has returned to the initial lattice site, but the gauge fields have evolved into an entangled state that is equivalent to a Bell pair in the Hadamard basis
\beq
\label{eq:Bell_state_1}
\ket{\Psi_{\rm Bell}^{-}}=\textstyle{\frac{1}{\sqrt{2}}}\left(\ket{-_{1,{\bf e}_1},-_{1,{\bf e}_2}}-\ii\ket{+_{1,{\bf e}_1},+_{1,{\bf e}_2}}\right).
\eeq
It is interesting to remark that this entangled state is the result of summing  over the two tunneling histories of the charged particle, leading to a linear superposition between two different  electric-field strings. In the first one, there is no electric field within the loop,  as the boson tunnels forth and back along the same link. Conversely,  in the second case,  a  t' Hooft electric-field loop winding around the plaquette has been created since the boson has enclosed a loop around the plaquette during the virtual process.
In Fig.~\ref{fig:z2_plaquette_entanglement_fidelities}, we represent the dynamics of such internal state as a function of time by representing the state fidelities of the time-evolved state with the $\ket{{\rm L}_1},\ket{{\rm L}_2}$ and the target Bell state $\ket{\Psi_-}$~\eqref{eq:Bell_state_1}. On  sees how the Bell-state fidelity approaches unity after half  the exchange duration $\Delta t_{\rm ex}/2$. Since the dynamics is i second order, we see that the timescale is larger than that of Figs.~\ref{fig:rabi_flopping_dynamics} and~\ref{fig:noon_flopping_dynamics}.

\subsection{ $\mathbb{Z}_2$ chain: synthetic dimensional reduction}

In the previous subsection, we have seen that introducing more ions and playing with collective vibrational modes allows us to extend the $\mathbb{Z}_2$ gauge-field toolbox towards  interesting and more complex real-time phenomena. We discussed how, by working with only two ions, the spectral crowding of collective modes can still be handled, and one can selectively address the gauge-invariant tunneling between a pair of collective modes along the two different transverse directions.  In this subsection, we present a scheme that exploits these collective modes with reduced crowding, together with  the idea of synthetic dimensional reduction, to scale the quantum simulator of $\mathbb{Z}_2$ gauge theories to chains of arbitrary size. We start the discussion by considering the generalisation of Eq.~\eqref{eq:link_ions} to  a  chain of $N$ trapped ions, namely 
\beq
\label{eq:link_ions1}
V_{1}(t)\approx \sum_i
\frac{\Omega_{\rm d}}{2}\ee^{\ii\phi_{i}}a^{{\dagger}}_{{i},y}\sigma^z_{i}a^{\phantom{\dagger}}_{i,x}+{\rm H.c.}, 
\eeq
 where we recall that the microscopic parameters were 
\beq
\label{eq:parameters_tunneling_synthetic}
\Omega_{\rm d}=|\Omega_{1,2}|\eta_{x}\eta_y, \hspace{2ex}\phi_{i}=\boldsymbol{k}_{\rm d}\cdot\boldsymbol{r}_i^0+{\arg}(-\Omega_{1,2}).
\eeq
If we align the laser wave-vectors such that  $\boldsymbol{k}_{\rm d}\cdot\boldsymbol{r}_i^0=0$, the driving phase becomes  homogeneous, and can be gauged away without loss of generality. The  Hamiltonian can still be described in terms of the tight-binding model of Eq.~\eqref{eq:trapped_ion_harmonic}, but one must introduce a state-dependent tunneling matrix 
\beq
\label{eq:tunneling_ions_synthetic_gauge}
\hat{t}_{(i,\alpha)(j,\beta)}={t}_{(i,\alpha)(j,\beta)}\mathbb{I}_2+\frac{\Omega_{\rm d}}{2}\sigma_i^z\ee^{\ii \epsilon_{\alpha,\beta}\phi_j}\delta_{i,j}(1-\delta_{\alpha,\beta}).
\eeq
 Using the same mapping to a synthetic ladder as the one in Eq.~\eqref{eq:Peierls_matter_mapping}, we would obtain an effective model as the one depicted in Fig.~\ref{fig:transverse_phonons_syntehtic_ladder}{\bf (a)}, where only the vertical synthetic links  in yellow contain a $\mathbb{Z}_2$ gauge field that mediates the tunneling. On the other hand, the horizontal tunnelings are still $c$-numbers, and the complete model is thus not consistent with the local gauge symmetry. This caveat  generalises  the situation in the   plaquette of Fig.~\ref{fig:z2_plaquette_ions_scheme} {\bf (a)} to the  full rectangular ladder: we would need  $3N -2$ gauge qubits to make the model gauge invariant, but only have $N$ at our disposal.
 
\begin{figure}[t]
	\centering
	\includegraphics[width=1\columnwidth]{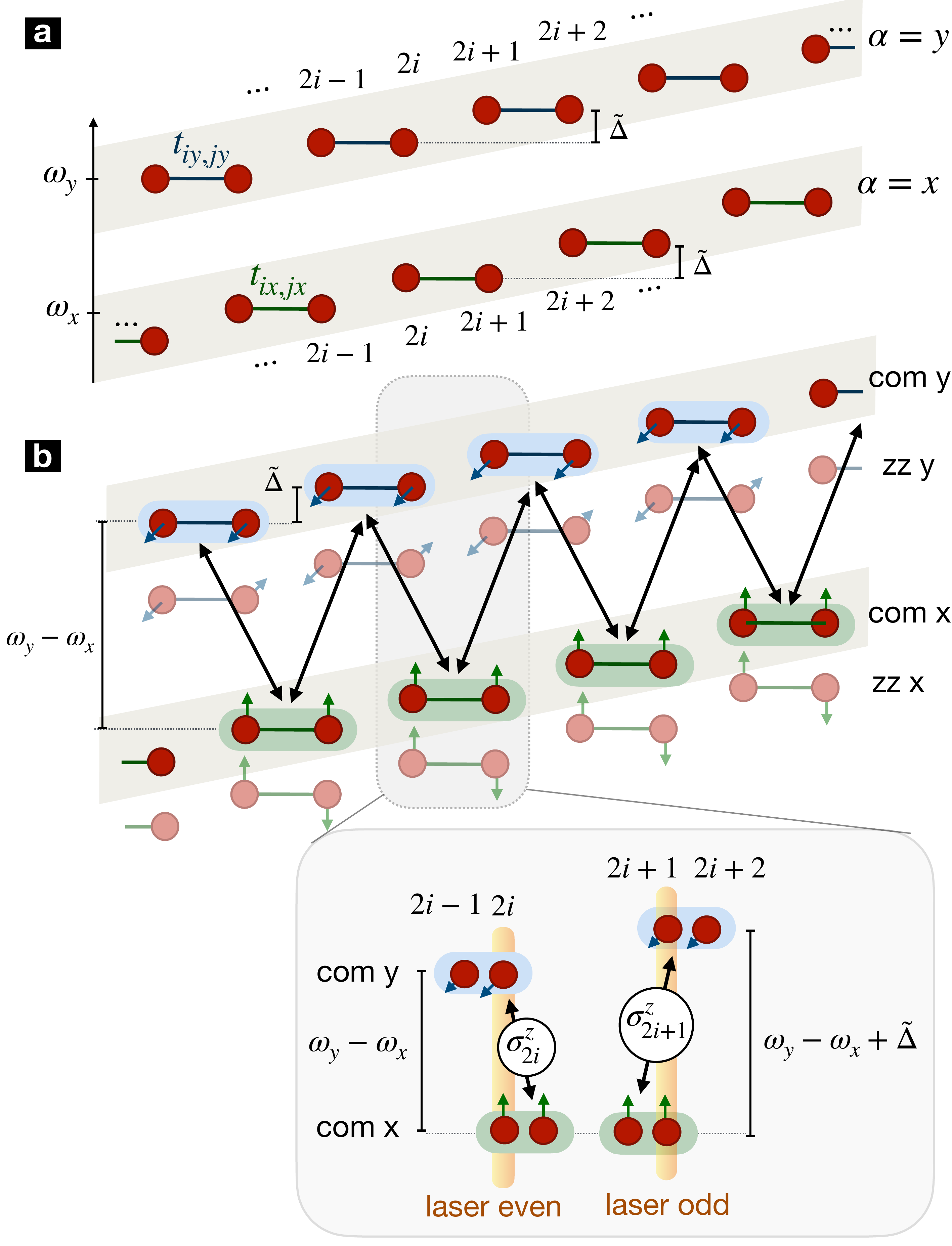}
	\caption{{\bf Synthetic dimensional reduction for a $\mathbb{Z}_2$ chain: } {\bf (a)} We represent the transverse vibrational degrees of freedom of a trapped-ion chain in a frequency scheme, where the corresponding trap frequencies $\omega_x< \omega_y$ can be resolved by external parametric drives. The introduction of the  site-dependent shift of the frequencies in Eq.~\eqref{dimer_gradient} leads to a two-site gradient here depicted by $\tilde{\Delta}$. For $|t_{ij}|\ll\tilde{\Delta}$, the exchange of vibrational quanta  {leads} to alternating dimers, here depicted by green solid  lines. {\bf (b)} As a consequence of this exchange, the vibrational states inside the dimers split into center of mass (com) and zigzag (zz) modes, which can also be resolved in energies. As shown in the inset, we apply a pair of state-dependent parametric drives addressed to even-odd or odd-even dimers for the com modes, respectively,  in order to induce the desired light-shift potential underlying the state-dependent tunneling of Eq.~\eqref{eq:link_ions2}.  }
	\label{fig:z2_chain_reduction}
\end{figure}

 Accordingly, the horizontal links cannot be gauged with the available $\mathbb{Z}_2$ fields, which are already in use to gauge the vertical ones. In order to obtain a gauge-invariant model, the idea is to make a synthetic dimensional reduction by exploiting the collective normal modes as follows. We first introduce a site-dependent shift of the effective on-site energies of   Eq.~\eqref{eq:trap_freq}
 \beq
 \label{dimer_gradient}
 \begin{split}
\omega_{2i-1,y}-\omega_{2i,y}=0,\hspace{2ex}\omega_{2i+1,y}-\omega_{2i,y}=\tilde{\Delta},\\
\omega_{2i-1,x}-\omega_{2i,x}=\tilde{\Delta},\hspace{2ex}\omega_{2i+1,x}-\omega_{2i,x}=0,
 \end{split}
\eeq
where we have introduced a  parameter for these shifts $\tilde{\Delta}$ that fulfills 
$|t_{(i,\alpha)(j,\alpha)}|\ll |\tilde{\Delta}|$. In this regime, the tunneling of phonons vibrating along the $x$-axis ($y$-axis) can only take place within dimers, i.e. pairs of neighbouring sites, composed of  even-odd (odd-even) sites (see Fig.~\ref{fig:z2_chain_reduction}{\bf (a)}). In analogy with ultracold atoms, this dimerisation could be obtained from the light shift of a tilted optical super-potential, or by working with micro-fabricated traps that allow one to design the local trap frequencies individually~\cite{https://doi.org/10.48550/arxiv.quant-ph/0501147,PhysRevA.73.032307,PhysRevLett.96.253003,PhysRevA.77.022324,PhysRevLett.100.013001,PhysRevLett.102.233002,Kumph_2011,Welzel2011,Sterling2014,Mielenz2016,Bruzewicz2016,Kumph_2016,PhysRevX.10.031027}. As depicted in Fig.~\ref{fig:z2_chain_reduction}   {\bf (b)}, the normal vibrational modes of the chain then break into a collection of  center-of-mass and zigzag modes~\eqref{eq:com_zz_modes} that only have support on the alternating dimers
\beq
\begin{split}
a_{{ \rm c}, 2i-1,y}&=\textstyle{\frac{1}{\sqrt{2}}}(a_{2i-1,y}+a_{2i,y}), \hspace{1ex}
a_{{\rm z},2i-1, y}=\frac{1}{\sqrt{2}}(a_{2i-1,y}-a_{2i,y}),\\ 
a_{{ \rm c}, 2i,x}&=\textstyle{\frac{1}{\sqrt{2}}}(a_{2i,x}+a_{2i+1,x}), \hspace{3ex}
a_{{\rm z},2i, x}=\frac{1}{\sqrt{2}}(a_{2i,x}-a_{2i+1,x}),
\end{split}
\eeq
where the index $i\in\{1,\cdots N/2\}$ labels the different dimers. 

The additional  ingredient of the proposed scheme is a pair of  light-shift optical potentials, each of which is  only addressed to the ions that sit on the even or odd sites of the chain, and can be described by Eq.~\eqref{eq:state_dep_parametric_ions} with the corresponding restrictions in the addressed ions. As depicted in the inset of Fig.~\ref{fig:z2_chain_reduction}   {\bf (b)}, the even (odd) optical potentials are tuned to required energy offsets $\omega_{\rm d}=\omega_y-\omega_x$ ($\omega_{\rm d}=\omega_y-\omega_x+\tilde{\Delta}$). In this expression,  we further assume that the even terms lie in the resolved-sideband limit of Eq.~\eqref{eq:constraints_param_tunneling_ions_plaquette} and, analogously, the odd ones lie in the resolved-sideband limit  taking into account the additional energy shift by $\tilde{\Delta}$. As a result of these conditions, using the lessons learned from the previous simpler setups, one can check that the corresponding state-dependent parametric couplings  will dress the tunneling between the center-of-mass modes of neighbouring $x$ and $y$  dimers  while minimising the tunneling between  the zigzag modes. In this way, we are halving the number of degrees of freedom, a precursor of the aforementioned synthetic  dimensional reduction, and the available trapped-ion qubits serve to simulate the $\mathbb{Z}_2$ gauge field defined on the links of a dimension-reduced  chain.

\begin{figure*}[t]
	\centering
	\includegraphics[width=1.2\columnwidth]{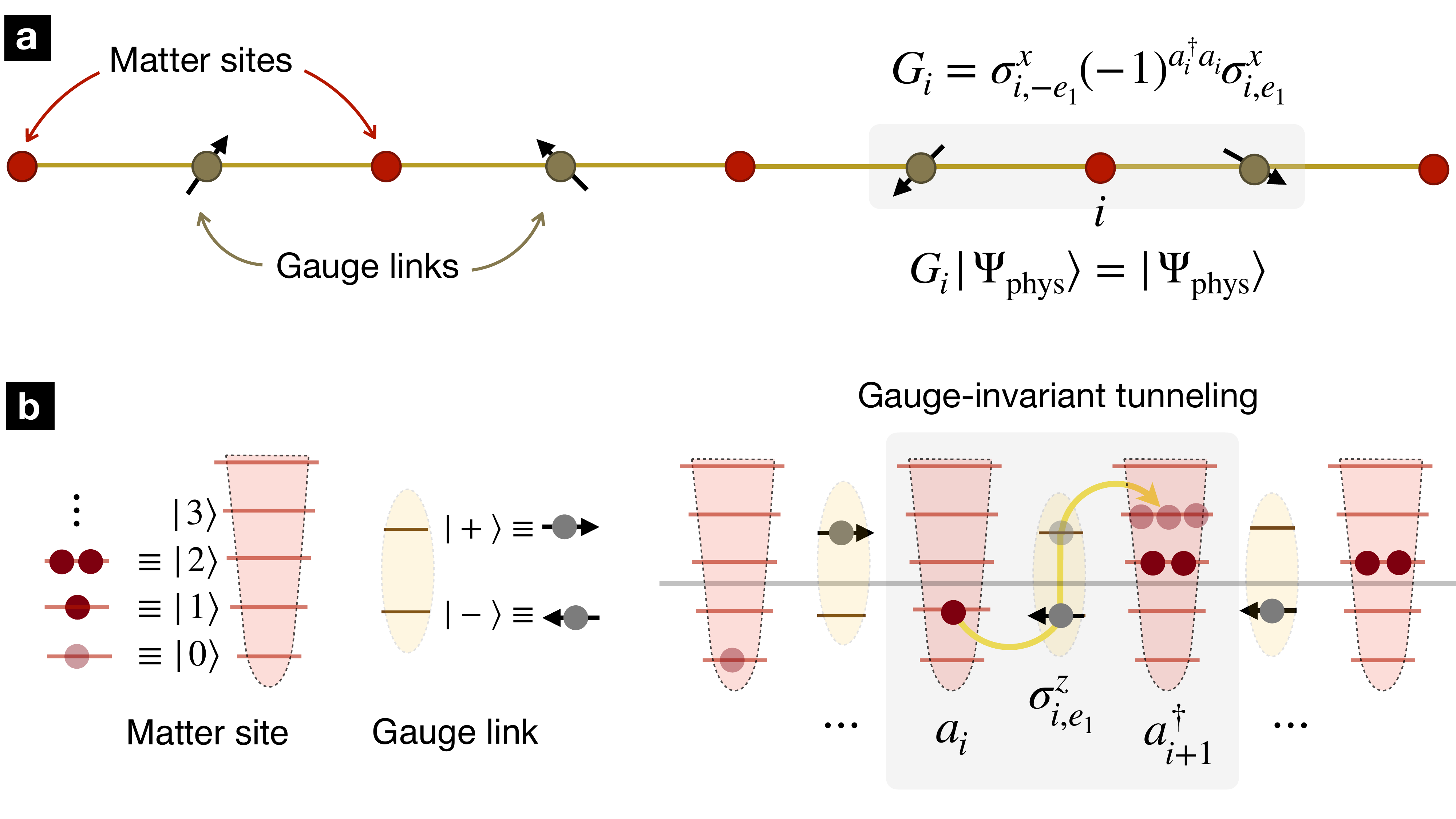}
	\caption{{\bf Gauss' law and  gauge-invariant tunneling in a bosonic $\mathbb{Z}_2$  chain:} {\bf (a)} Schematic representation of a $\mathbb{Z}_2$ gauge theory of bosonic matter on a one-dimensional chain. The bosons sit on the lattice sites (red circles), whereas the gauge fields reside on the links and correspond to qubits. The shaded area represents the Gauss' law in the bulk. {\bf (b)} At each matter site, we can have any integer number of bosons, whereas the gauge fields in the links have two possible states in a given basis, here the electric-field Hadamard basis. On the right panel, we represent schematically the gauge-invariant tunneling of a boson towards a neighbouring site, which is already occupied by two bosons. As depicted with the yellow arrow,  this requires the gauge qubit to be flipped  $\ket{-_{i,{\bf e}_1}}\to\ket{+_{i,{\bf e}_1}}$, such that the electric  string grows.  }
	\label{fig:z2_chain_scheme}
\end{figure*}

As depicted in Fig.~\ref{fig:z2_chain_reduction} {\bf (b)}, this allows us to derive an effective dressed tunneling 
\beq
\label{eq:link_ions2}
\begin{split}
V_{1}(t)\approx \sum_{i=1}^{N/2}
\frac{\Omega_{\rm d}}{4}\bigg(&\ee^{+\ii\phi_{2i}}a^{{\dagger}}_{{\rm c}, 2i-1,y}\sigma^z_{2i}a^{\phantom{\dagger}}_{{\rm c},2i,x}\\
&+\ee^{-\ii\phi_{2i+1}}a^{{\dagger}}_{{\rm c}, 2i,x}\sigma^z_{2i+1}a^{\phantom{\dagger}}_{{\rm c},2i+1,y}\bigg)+{\rm H.c.}, 
\end{split}
\eeq
 where the microscopic parameters are again those of Eq.~\eqref{eq:parameters_tunneling_synthetic}.
Note that the alternation in the sign of the tunneling phases $\ee^{+\ii\phi_{2i}}$ and $ \ee^{-\ii\phi_{2i+1}},$ is caused by the fact that the light-shift  potential  provides (absorbs) the missing (excess) energy to activate the tunneling against the corresponding energy offset $\tilde{\Delta}$, making the dressed tunneling resonant.
In this case, in order to make this complex phase irrelevant, we should align the laser wave-vector in a direction orthogonal to the chain $\boldsymbol{k}_{\rm d}\cdot\boldsymbol{r}_i^0=0$. 

In addition to these ingredients, we would again need a second tone that drives the qubit transition~\eqref{eq:second_tone}, corresponding to a carrier term with a resonance condition that includes the ac-Stark shifts~\eqref{eq:constraints_carrier_ac_stark}. We can then perform the aforementioned dimensional reduction by considering the  formal mapping 
\beq
a^{\phantom{\dagger}}_{{\rm c},2i,x},a^{{\dagger}}_{{\rm c},2i,x}\mapsto a^{\phantom{\dagger}}_{2i},a^{{\dagger}}_{2i},\hspace{1ex} a^{\phantom{\dagger}}_{{\rm c},2i-1, y},a^{\dagger}_{{\rm c},2i-1,y}\mapsto a^{\phantom{\dagger}}_{2i-1}, a^{{\dagger}}_{2i-1},
\eeq
for $i\in\{1,\cdots,N/2\}$, such that the odd (even) matter sites correspond to the $y$-axis ($x$-axis) center-of-mass modes. Besides,  the gauge qubits are identified via the standard mapping
\beq
\sigma_{i}^x,\sigma_{i}^z\mapsto \sigma_{i,{\bf e}_1}^x,\sigma_{i,{\bf e}_1}^z,
\eeq
where the index now covers all links $i\in\{1,\cdots,N-1\}$. 

In this way, we have reduced the synthetic two-leg ladder onto a  chain (see Fig.~\ref{fig:z2_chain_scheme}{\bf (a)}), halving the number of matter sites and reducing the required  links. 
Accordingly,  the available number of physical qubits suffice to gauge all synthetic tunnelings, obtaining an effective gauge-invariant model that generalises the single-link case~\eqref{eq:tunneling_gauge} to the full chain
\beq
\label{eq:gauge_chain}
H_{\rm eff}=\sum_{i=1}^{N-1}\!\!\left(\left(t^{\phantom{\dagger}}_{i,\textbf{e}_1}a^{{\dagger}}_{i+1}\sigma^z_{i,{\bf e}_1}a^{\phantom{\dagger}}_{i}+{\rm H.c.}\right)+h\sigma^x_{i,{\bf e}_1}\right).
\eeq
Here, the gauge-invariant tunneling (see Fig.~\ref{fig:z2_chain_scheme} {\bf (b)}) has a strength  that is homogeneous along the chain, and gets halved with respect to the single-link case~\eqref{eq:tunneling_gauge},  while the electric field remains to be the same 
\beq
t^{\phantom{\dagger}}_{i,\textbf{e}_1}=\frac{\Omega_{1,2}}{4}{\eta_{x}\eta_y},\hspace{2ex} h=\frac{\tilde{\Omega}_{\rm d}}{2}.
\eeq
Paralleling  our discussion of the single-link case, the invariance under local $\mathbb{Z}_2$ transformations of this gauge theory is generated by the following operators
\beq
\label{eq:generators_plaquette1}
G_i=\ee^{\ii\pi a_i^\dagger a_i^{\phantom{\dagger}}}\!\!\!\!\!\prod_{\ell\in\mathcal{L}\{i\}}\!\!\sigma_{i,\ell{\bf e}_1}^x,
\eeq
where $j\in \mathcal{L}\{i\}$ is the set of links that surround a given matter site labelled by $i$, namely $\pm \textbf{e}_1$ in the bulk (Fig.~\ref{fig:z2_chain_scheme}{\bf (a)}), and $+ \textbf{e}_1$ ($- \textbf{e}_1$) at the leftmost (rightmost) boundary sites $i=1$ ($i=N$).

Once this method has been presented, let us discuss  neat dynamical effects that go beyond the previous periodic oscillations in the link/plaquette limits. Let us recall that, as depicted in Fig.~\ref{fig:rabi_flopping_dynamics} {\bf (b)}, the gauge-invariant tunneling gets inhibited as one increases the electric field $h$, which we referred to as a precursor of  confinement in larger lattice gauge theories. In the following subsections, we discuss how the trapped-ion quantum simulator~\eqref{eq:gauge_chain} would allow for a clear manifestation of this confinement, focusing particularly  in the one- and two-particle sectors. We will then move to half-filling, where minor modifications of the quantum simulator will allow to explore if the phenomenon of string breaking in real-time dynamics~\cite{PhysRevLett.111.201601,PhysRevX.6.011023,Kuehn_2019,Magnifico2020realtimedynamics,PhysRevLett.124.180602} also occurs in this  gauge theory.

We consider a chain of $N$ bosonic matter sites, and $(N-1)$ gauge fields (see Fig.~\ref{fig:z2_chain_scheme} {\bf (a)}), such that the full Hilbert space is  $\mathcal{H}=\mathcal{F}\otimes\mathbb{C}^{2(N-1)}$, with $\mathcal{F}=\oplus_{n=0}^\infty\mathcal{F}_{n}$. Here, each subspace is  $ \mathcal{F}_{n}={\rm span}\{\ket{n_1}\otimes\ket{n_2}\otimes\cdots \otimes\ket{n_N}:n_1+n_2+\cdots n_N=n\}$. Due to the global $U(1)$ symmetry in the matter sector, the dynamics will take place within one of these subspaces  depending on the number of bosons of the initial state. In addition, due to the gauge symmetry, the physical states are further restricted  via Gauss' law, which now imposes $N$ constraints
\beq
\label{eq:gauss_law_chain}
G_i\ket{\Psi_{\rm phys}}=\ee^{\ii\pi q_i} \ket{\Psi_{\rm phys}}, \,\,\forall i\in\{1,\cdots,N\}.
\eeq
We recall that $q_i\in\{0,1\}$ are  the background $\mathbb{Z}_2$ charges that specify the super-selection sector where the dynamics occurs. This Gauss' law at the bulk  has  also been depicted in Fig.~\ref{fig:z2_chain_scheme} {\bf (a)}. In Fig.~\ref{fig:z2_chain_scheme} {\bf (b)}, we represent the Hilbert spaces of the bosonic matter sites and qubit links, as well as the transitions involved in a gauge-invariant tunneling.

\subsubsection{ One-boson sector:Wannier-Stark localisation}

\begin{figure*}[t]
	\centering
	\includegraphics[width=1.85\columnwidth]{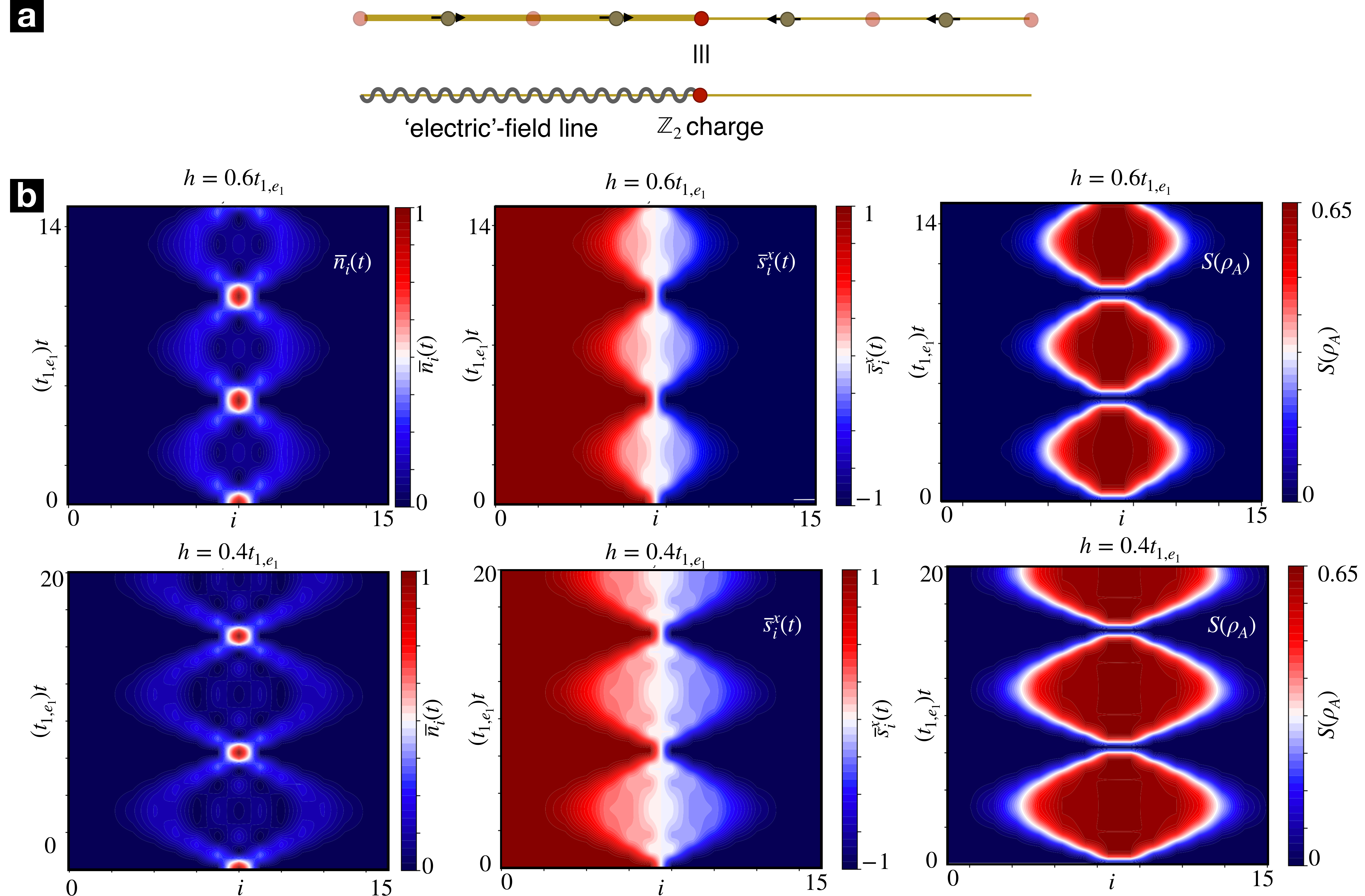}
	\caption{{\bf Wannier-Stark localisation of a single $\mathbb{Z}_2$ charge in the  chain: }  {\bf (a)} Single bosonic particle (filled red circle), with an attached electric-field string that connects it to the background charge $q_1=1$ at the left boundary. {\bf (b)} Contour plots for the dynamics of the boson distribution $ \overline {n}_{i}(t)=\langle a_i^\dagger a_i^{\phantom{\dagger}}(t)\rangle$, the electric field on the links  $ \overline {s}^x_{i}(t)=\langle \sigma^x_{{i},{\bf e}_1}(t)\rangle$, as well as the block entanglement entropy $S(\rho_A)=-{\rm Tr}\{\rho_A\log\rho_A\}$. We consider a chain of $N=16$ sites, and set the transverse electric field to $h=0.6t_{1,{\bf e}_1}$ (upper row), and $h=0.4t_{1,{\bf e}_1}$ (lower row). The initial state is $\ket{\Psi(0)}=\ket{	\sim\!\!\sim\!\!\!\!\!\!\bullet_{7}}$ in the notation of Eq.~\eqref{eq:phys_subspace_basis}, which corresponds to a product state for the boson being localised at the center of the chain, and a domain wall of the gauge qubits with respect to the Hadamard $x$-basis magnetisation.}
	\label{fig:z2_chain_localization_charges}
\end{figure*}

In this case, the physical subspace $\mathcal{V}_{\textrm{phys}}=\textrm{span}\{\ket{	\sim\!\!\sim\!\!\!\!\!\!\bullet_i}:i\in\{1,\cdots, N\}\}	\subset\mathcal{F}_1\otimes\mathbb{C}^{2(N-1)}$ can be be spanned by 
\beq
\label{eq:phys_subspace_basis}
\begin{split}
\ket{	\sim\!\!\sim\!\!\!\!\!\!\bullet_i}&=\left(\prod_{\ell< i}\sigma^z_{\ell,\textbf{e}_1}\right)a_i^\dagger\ket{\rm vac}.
\end{split}
\eeq
Here, as  the lattice starts and finishes with a matter site, the vacuum  $\ket{\rm vac}=\ket{0_1,-_{1,\textbf{e}_1},\cdots,0_{N-1}, -_{N-1,\textbf{e}_1},0_{N}}$  belongs to the  super-selection sector with a single background charge  at the leftmost boundary $q_1=1$, and zero elsewhere $q_j=0,\, \forall j\neq 1$. As a result of the composite operator in Eq.~\eqref{eq:phys_subspace_basis}, the basis states represent a boson at site $i$, together with a domain-wall configuration of the gauge qubits in the Hadamard basis $\ket{\sim\!\!\sim\!\!\!\!\!\!\bullet_i}=\ket{0_1,+_{1,\textbf{e}_1},\cdots, +_{i-1,\textbf{e}_1},1_i,-_{i,\textbf{e}_1},\cdots, -_{N-1,\textbf{e}_1},0_{N}}$. This
 has been depicted in  Fig.~\ref{fig:z2_chain_localization_charges} {\bf (a)}, where the  boson  at site $i$ is connected to an electric field line that extends towards the left boundary, connecting the static charge and the dynamical one. We thus see that the dynamics, which in principle is defined in an exponentially-large Hilbert space, gets restricted to a much smaller subspace whose size only grows linearly with the length of the chain. Up to an irrelevant shift of the zero of energies, the Hamiltonian~\eqref{eq:gauge_chain} can be rewritten as
\beq
\label{eq:gauge_chain1}
H_{\rm eff}=\sum_{i=1}^{N-1}\left(t^{\phantom{\dagger}}_{i,\textbf{e}_1}\ket{\sim\!\!\sim\!\!\!\!\!\!\bullet_{i+1}}\bra{\sim\!\!\sim\!\!\!\!\!\!\bullet_i}+{\rm H.c.}\right)+2h\sum_{i=1}^N i\ket{\sim\!\!\sim\!\!\!\!\!\!\bullet_i}\bra{\sim\!\!\sim\!\!\!\!\!\!\bullet_i},
\eeq
which corresponds to a tight-binding problem where a single composite particle built from the dynamical $\mathbb{Z}_2$ charge with the attached electric string,  tunnels in the background of a linear potential. This problem maps exactly to the tight-binding Wannier-Stark ladder~\cite{RevModPhys.34.645,PhysRevB.8.5579,GRIFONI1998229,doi:10.1080/01418639608240331,Hartmann_2004}. In contrast to the analogous problem in classical physics, where the particle would simply fall down the linear slope until reaching the  bottom, the quantum particle can only oscillate around the initial position, leading to the so-called Wannier-Stark localisation. In the present context of the $\mathbb{Z}_2$ gauge theory~\eqref{eq:gauge_chain}, these oscillations  will be accompanied by the periodic stretching and compressing of the attached electric-field line, as we now discuss in detail.

In Appendix~\ref{sec:WS_solution}, we present a detailed discussion of an exact solution of the real-time dynamics in the boson sector. These analytical expressions serve as a benchmark for a numerical method that can be easily adapted to other matter contents in which an exact solution no longer exists:  matrix product states  (MPSs) \cite{schollwock2011density,orus2014practical}. The MPSs are major tools for the classical numerical simulation of 1D strongly correlated models. These methods  capture the interplay of locality and entanglement by expressing an entangled many-body wave function in terms of local tensors.
For static calculations, the MPS-based density matrix renormalization
group (DMRG) \cite{white1992density} has become a common choice for obtaining 
the ground and a few low-lying excited states of many-body Hamiltonians, 
as it can reach remarkable  accuracy and reliability. In the case of real-time 
evolution, the breakthrough came with the development of the time-evolving-
block-decimation (TEBD) algorithm \cite{vidal2003efficient}, but it can also be 
treated using a variety of methods \cite{paeckel2019time}.
Among these, the time-dependent variational principle (TDVP) \cite{haegeman2011time}, which uses a Lie-Trotter decomposition to integrate a train of tensors sequentially, is less error prone and more accurate than other available methods. We thus select  it  as our method of choice for the current work.

In TDVP, the MPS ansatz $|\psi (t) \rangle$ can be understood as a variational  manifold of reduced dimensionality within the full many-body Hilbert space. The time evolution of the MPS is obtained by computing the action of the Hamiltonian $H$  along the tangent direction to this variational manifold, which we recall is described by the MPS bond dimension $\chi$. This approach
leads to an effective Schrödinger equation for states constrained to the MPS manifold that reads as follows 
\beq 
\ii \frac{d}{dt} |\psi(t)\rangle = P_{T_{|\psi \rangle}} {H} |\psi (t) \rangle
\eeq
where $P_{T_{|\psi \rangle}}$ is an orthogonal projector onto the tangent space of $|\psi(t)\rangle$. In our work, we follow the prescription described in Ref.~\cite{haegeman2016unifying}  to implement a one-site version of TDVP. Moreover, we use the time-step $\delta t=0.05/t_{i,{\bf e}_1}$ and $\chi=100$ for the TDVP calculations. In the Appendix, we present a detailed benchmark of the MPS numerics using   this specific expectation value. We can now move beyond it, and look into other observables that give further insight in the localisation phenomenon. 
As depicted in the first two panels of Fig.~\ref{fig:z2_chain_localization_charges} {\bf (b)}, the boson density  and the attached electric-field line remain localised around the initial position. On the left of this figure, we present two contour plots for  the expectation value of the  boson number operator $\overline {n}_{i}(t)=\langle a_i^\dagger a_i^{\phantom{\dagger}}\!\!(t)\rangle$ as a function of time and the site index of the lattice. The two plots correspond to different values of the electric field $h=0.6t_{1,{\bf e}_1}$ and $h=0.4t_{1,{\bf e}_1}$, and one can see how the spread of the breathing-type oscillations of the boson decreases as the value of the electric field $h$ grows. The next column shows the corresponding contour plots of the electric field sustained by the gauge qubits $\overline{s}^x_i(t)=\langle \sigma^x_{i,{\bf e}_1}(t)\rangle$. In these two plots, one can see how the   electric-field string oscillates periodically instead of spreading ballistically, which distorts the initial domain-wall correlations.
In the third column of Fig.~\ref{fig:z2_chain_localization_charges} {\bf (b)}, we also represent the block entanglement entropy  $S(\rho_A)=-{\rm Tr}\{\rho_A\log\rho_A\}$, showing that the region where the stretching and compressing of the electric-field line  takes place coincides with the region where entanglement is built up in the real-time dynamics.

Note that, after multiples of the exchange period $\Delta t_{\rm ex}=2\pi/h$, the boson and the domain wall on the qubits fully refocus to the initial position. The resulting state becomes a product state, as can be inferred from the vanishing entanglement entropy at those instants of time.  This is different from the trend in the limit of a single link~\eqref{eq_Rabi flops_link}, where the time it takes to the boson to return to the initial site depends on the ratio of the tunneling and the electric field (see Fig.~\ref{fig:rabi_flopping_dynamics} {\bf (b)}).  As we increase the value of the electric field $h$, the oscillations become faster and the breathing-type dispersion is more localised. This so-called Wannier-Stark localisation is particularly transparent in the regime $2\gamma\ll1$, where the asymptotic of Bessel functions allows us to show that the boson remains exponentially localised around the center   $\overline{ n}_{{N}/{2}+ r}(t)\leq J_r^2(2\gamma)\approx \textrm{exp}\{-r/\xi_{\rm loc}\}$ with $\xi_{\rm loc}=-1/2\log\gamma$. By using the maximum of the first-order Bessel function, one can also predict the short-time dynamics of the boson  $\overline{ n}_{{N}/{2}\pm 1}(t)\approx J_1^2(2\gamma ht)$ to be ballistic, displaying an initial linear light-cone-like spreading. However, as time elapses, the effects of the stretching/compressing electric-field string start becoming manifest, and the dynamics is no longer ballistic but, instead, displays a breathing-type periodic behaviour. This also contrast with the dynamics of disordered one-dimensional systems~\cite{doi:10.1080/00018736100101271}, where an initially-localised particle tends to a stationary exponentially-localised solution characteristic of Anderson localisation, which can also be observed with trapped ions~\cite{Bermudez_2010}. In such Anderson localised system, there  is no breathing-like behaviour as displayed in this case. Let us now move to a two-particle sector, where one can see how the Wannier-Stark localisation develops into a specific confinement phenomenon.

 \begin{figure*}[t]
	\centering
	\includegraphics[width=1.85\columnwidth]{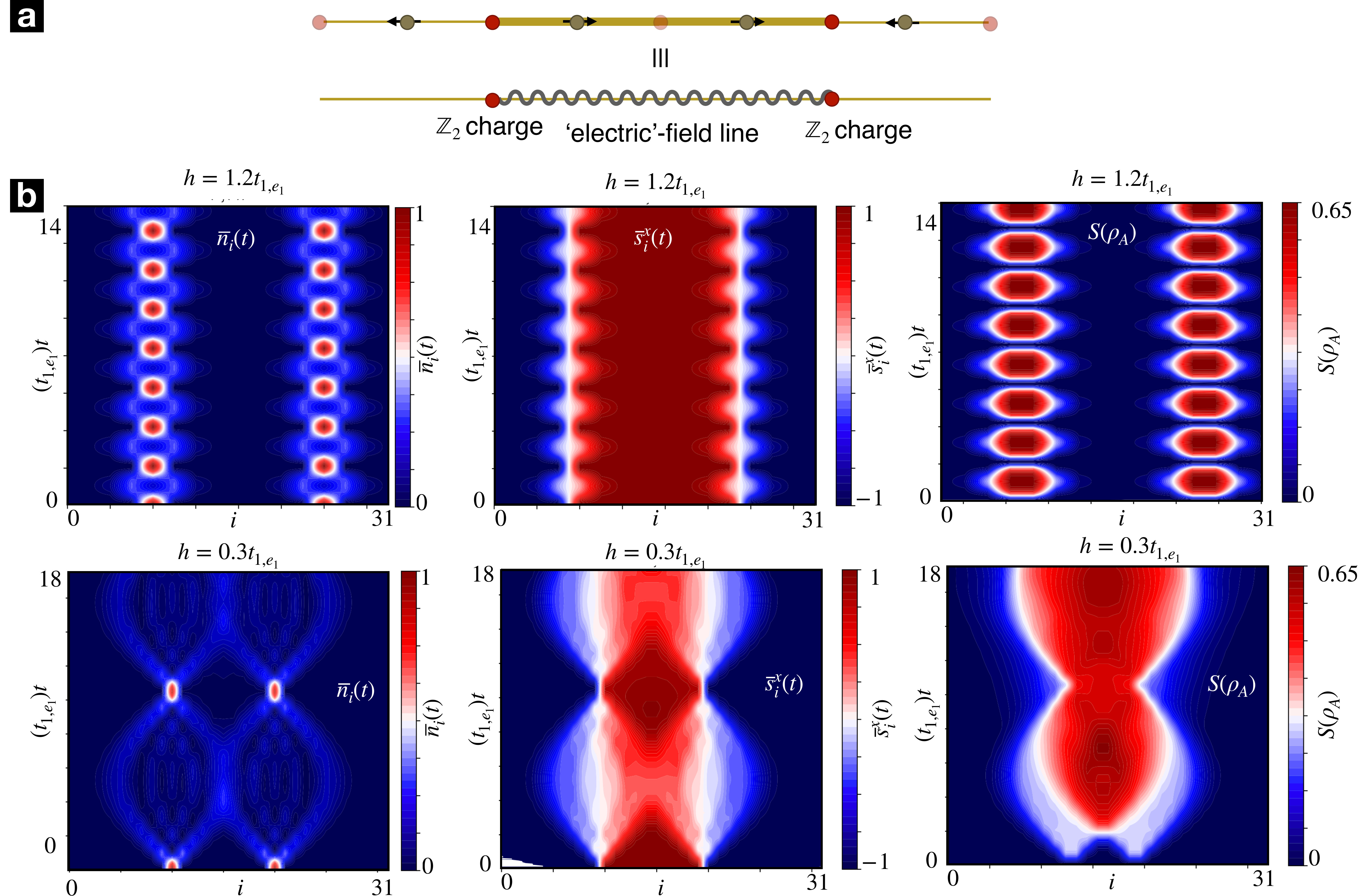}
	\caption{{\bf Wannier-Stark  confinement of a pair of $\mathbb{Z}_2$ charges in the  chain: } {\bf (a)} Two bosonic particles (filled red circles), with an attached electric-field string  connecting them. {\bf (b)} Contour plots for the dynamics of the boson distribution $ \overline {n}_{i}(t)=\langle a_i^\dagger a_i^{\phantom{\dagger}}(t)\rangle$, the electric field on the links  $ \overline {s}^x_{i}(t)=\langle \sigma^x_{{i},{\bf e}_1}(t)\rangle$, as well as the block entanglement entropy $S(\rho_A)=-{\rm Tr}\{\rho_A\log\rho_A\}$. We consider a chain of $16, 32$ sites, and set the  electric field to $h=1.2t_{1,{\bf e}_1}$ (upper row), and $h=0.3t_{1,{\bf e}_1}$ (lower row). The initial state is $\ket{_i\bullet\!\!\!\!\sim\!\!\sim\!\!\!\!\!\!\bullet_j}$ in the notation of Eq.~\eqref{eq:phys_subspace_basis_2_bosons}, which corresponds to a product state for the bosons being localised at sites $i.j$, and a domain wall of the gauge qubits for the links in between.  }
	\label{fig:z2_chain_localization_2_phonons}
\end{figure*}

\subsubsection{Two-boson sector:Wannier-Stark$\,$confinement}

In the two-boson sector, the physical subspace $\mathcal{V}_{\textrm{phys}}=\textrm{span}\{\ket{{}_i\bullet\!\!\!\!\sim\!\!\sim\!\!\!\!\!\!\bullet_j}:i,j\in\{1,\cdots, N\}, \,\, {\rm and}\,\, j\geq i\}	\subset\mathcal{F}_2\otimes\mathbb{C}^{2(N-1)}$ can be be spanned by the following states
\beq
\label{eq:phys_subspace_basis_2_bosons}
\begin{split}
\ket{_i\bullet\!\!\!\!\sim\!\!\sim\!\!\!\!\!\!\bullet_j}=a_i^{\dagger}\left(\prod_{i\leq \ell< j}\sigma^z_{\ell,\textbf{e}_1}\right)a_j^\dagger\ket{\rm vac}.
\end{split}
\eeq
 These states contain a pair of bosons connected by an electric-field line (see  Fig.~\ref{fig:z2_chain_localization_2_phonons} {\bf (a)}). In analogy to the single-boson sector~\eqref{eq:sing_part_recurssion}, one can expand  these two-particle solutions as
 \beq
\ket{\epsilon_{m,P}}=\sum_{i=1}^N\sum_{j\geq i}c_{i,j}^{\phantom{*}}\ket{_i\bullet\!\!\!\!\sim\!\!\sim\!\!\!\!\!\!\bullet_j}=\sum_{i,j}{}^{\!\!'}c_{i,j}^{\phantom{*}}\,\,a_i^{\dagger}\left(\prod_{i\leq \ell< j}\sigma^z_{\ell,\textbf{e}_1}\right)a_j^\dagger\ket{\rm vac},
 \eeq
 where the matrix of coefficients is symmetric $c_{i,j}=c_{j,i}$, and is further constrained by normalisation $\langle{\epsilon_{m,P}}\ket{\epsilon_{m,P}}=1$. This type of states have been used as Bethe-type ansatz for the two-body problem of a Bose-Hubbard model~\cite{bose-hubbard}, where they allow one to determine the scattering and bound states~\cite{Valiente_2008,PhysRevA.81.042102,PhysRevA.90.043606}.  In our case, the  recurrence relation obtained after applying the $\mathbb{Z}_2$ gauge-theory Hamiltonian reads  
\beq
\label{eq:2_part_recurssion}
\begin{split}
 t^{\phantom{*}}_{i-1,\textbf{e}_1}c_{i-1,j}^{\phantom{*}}+t_{i,\textbf{e}_1}^*c_{i+1,j}^{\phantom{*}} +t^{\phantom{*}}_{j-1,\textbf{e}_1}c_{i,j-1}^{\phantom{*}}+t_{j,\textbf{e}_1}^*c_{i,j+1}^{\phantom{*}}\\
 +2h(j-i)\,c_{i,j}^{\phantom{*}}=\epsilon_{m,P} c_{i,j}^{\phantom{*}}.
 \end{split}
\eeq
In addition to the tunneling, which resembles a 2D tight-binding problem, we see that the potential only depends on the relative distance of the two bosons. Hence, the problem is  different from the Wannier-Stark case of  a pair of tight-binding charges subjected to a constant background electric field. In Appendix~\ref{sec:WS_solution}, we present a detailed discussion of the exact solution in this two-boson sector, and use the analytical expressions for the density to    benchmark  the MPS numerics.

In order to  highlight other aspects of the two-charge confinement, we present in Fig.~\ref{fig:z2_chain_localization_2_phonons} {\bf (b)} the MPS numerical results for other observables.
As shown in the first column, where we present a contour plot for the boson distribution $ \overline {n}_{i}(t)=\langle a_i^\dagger a_i^{\phantom{\dagger}}(t)\rangle$ for $h=1.2t_{1,{\bf e}_1}$ (upper row) and  $h=0.3t_{1,{\bf e}_1}$ (lower row), the center of mass of the two bosons remains  at the centre of the chain as time evolves. In this dynamics,   the two particles disperse developing a pair of  breathing-type oscillations similar to the single-boson case discussed  in Fig.~\ref{fig:z2_chain_localization_charges} {\bf (b)}. For sufficiently-large electric field (upper row), the pair of bosons only perform small oscillations about the initial position, but do not interfere. For weaker electric fields (lower row), the width of the breathers  is big enough such that their oscillations overlap, and we find  interference effects through which the probability to find a boson at the center of the chain adds constructively for given instants of time.  As a result, new breathing-type oscillations are superposed. In the center and rightmost columns of Fig.~\ref{fig:z2_chain_localization_charges} {\bf (b)}, we represent the corresponding electric field $ \overline {s}^x_{i}(t)=\langle \sigma^x_{{i},{\bf e}_1}(t)\rangle$, as well as the block entanglement entropy $S(\rho_A)$. One can clearly see that the interference also appears in the gauge degrees of freedom via the structure of the initial domain wall, i.e. electric-field line. Regarding the evolution of the block entanglement, we see that quantum correlations build up at the edges of the initial electric-field line, and then grow defining a light-cone like dispersion. After this expansion, inside the mutual effective light cone, entanglement can grow further and show a characteristic interference pattern that coincides with the region where the electric field string stretches and compresses.

After the same periodic exchange durations found in the single-particle case, namely multiples of $\Delta t_{\rm ex}=2\pi/h$, the dispersion and interference of the bosons  refocuses completely, and we come back to the original situation in which the boson pair and the intermediate electric-field line  have a distance of $2r_0$. We thus see that, for sufficiently-large systems in any non-zero electric field $h$, the pair or bosons do not spread ballistically but, instead, disperse up to a maximal distance and then refocus periodically.  This unveils an additional aspect   of the  confinement discussed previously in terms of the bound-state solutions of Eq.~\eqref{eq:eigsnetates_two_body_problem}. The Bessel-function envelope of these solutions, which decays rapidly with the inter-boson distance, underlies the lack of  excitations with a non-zero $\mathbb{Z}_2$ charge from the energy spectrum of the model, which is the ultimate smoking gun for confinement. Given  the connection to the Wannier-Stark physics, we refer to this neat manifestation of confinement   as Wannier-Stark confinement.  

At this point, it should be pointed out that similar real-time signatures of confinement have also been explored numerically for fermionic~\cite{PhysRevLett.111.201601, PhysRevX.6.011023,PhysRevD.96.114501,PhysRevD.98.034505, Magnifico2020realtimedynamics} and bosonic~\cite{PhysRevLett.124.180602} versions of the one-dimensional Schwinger model,  {including more recent results on jet production~\cite{PhysRevLett.131.021902} and particle collisions~\cite{papaefstathiou2024realtime,su2024coldatom}}. The dynamics in these cases is richer, as there can also be spontaneous production of particle-antiparticle pairs  due to the presence of the electric field associated to the electric-field string, i.e. Schwinger pair production mechanism~\cite{PhysRev.82.664}. In the fermionic case~\cite{PhysRevLett.111.201601, PhysRevX.6.011023,PhysRevD.96.114501,PhysRevD.98.034505, Magnifico2020realtimedynamics}, after this pair production and subsequent  recombination,  there is    a string-breaking mechanism, whereby the intermediate electric field relaxes and there is a screening effect for the outer charges, which form new particle-antiparticle pairs of zero net charge that can move freely as a bound meson. Then, the process is reversed by creating  an inverted electric-field line (anti-string) in the bulk that connects an antiparticle-particle pair (anti-pair), which can then be screened again creating new mesons that travel freely and so on.  In order to explore if a string breaking mechanism can occur in our model, we explore the half-filled sector for our simpler $\mathbb{Z}_2$ gauge theory in the following section, and  introduce the local term in Eq.~\eqref{eq:gauge_chain}, to account for the energy cost  of the pair production.

\subsubsection{Half-filled sector: Partial string breaking}

Let us now consider  a charge-density-wave distribution of the bosons. In the case of half filling, one can distribute $N/2$ bosons to populate the odd sites of the chain, whereas the link spins all point in the opposite direction of the transverse term, such that there is no  electric field in the initial state. This state can be considered as a metastable ground state 
\beq
\label{eq:vacuum_state}
\ket{\overline{\rm vac}}=\ket{1_1,-_{1,\textbf{e}_1},0_2,-_{2,\textbf{e}_1},1_3,-_{3,\textbf{e}_1},\cdots,1_{N-1}, -_{N-1,\textbf{e}_1},0_{N}}
\eeq
of a new gauge-invariant Hamiltonian that contains an additional staggered mass term with respect to Eq.~\eqref{eq:gauge_chain}, namely
\beq
\label{eq:z2_staggered}
H_{\rm eff}\!=\!\!\sum_{i=1}^{N-1}\!\!\!\left(\!\left(t^{\phantom{\dagger}}_{i,\textbf{e}_1}a^{{\dagger}}_{i+1}\sigma^z_{i,{\bf e}_1}a^{\phantom{\dagger}}_{i}+{\rm H.c.}\right)\!+h\sigma^x_{i,{\bf e}_1}\right)+\mu\!\sum_{i=1}^N\!(-1)^ia^{{\dagger}}_{i}a^{\phantom{\dagger}}_{i}.
\eeq
Let us note that this additional term is a specific case of the generic detunings in the state-dependent parametric tunneling, which were already discussed around Eq.~\eqref{eq:gauge_plaquette_total2}.
In the limit of a very large ``mass'' $\mu\gg t_{i,{\bf e}_1}$, and for hard-core bosons,  this state~\eqref{eq:vacuum_state} is the gauge-invariant ground state of the Hamiltonian~\eqref{eq:z2_staggered} in a super-selection sector in which Gauss' law~\eqref{eq:gauss_law_chain} has a staggered distribution of static $\mathbb{Z}_2$ charges
\beq
q_i=\half(1-(-1)^i),\hspace{2ex}\forall i\in\{1,\cdots,N\}.
\eeq
In the case of standard bosons, the odd sites are not restricted to single occupancy by the hardcore constraint, and the ground state becomes a highly degenerate manifold. In any case, since we are interested in real-time dynamics, we can always consider this configuration as a reference to build a specific initial state, and then study its real-time dynamics.

 \begin{figure*}[t]
	\centering
	\includegraphics[width=1.5\columnwidth]{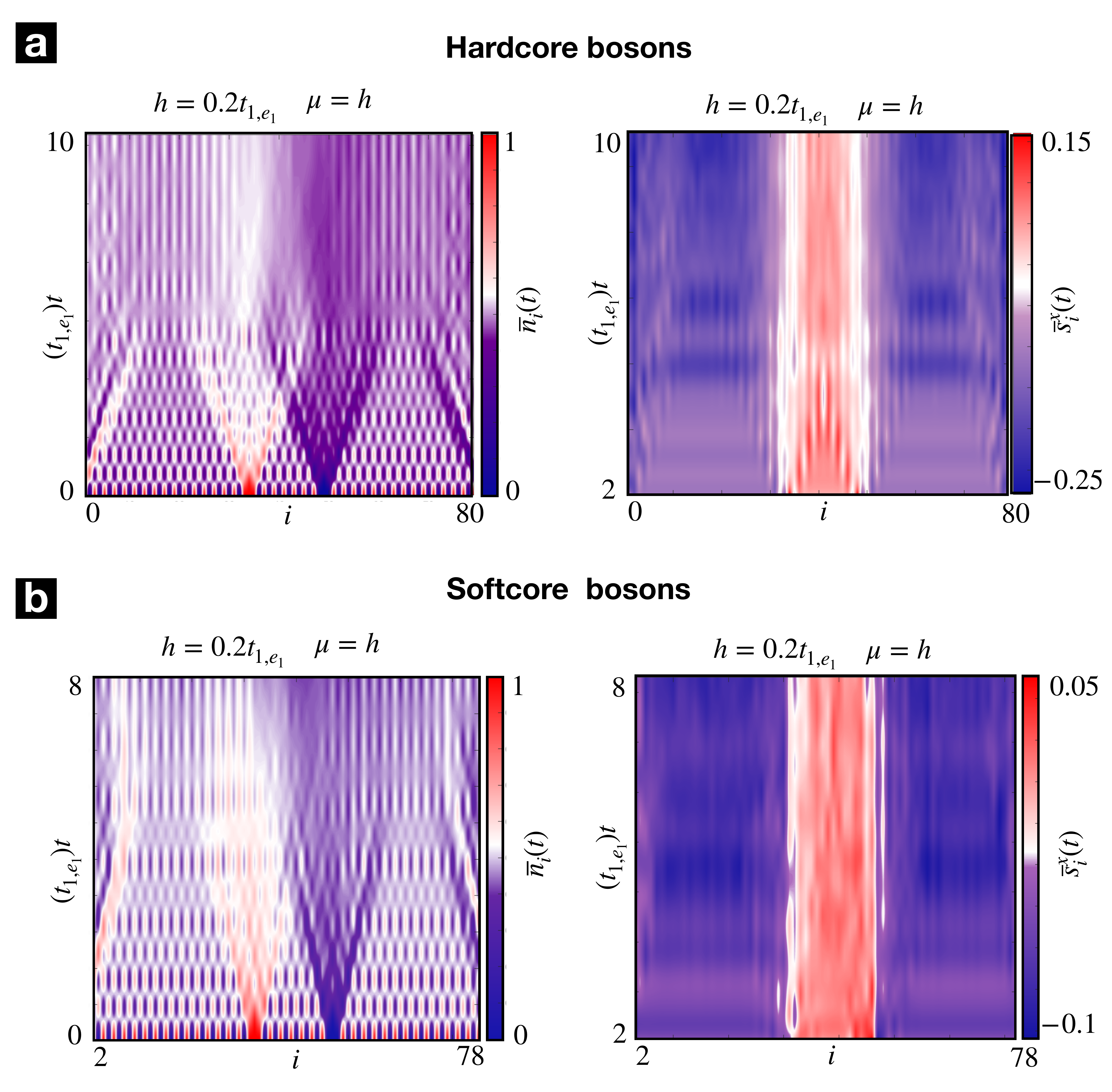}
	\caption{{\bf Partial String breaking in the bosonic $\mathbb{Z}_2$ gauge chain: } Contour plots for the dynamics of the boson distribution $ \overline {n}_{i}(t)=\langle a_i^\dagger a_i^{\phantom{\dagger}}(t)\rangle$, the electric field on the links  $ \overline {s}^x_{i}(t)=\langle \sigma^x_{{i},{\bf e}_1}(t)\rangle$. We consider a chain of $N=80$ sites, and set the transverse electric field to $h=0.2t_{1,{\bf e}_1}$ and a mass $\mu=h$. The initial state is $\ket{_{2i}\bullet\!\!\!\!\sim\!\!\sim\!\!\!\!\hspace{0.1ex}\circ_{2j+1}}$ in the notation of Eq.~\eqref{eq:phys_subspace_string_state}, which corresponds to a product state for the bosons being localised at even sites $i$, and a domain wall of the gauge qubits for the links in between the sites $i_0=32$ and $j_0=48$. In {\bf (a)} we show the results for the hard core bosons case instead in {\bf (b)} we present the results for the soft core bosons where we fixed the local dimension $n_{\rm max}=4$.}
	\label{fig:z2_chain_localization}
\end{figure*}

We can now start from this state, and consider a meson-like excitation by adding a particle-hole excitation in which a single even (odd) site is populated (emptied) by  moving a single boson between two nearest-neighbour sites. To comply with Gauss' law, an electric field must be established at the  link in between, leading to
\beq
\label{eq:phys_subspace_meson}
\begin{split}
\ket{_{2i}\bullet\!\!\!\!\sim\!\!\sim\!\!\!\!\hspace{0.1ex}\circ_{2i+1}}=a_{2i}^{\dagger}\left(\sigma^z_{2i,\textbf{e}_1}\right)a_{2i+1}^{\phantom{\dagger}}\ket{\overline{\rm vac}}.
\end{split}
\eeq
In the large-mass limit,  this  state corresponds to an excitation with an energy of $\epsilon^{\rm 1m}_{\rm ex}\approx 2\mu+2h$ with respect to the vacuum state~\eqref{eq:vacuum_state}. The half-filling and the staggered mass change the physics considerably, as the tunneling dynamics can now generate more of these meson states by pair production, even when the total number of bosons is conserved and fixed to $N/2$. There can be thus production of particle-antiparticle pairs within this interpretation, making the lattice model~\eqref{eq:z2_staggered} closer to a discretisation of a quantum field theory of gauge and matter fields. This connection can be pushed further in the staggered-fermion approach~\cite{PhysRevD.11.395} to lattice gauge theories. In the following, we stick to the simpler bosonic version, and study the analog of the aforementioned string breaking.

By analogy to the two-boson state~\eqref{eq:phys_subspace_basis_2_bosons}, one could indeed separate the particle and the hole to a couple of distant sites by  creating an electric-field string in between, namely  
\beq
\label{eq:phys_subspace_string_state}
\begin{split}
\ket{_{2i}\bullet\!\!\!\!\sim\!\!\sim\!\!\!\!\hspace{0.1ex}\circ_{2j+1}}=a_{2i}^{\dagger}\left(\prod_{2i\leq \ell< 2j+1}\sigma^z_{\ell,\textbf{e}_1}\right)a_{2j+1}^{\phantom{\dagger}}\ket{\overline{\rm vac}}.
\end{split}
\eeq
 This string state  has an excitation energy of $\epsilon^{\rm 1s}_{\rm ex}\approx 2\mu+2hr$ in the large-mass limit, where $r$ is the distance between the particle and the hole. The difference with respect to the previous two-boson case is that, due to the dynamics of the gauge-invariant Hamiltonian~\eqref{eq:z2_staggered}, these string states do not span the complete physical subspace where the dynamics takes place, as occurred previously with Eq.~\eqref{eq:phys_subspace_basis_2_bosons}. In fact, if one abandons  the large-mass condition, the gauge invariant tunneling can lead to further meson-type excitations by itself, and the problem is no longer exactly solvable by mapping it to a Wannier-Stark ladder. Moreover, since there are $N_{\rm m}\leq N/2$ mesons that can be formed by a redistribution of the original bosonic particles in the half-filled vacuum~\eqref{eq:phys_subspace_meson}, and these mesons can move and be created-annihilated, the problem is no longer numerically tractable by a brute force approach that aims at finding the full spectrum in analogy to Eq.~\eqref{eq:eigsnetates_two_body_problem}.
 
 Nonetheless, we can use the same MPS algorithm as before, and explore how the real-time dynamics differs from the previous Wannier-Stark confinement. The possible novelty is that, in analogy to lattice gauge theories with fermionic matter~\cite{PhysRevLett.111.201601, PhysRevX.6.011023,PhysRevD.96.114501,PhysRevD.98.034505, Magnifico2020realtimedynamics}, mesons can be created, distorting the initial electric-field string, which can lead to the screening of charges and the emission of pairs of mesons that propagate freely. This leads to the aforementioned string breaking. We need to set the parameters in such a way that it could be energetically favourable for the above string can decay into a pair of meson-like states~\eqref{eq:phys_subspace_meson}, which may then travel freely towards the edges. Since a 2-meson state has the excitation energy $\epsilon^{\rm 2m}_{\rm ex}\approx 4\mu+4h$ in the large-mass limit, we see that $r>\mu/h+2$ for the meson configuration to be energetically favourable with respect to the string state.

 In Fig.~\ref{fig:z2_chain_localization}, we consider the initial string state $\ket{\Psi(0)}=\ket{_{2i_0}\bullet\!\!\!\!\sim\!\!\sim\!\!\!\!\circ_{2j_0+1}}$ for $i_0=32,j_0=48$ and a chain for $N=80$ lattice sites, such that $r=16$. We solve numerically for the time evolution using our MPS algorithm with $\mu/t_{1,{\bf e}_1}=0.2$, and $h/t_{1,{\bf e}_1}=0.2$, such that the 2-meson state can be favourable. As can be seen in Fig. \ref{fig:z2_chain_localization} {\bf (a)},  for hardcore bosons, the dynamics is reminiscent of previous studies on the fermionic Schwinger model. We find that the initial pair production distorts the intermediate electric-field string, and there is some partial screening leading to 2 meson-like excitations that initially spread from the edges of the string towards the boundaries of the chain. However, as can be seen in the second panel of Fig. \ref{fig:z2_chain_localization} {\bf (a)}, there is no perfect screening and no string inversion, such that these meson-like states bend and finally refocus in a breathing-type dynamics. This partial string breaking has been previously found in a bosonic Schwinger model~\cite{PhysRevLett.124.180602}. In the present case, we believe that the lack of perfect screening  is likely caused by the different nature of the electric field term in the $\mathbb{Z}_2$ gauge theory with respect to the Schwinger model. In our model, the two possible electric field eigenstates, i.e. the $\pm$ Hadamard basis, have a very different electric-field energy. On the contrary, in the Schwinger model, the electric energy is quadratic in the electric field, and  would thus be  the same for these two states, such that the string inversion seems energetically more plausible. This type of electric-field energy is however  not possible in light of the underlying Pauli-matrix representation of our $\mathbb{Z}_2$ gauge theory. It is likely this difference which is responsible for the lack of screening and string inversion in our model, and thus leads to the final refocusing.

Moving away from the artificial hardcore constraint, which is the relevant case for the trapped-ion system, we find that the evolution is now described by Fig.  \ref{fig:z2_chain_localization} {\bf (b)}. In the left panel, we see that the dynamics in the charge sector is very similar to the hardcore case. The main differences, however, appear when looking into the gauge-field sector. As shown in the right panel,  Here, t the possibility of populating the sites with more than one boson changes appreciably the dynamics, as the mirror symmetry of the dynamics is broken by a boson-enhanced tunneling.    The electric string can get distorted by   particle-hole pair creation, but this distortion is asymmetric and no longer resembles the hardcore case of Fig.  \ref{fig:z2_chain_localization} {\bf (b)}.

\section{\bf  Conclusions and Outlook}\label{sec:conclusion_outlook}

In this article, we have presented a rich toolbox for the quantum simulation of $\mathbb{Z}_2$ gauge theories using trapped ions that spans several levels of complexity. In this toolbox, the matter particles are simulated by the vibrational excitations of the ions, and the gauge field corresponds to a qubit encoded in two electronic states. In general, we have shown how to exploit a state-dependent parametric tunneling, which arises from a specific laser-ion interaction,  to induce the desired gauge-invariant tunneling  of a $\mathbb{Z}_2$ gauge theory in the ion dynamics. Furthermore, we have shown that it is possible to explore the competition of this term with a confining electric-field term, which can be readily implemented by a direct resonant driving of the qubit transition.

At the simplest-possible level, that of a $\mathbb{Z}_2$   gauge theory on a single link, we have presented two quantum simulation schemes that take into account realistic numbers, and are at reach for various  experiments  working with a single ion. Here, by exploiting the idea of synthetic dimensions,  two vibrational modes of the ion encode the matter bosons, whereas the $\mathbb{Z}_2$ gauge field is represented by the ion qubit, which sits in a synthetic link and effective mediates frequency conversion between such modes.  We have discussed several manifestations of gauge invariance, which have neat quantum-optical counterparts, and are at reach of current trapped-ion experiments: observing the correlated dynamics of a single matter boson, and the attached electric-field string, which would go beyond the available measurement capabilities of~\cite{Schweizer2019}. Moreover, we have also explored the two-boson dynamics, unveiling interesting connections to dark states in $\Lambda$-systems and  entanglement between the matter bosonic modes that could also be explored in the trapped-ion experiment. 

Increasing in complexity, we have discussed a scheme with a two-ion crystal that can be used to simulate a $\mathbb{Z}_2$   gauge theory on the simplest plaquette. Provided that the parametric tunnelings can resolve the structure of the collective vibrational modes along two directions of the crystal, we have shown that the qubits of the two ions can be arranged in the links of a circular plaquette that can be pierced by a gauge-invariant $\mathbb{Z}_2$ flux known as a Wegner-Wilson loop. The related   $\mathbb{Z}_2$-plaquette dynamics would be slower, requiring further improvements to  minimise noise sources below this timescale. For a single boson in the matter sector, we have shown that the gauge-invariant tunneling can lead to a 't Hooft loop of the electric-field variables and that this can give rise to entanglement between the gauge qubits. Once again, the gauge-invariant dynamics of the  plaquette have a quantum-optical analog in terms of a double-$\Lambda$ system.  

Finally, we have shown how one can exploit the parametric excitations in the resolved-mode regime for a larger $N$-ion crystal. We have introduced a generic idea of synthetic dimensional reduction, by means of which, it is possible to obtain a trapped-ion quantum simulator of a $\mathbb{Z}_2$   gauge theory on a full chain. This will require further developments in which the mode frequencies used to encode the matter particles can be tailored in an inhomogeneous fashion. We have shown that a single phonon in the trapped-ion chain will evolve under this  $\mathbb{Z}_2$   gauge theory in complete analogy to the problem of Wannier-Stark ladders, showing in this way localisation and breathing dynamics due to a periodically stretching electric-field string. By going to the two-phonon sector, we have presented quantitative expressions for the confinement of the simulated $\mathbb{Z}_2$ charges, which have been benchmarked with exhaustive numerical simulations based on matrix product states (MPS).  Finally, we have also explored the half-filled sector, and shown that our analog quantum simulator can host a string-breaking mechanism, contributing in this way to the  initial  progress in the digital approach~\cite{Martinez2016}. 

Future work will include the generalisation of the presented toolbox towards the two-dimensional case. As a starting point, it would be interesting to develop schemes that allow for the quantum simulation of one-dimensional arrays of the simple $\mathbb{Z}_2$ plaquettes studied in this work. This would allow to explore the interplay of the electric confining term and the magnetic flux term that would here reduce to a two-spin interaction that is also at reach of trapped-ion quantum simulators. This  playground  is simple enough such that analytical results and  numerical simulations based on MPS could be  developed. Going beyond this limiting case, it would also be interesting to explore full two-dimensional models coupled to matter, even if the Wegner-Wilson higher-weight plaquette terms cannot not be realized in the experiment. There are known examples where the intertwining of the matter particles with the gauge fields can actually lead to deconfined phases~\cite{PhysRevX.10.041007}.

\acknowledgements

A.B. thanks D. Gonz\'alez-Cuadra, S.J. Hands,  D. Leibfried, and G. Magnifico for useful discussions. 
A. B. acknowledges support from PID2021-127726NB-I00 (MCIU/AEI/FEDER, UE), from the Grant IFT Centro de Excelencia Severo Ochoa CEX2020-001007-S, funded by MCIN/AEI/10.13039/501100011033,  from the CSIC Research Platform on Quantum Technologies PTI-001,  from the MINECO through the QUANTUM ENIA project call - QUANTUM SPAIN project,
and from the EU through the RTRP-NextGenerationEU within the framework of the Digital Spain 2025 Agenda.  The project leading to this application/publication has re- ceived funding from the European Union’s Horizon Europe research and innovation programme under grant agreement No 101114305 (“MILLENION-SGA1” EU Project).
O.B., S.S. G.A and R.S thank D.M. Lucas and C.J. Ballance and A.C. Hughes for useful discussions and acknowledge support from the US Army Research Office (W911NF-20-1-0038) and the UK EPSRC Hub in Quantum Computing and Simulation (EP/T001062/1).
G.A. acknowledges support from Wolfson College, Oxford. R.S. acknowledges support from the EPSRC Fellowship EP/W028026/1 and Balliol College. E.T. acknowledges support from the MIUR Programme FARE (MEPH), and from  QUANTERA DYNAMITE PCI2022-132919.

\vspace{1ex}
{\bf Author contributions.--}
A.B. conceived the idea with useful discussions with  G.A., O.B., S.S. and R.S.   O.B. and S.S.  devised the experimental schemes presented in Sec.~\ref{sec:trapped-ion_toolbox} and Appendix~\ref{sec:trapped_ion_dg} with input from G.A. and R.S., and performed  the corresponding numerical simulations.  E.T.  performed the TDVP numerical simulations of Sec.~\ref{sec:dim_reduction} and Appendix~\ref{sec:WS_solution}. A.B. wrote the bulk of the manuscript with contributions from all of the authors.
All authors discussed the results and conclusions presented in  the manuscript.

\vspace{1ex}
{\bf Data and code availability.--}
The data that support the findings of this study are available from the corresponding author, A.B., upon reasonable request. Likewise, requests for the numerical code of various parts of the manuscript can be considered.

\begin{appendix}

SUPPLEMENTAL MATERIAL: SYNTHETIC Z2 GAUGE THEORIES BASED ON PARAMETRIC EXCITATIONS OF TRAPPED IONS
\section{\bf Background gauge fields and trapped ions}
\label{sec_bg:gauge}

\subsection{ Parametric  tunnelling and  synthetic dimensions}
\label{sec:param_tunneling}

Given the importance of the idea of  parametric tunnelling for the quantum simulation  schemes  of dynamical gauge theories  presented in the main text, we  present in this Appendix a detailed discussion, an its application to ion crystals. We start in a general setup by considering a set of bosonic/fermionic particles that can be created and annihilated  by  operators $a^{\dagger}_{d},a^{\phantom{\dagger}}_{d}$ with the corresponding commutation/anti-commutation algebra, where $d\in\mathcal{D}$ labels a specific degree of freedom of these particles. In the context of quantum many-body models and lattice field theories, the indexing set $\mathcal{D}$ typically contains the positions of a microscopic lattice  in which the particles can reside, as already mentioned in the introduction. In this context,  the geometry of the lattice determines the kinetic energy of the microscopic  Hamiltonian, which is described by a tunnelling term $t^{\phantom{\dagger}}_{dd'}a^{\dagger}_{d}a^{\phantom{\dagger}}_{d'}$, where $t^{\phantom{\dagger}}_{dd'}$ is the hopping matrix element between a pair of sites labeled by $d\neq d'$. Typically, these tunnelings decay very fast with the distance, and one only considers nearest neighbours, such that the connectivity of the lattice, i.e. the edges/links of a   graph, gets directly in-built in the tunnelling matrix. Additionally, in condensed matter,  $d\in\mathcal{D}$ can  contain other internal degrees of freedom, e.g. spin of the valence electrons. In the context of synthetic dimensions, it is these extra degrees of freedom that provide us with a new means to engineer a synthetic dimension.

The idea of synthetic dimensions  is that  the effective connectivity of the tunnelling matrix can be externally designed by introducing additional periodic drivings. These, in fact, induce new couplings that can be interpreted as effective edges/links even when the corresponding degrees of freedom are not related to any Bravais lattice at all.  A possible scheme uses a parametric tunnelling, as illustrated now with a simple example. We consider   two  modes $d\in\mathcal{D}=\{1,2\}$ of  energies $\omega_d$ ( $\hbar=1$ henceforth), such that the bare Hamiltonian is 
\beq
\label{eq:H_0}
H_0=\omega_1 a^{{\dagger}}_{1}a^{\phantom{\dagger}}_{1}+\omega_2 a^{{\dagger}}_{2}a^{\phantom{\dagger}}_{2}.
\eeq
One now adds the following parametric excitation
\beq
\label{eq:parametric_tunneling}
V(t)=\Omega_{\rm d} a^{{\dagger}}_{2}a^{\phantom{\dagger}}_{1}\cos(\phi_{\rm d}-\omega_{\rm d}t)+{\rm H.c.},
\eeq
where $\Omega_{\rm d}$, $\omega_{\rm d}$, and $\phi_{\rm d}$ are the amplitude, frequency, and phase of the  drive, respectively. In the parametric regime, i.e.,  
\beq
\label{eq:conditions_parametric}
\omega_{\rm d}=\omega_2-\omega_1,\hspace{2ex}|\Omega_{\rm d}|\ll 4|\omega_2-\omega_1|,
\eeq
one can show that an effective tunnelling term between both modes is induced by  the drive. Going to the interaction picture with respect to Eq.~\eqref{eq:H_0}, it is straightforward to recognise that the resonance condition of Eq.~\eqref{eq:conditions_parametric}     provides the required energy to bridge the  gap between the modes and couple them. Additionally, when the driving amplitude is constrained by Eq.~\eqref{eq:conditions_parametric}, a rotating-wave approximation shows that the mode coupling $V_{\rm I}(t)\approx H_{\rm eff}$ becomes a time-independent effective Hamiltonian with  a simple frequency-conversion term
\beq
\label{eq:tunneling}
H_{\rm eff}=t^{\phantom{\dagger}}_{1,\textbf{e}_1}a^{{\dagger}}_{2}a^{\phantom{\dagger}}_{1}+{\rm H.c.},\hspace{2ex}\text{with}\hspace{1ex}t^{\phantom{\dagger}}_{1,\textbf{e}_1}=\frac{\Omega_{\rm d}}{2}\ee^{\ii\phi_{\rm d}}.
\eeq
In the context of synthetic dimensions, one finds   that a non-zero tunnelling has been established, which could be understood as a new connectivity link of a synthetic lattice. This tunnelling $t^{\phantom{\dagger}}_{1,\textbf{e}_1}$ is labelled by the  synthetic lattice site index 1 from which the particle departs, and the unit link vector $\textbf{e}_1$ that connects it to the lattice site 2, into which the particle tunnels. In this  simple  case, accordingly, the synthetic lattice is just composed of two sites labelled by the indexes of the mode frequencies. We note that,  for a single link, the complex phase of the tunnelling is trivial and has no dynamical consequences, i.e. it can be readily gauged away by a local $U(1)$ transformation acting on the modes.  However, this parametric scheme  can be  generalised to a larger set $\mathcal{D}$, in which the complex phase of the effective tunnelling~\eqref{eq:tunneling} may have non-trivial consequences. As discussed in~\cite{Fang2012,Roushan2017}, one can create synthetic lattices in a way that, when the particle tunnels around a closed path $\gamma$, it gains a non-zero phase $\sum_{\ell\in\gamma}\phi_\ell=\Phi_{\rm AB}$ that simulates a synthetic Aharonov-Bohm phase. Even if the particles have a vanishing  charge, their tunnelling resembles that of a charged particle in an external magnetic field via the so-called  Peierls' substitution~\cite{Peierls1933}, which originally concerned  electrons in a narrow-band material subjected to a perpendicular magnetic field~\cite{PhysRevB.14.2239}. The quadratic lattice models with Peierls' phases  provide a  playground for studying the integer quantum Hall effect and topological band theory~\cite{PhysRevLett.49.405}. We discuss this point in detail below.

\subsection{Peierls ladders with trapped-ion chains}
\label{sec:Peierls:ions}

So far, we have not yet discussed  how  the parametric term~\eqref{eq:parametric_tunneling} can be created and controlled in a specific physical system.  The  parametric scheme  has been implemented in arrays of superconducting circuits~\cite{Roushan2017} but, to the best of our knowledge, its realisation in trapped-ion crystals has not been discussed so far. We  now describe how to exploit this method to build a quantum simulator of a bosonic quantum Hall ladder using the transverse vibrations of a  chain of $N$ trapped ions in a linear Paul trap~\cite{PhysRevA.45.6493}. This will prepare the ground for the scheme of dynamical $\mathbb{Z}_2$ gauge fields discussed in the main text, which  exploit similar concepts with a   new twist.

\begin{figure}[t]
	\centering
	\includegraphics[width=0.95\columnwidth]{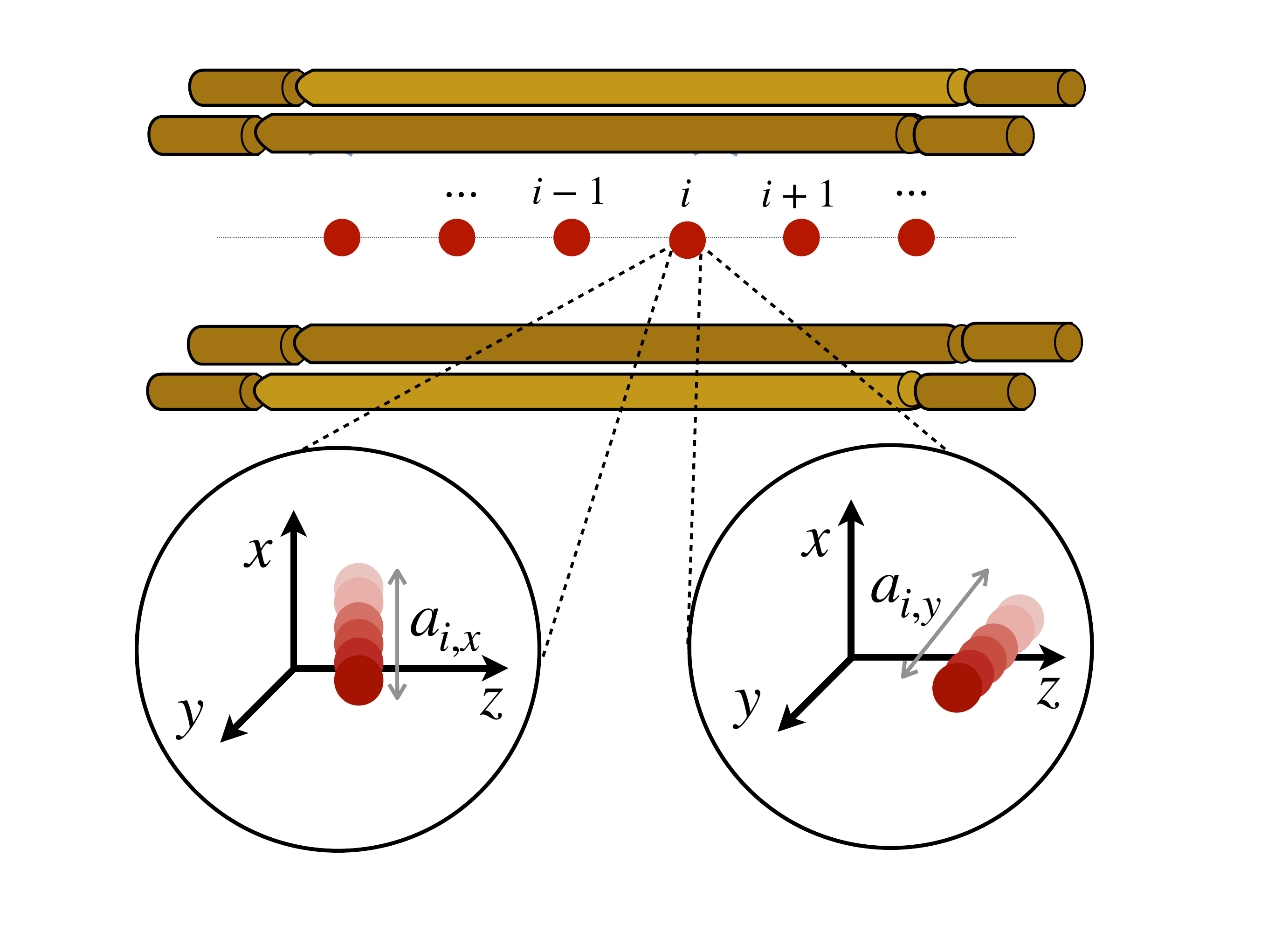}
	\caption{{\bf Transverse vibrational excitations:} Schematic representation of an ion chain in a linear Paul trap. In the insets, we represent the transverse local vibrational excitations of a single ion in the chain.   }
	\label{fig:transverse_phonons}
\end{figure}

Following~\cite{Steane1997,James1998,Marquet2003}, for a linear Paul trap with trap frequencies $\omega_z\ll\omega_x,\omega_y$, the ions form  a linear chain along the $z$-axis (see Fig.~\ref{fig:transverse_phonons}). The  transverse  vibrations of each ion~\cite{PhysRevLett.93.263602} around  its equilibrium  position are described by  
\beq
\label{eq:trapped_ion_harmonic}
H_0=\sum_{d}\omega^{\phantom{\dagger}}_d a^{{\dagger}}_{d}a^{\phantom{\dagger}}_{d}+\sum_{d\neq d'}t^{\phantom \dagger}_{dd'} a^{{\dagger}}_{d}a^{\phantom{\dagger}}_{d'}.
\eeq
Here, the labelling index reads $d=(i,\alpha)\in\mathcal{D}$, and the  set $\mathcal{D}$ contains the label for the ions in the chain  $i\in\{1,\cdots,N\}$, and the label  for the two possible   directions of the vibrations transverse to the chain $\alpha\in\{x,y\}$. In addition, $a^{{\dagger}}_{d}, a^{\phantom{\dagger}}_{d}$ are the bosonic creation-annihilation operators for the corresponding local vibrations around the equilibrium positions of the ions $\boldsymbol{r}_i^0$, which we have assumed to be aligned along the null of the radio-frequency (rf) pseudo-potential of the linear Paul trap, such that excess micromotion can be neglected~\cite{doi:10.1063/1.367318}. Additionally, the modulation frequencies to be introduced below must be much lower than the rf driving of the trap to also neglect the intrinsic quantum-mechanical micromotion~\cite{Bermudez_2017}.  As discussed in~\cite{Marquet2003}, the expansion of the Coulomb interaction to second order leading to Eq.~\eqref{eq:trapped_ion_harmonic} does not mix the $x,y$  modes, and one finds that the tunnelling matrix decay with the inter-ion distance  following a dipolar law
\beq
\label{eq:tunneling_ions}
t_{(i,\alpha)(j,\beta)}=\frac{1}{2m\omega_\alpha}\frac{e^2}{8\pi\epsilon_0|\boldsymbol{r}^0_i-\boldsymbol{r}_j^0|^3}\delta_{\alpha,\beta},
\eeq
where $m$ is the ion mass,  $\epsilon_0$ the vacuum permittivity, and $\delta_{\alpha,\beta}$ is the Kronecker delta. The on-site energies 
are related to the effective trap frequencies $\omega_\alpha$ of the time-averaged pseudo-potential~\cite{PhysRevLett.93.263602}  by the following expression
\beq
\label{eq:trap_freq}
\omega^{\phantom{\dagger}}_{i,\alpha} =\omega_\alpha-\sum_{j\neq i}\frac{1}{2m\omega_\alpha}\frac{e^2}{8\pi\epsilon_0|\boldsymbol{r}^0_i-\boldsymbol{r}_j^0|^3}.
\eeq

 Since the Hamiltonian of Eq.~\eqref{eq:trapped_ion_harmonic} has a global $U(1)\times U(1)$ symmetry under  $a^{\phantom{\dagger}}_{d}\mapsto\ee^{\ii\varphi}a^{\phantom{\dagger}}_{d}$,  $a^{\dagger}_{d}\mapsto\ee^{-\ii\varphi}a^{\dagger}_{d}$,  the total  number of transverse vibrational excitations along each trap axis is individually conserved. Although  phonons in crystals typically refer to the excitations of the collective vibrational modes, it is customary to refer to these local vibrational modes also as phonons in the trapped-ion community, and we have followed this convention in the main text. The novelty with respect to the crystal phonons underlying the transverse sound waves in elastic solids~\cite{PRXQuantum.3.020352} is that, since the phonon number is conserved when $|t_{dd'}|\ll2\omega_\alpha$~\cite{PhysRevLett.93.263602}, we can thus think of these transverse vibrational excitations as  particles localised to each of the ions. Just like electrons on a solid, the phonons tend to spread over the chain due to the dipolar tunnelling of Eq.~\eqref{eq:tunneling_ions}. We note that the dynamics of these local phonons due to the effective tight-biding Hamiltonian of Eq.~\eqref{eq:trapped_ion_harmonic} has been  observed in various trapped-ion experiments~\cite{Brown2011,Harlander2011,Wilson2014,Ramm_2014,Toyoda2015,PhysRevLett.120.073001,PhysRevLett.123.100504,PhysRevLett.123.213605,PhysRevLett.124.200501}.  

 
 Let us now discuss how to exploit parametric excitations to realise synthetic phonon ladders subjected to an  effective background gauge field. There have been some prior works on trapped-ion parametric drivings which, to the best of our knowledge,   have focused on  rather different goals. For instance, in~\cite{PhysRevA.42.2977}, parametric modulations are used to design cooling and detection methods for the spectroscopy of a single trapped electron and proton, as well as for squeezing and linear amplification. The former requires a parametrically-modulated quadrupole potential that couples two different vibrational directions~\cite{PhysRevA.41.312}, whereas the latter employs a periodic modulation of the trap frequencies, and can thus be achieved by applying an additional oscillating potential to the rf  trap electrodes. Parametric modulations can also be obtained optically, exploiting  the cross-beam ac-Stark shift of  a pair of far-detuned laser beams~\cite{3981,PhysRevLett.76.1796}. Although the  parametric modulations obtained through the electronic equipment have led to larger amplification in recent experiments~\cite{doi:10.1126/science.aaw2884}, we  stick to  optical ones in our work, as they are more flexible for the generation of synthetic gauge fields.

Our goal is to interpret the transverse vibrational directions of Fig.~\ref{fig:transverse_phonons} as a new synthetic ``dimension''. Note that, however, the $x$ and $y$ directions are decoupled at this quadratic order (Eq.~\eqref{eq:tunneling_ions}), such that the Hamiltonian of Eq.~\eqref{eq:trapped_ion_harmonic} describes two decoupled dipolar chains. We now discuss how a parametric excitation of the tunnelling can be induced, and how this term can be used to derive a model  with couplings between the two chains, such that the global symmetry reduces to $U(1)\times U(1)\mapsto U(1)$, and only the total number of transverse vibrational quanta is conserved. We will see that, from this perspective, the phonons move in a synthetic two-leg ladder  and, moreover, argue that they can also be subjected to an effective Peierls' phase  mimicking the microscopic model for charged particles under external magnetic fields. We consider that each ion is illuminated by a global two-beam laser field, the beat note of which is far detuned from any electronic transition~\cite{3981}. The ions, all  prepared in the same internal state of the ground-state manifold~\footnote{A similar argument can be made for a qubit encoding formed by two sub-levels of a metastable manifold.}, thus experience an ac-Stark shift  that yields the following  optical potential
\beq
\label{eq:ac_beams}
V(t)=\sum_{n,n'=1,2}\sum_{i=1}^N\Omega_{n,n'}\ee^{\ii(\boldsymbol{k}_{{\rm L},n}-\boldsymbol{k}_{{\rm L},n'})\cdot\boldsymbol{r}_i-(\omega_{{\rm L},n}-\omega_{{\rm L},n'})t}+{\rm H.c.}.
\eeq
Here, $\omega_{{\rm L},n}$($\boldsymbol{k}_{{\rm L},n}$) is the   frequency (wave-vector) of each beam $n\in\{1,2\}$ of the global laser field, 
 and $\Omega_{n,n'}=-\Omega_{{\rm L},n}^{\phantom{*}
}\Omega_{{\rm L},n'}^*/4\Delta$ is the ac-Stark shift arising from two-photon processes. In those processes,   a photon is absorbed from the $n$-th  beam by the $i$-th ion with the Rabi frequency $\Omega_{{\rm L},n}$ and a large  detuning $\Delta$, such that the ion is only virtually excited.  Subsequently, the ion is  de-excited to the same internal state by emitting a photon onto the $n'$-th beam~\cite{3981}. 

In addition to the standard ac-Stark shifts in Eq.~\eqref{eq:ac_beams}, i.e. terms with $n=n'$ that contribute with an energy shift  $\Delta E_{\rm ac}=\sum_n\Omega_{n,n}$, one also obtains  crossed beat note terms that lead to periodic modulations in  space (time) when the laser wave-vectors (frequencies) are not co-linear (equal). In this case, one defines the beat note wave-vector and frequency as 
$
\boldsymbol{k}_{\rm d}=\boldsymbol{k}_{{\rm L},1}-\boldsymbol{k}_{{\rm L},2}$,
 and $\omega_{\rm d}=\omega_{{\rm L},1}-\omega_{{\rm L},2}$, respectively.   Noting that the ion positions can be expanded in terms of the local phonon operators via $\boldsymbol{r}^{\phantom{0}}_i=\boldsymbol{r}^0_i+\sum_\alpha\textbf{e}_\alpha\frac{1}{\sqrt{2m\omega_{\alpha}}}(a^{\phantom{\dagger}}_{i,\alpha}+a^{\dagger}_{i,\alpha})$, where $\textbf{e}_\alpha$ is the unit vector in the direction of the transverse ion vibration $\alpha$, one can substitute and expand the optical potential~\eqref{eq:ac_beams} in the so-called Lamb-Dicke regime 
\beq
\label{eq:lamb_dicke_regime}
\eta_\alpha={\boldsymbol{k}_{\rm d}\cdot\textbf{e}_\alpha}/{\sqrt{2m\omega_{\alpha}}}\ll 1.
\eeq
The Taylor expansion of Eq.~\eqref{eq:ac_beams} then  leads to a sum of terms with all possible powers of the phonon operators. By choosing the correct beat note frequency, it is possible to select which of them brings in the leading contribution. In particular, for 
\beq
\label{eq:constraints_param_tunneling_ions}
\omega_{\rm d}=\omega_y-\omega_x,\hspace{2ex}|\Omega_{\rm d}|\ll|\omega_{y}-\omega_{x}|,
\eeq
where we assume that $|\omega_x-\omega_y|\ll\omega_x,\omega_y$, a rotating-wave approximation shows that the optical potential  contains the desired parametric excitation of a tunnelling term that generalises Eq.~\eqref{eq:parametric_tunneling} to an arbitrary number of lattice sites, namely
\beq
\label{eq:trapped_ion_par_tunn}
V(t)\approx\sum_i\Delta E_{\rm ac}+\sum_{i}\Omega_{\rm d}\cos(\phi_i-\omega_{\rm d}t)a^\dagger_{i,y}a^{\phantom{\dagger}}_{i,x}+{\rm H.c.},
\eeq
where we have introduced the parameters
\beq
\label{eq:parameters_parametric_ions}
\Omega_{\rm d}=|\Omega_{1,2}|\eta_{x}\eta_y, \hspace{2ex}\phi_i=\boldsymbol{k}_{\rm d}\cdot\boldsymbol{r}_i^0+{\arg}(-\Omega_{1,2}).
\eeq
 Note that, due to the constraints in Eq.~\eqref{eq:constraints_param_tunneling_ions}, we have neglected other contributions in the Lamb-Dicke expansion.

 \begin{figure}[t]
	\centering
	\includegraphics[width=1\columnwidth]{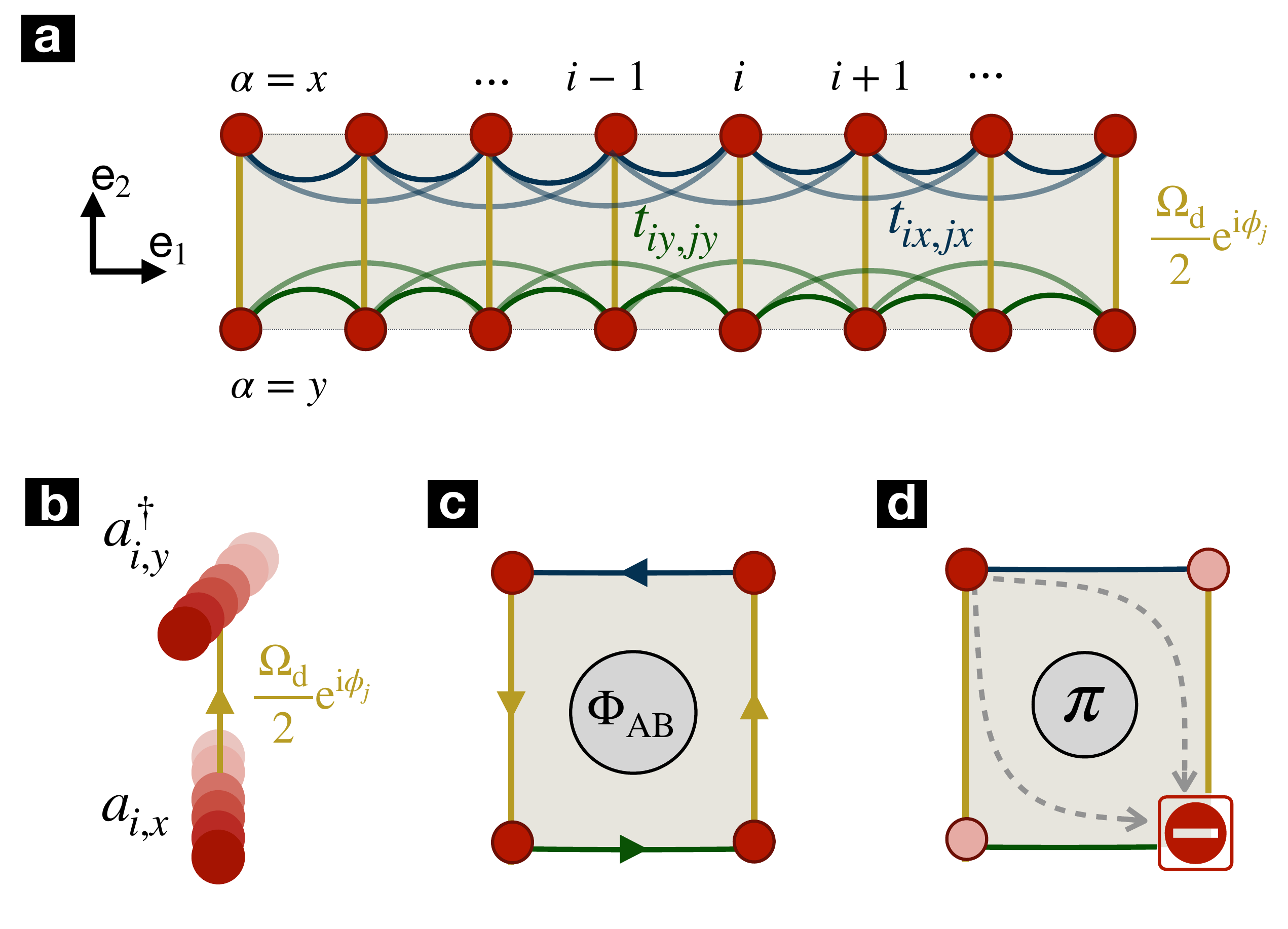}
	\caption{{\bf Synthetic dimensions in trapped-ion chains:} {\bf (a)} Schematic representation of a synthetic Peierls ladder. The sites of the upper and lower legs of the ladder represent the local vibrations of the ions along the $x$ and $y$ transverse directions, respectively. The dipolar tunnelings~\eqref{eq:tunneling_ions} are represented by intra-leg links that connect distant ions. The resulting parametric tunnelings in Eq.~\eqref{eff_tunneling_ion_ladder} are depicted by  the vertical inter-leg links, and correspond to the frequency-conversion process of {\bf (b)}. {\bf (c)} For a pair of ions, the effective rectangular plaquette can lead to a net Aharonov-Bohm  phase $\Phi_{\rm AB}$~\eqref{eq:AB_flux} for a phonon that tunnels along the corresponding synthetic links. {\bf (d)} For $\Phi_{\rm AB}=\pi$, there can be perfect destructive interference for the phonon, which mimics the Aharonov-Bohm interference of an electron that travels around an infinitely-thin solenoid.}
	\label{fig:transverse_phonons_syntehtic_ladder}
\end{figure}

 One can readily see that, in addition to the irrelevant ac-Stark shift $\Delta E_{\rm ac}$, we have obtained a   parametric modulation like Eq.~\eqref{eq:parametric_tunneling} that involves simultaneously  all of the ions in the chain. Repeating the same arguments as in the simple two-mode case (Eq.~\eqref{eq:parametric_tunneling}), one finds that the parametric drive can activate the tunnelling of a phonon along the new synthetic direction (see Fig.~\ref{fig:transverse_phonons_syntehtic_ladder} {\bf (a)}), such that the tunnelling matrix of Eq.~\eqref{eq:tunneling_ions} becomes $t_{dd'}\mapsto\tilde{t}_{dd'}$ with
\beq
\label{eq:tunneling_ions_synthetic}
\tilde{t}_{(i,\alpha)(j,\beta)}={t}_{(i,\alpha)(j,\beta)}+\frac{\Omega_{\rm d}}{2}\ee^{\ii \epsilon_{\alpha,\beta}\phi_j}\delta_{i,j}(1-\delta_{\alpha,\beta}),
\eeq
Here, in addition to the Kronecker delta,  we have used the fully anti-symmetric tensor defined as $\epsilon_{x,y}=-\epsilon_{y,x}=1$,  $\epsilon_{x,x}=\epsilon_{y,y}=0$. By making the following identification 
\beq
\label{eq:Peierls_matter_mapping}
\begin{split}
a^{\phantom{\dagger}}_{i,x},a^\dagger_{i,x}\mapsto a^{\phantom{\dagger}}_{i\textbf{e}_1},a_{i\textbf{e}_1}^\dagger,\hspace{2ex} a^{\phantom{\dagger}}_{i,y},a^{\dagger}_{i,y}\mapsto a^{\phantom{\dagger}}_{i\textbf{e}_1+\textbf{e}_2}, a_{i\textbf{e}_1+\textbf{e}_2}^\dagger,
\end{split}
\eeq
we obtain, in the interaction picture,  a  {tight-binding} model for bosons in a synthetic two-leg ladder 
\beq
\label{eq:trapped_ion_ladder}
H_{\rm eff}=\sum_{\boldsymbol{i}}\sum_{\boldsymbol{\ell}\in\mathcal{L}\{\boldsymbol{i}\}}t^{\phantom \dagger}_{\boldsymbol{i},\boldsymbol{\ell}} a^{{\dagger}}_{\boldsymbol{i}+\boldsymbol{\ell}}a^{\phantom{\dagger}}_{\boldsymbol{i}}.
\eeq
 Here, a boson at the synthetic lattice site $\boldsymbol{i}$ can tunnel horizontally or vertically to the site $\boldsymbol{i}+\boldsymbol{\ell}$ along the synthetic links  labelled by $\boldsymbol{\ell}\in\mathcal{L}\{\boldsymbol{i}\}$. The  tunnelling amplitudes read
\beq
\label{eff_tunneling_ion_ladder}
\begin{split}
t_{i\textbf{e}_1,\ell\textbf{e}_1}&=\tilde{t}_{(i+\ell,x)(i,x)},\hspace{5ex} t_{i\textbf{e}_1,\textbf{e}_2}=\frac{\Omega_{\rm d}}{2}\ee^{+\ii\phi_i},\\ t_{i\textbf{e}_1+\textbf{e}_2,\ell\textbf{e}_1}&=\tilde{t}_{(i+\ell)(y;i,y)},\hspace{1ex}t_{i\textbf{e}_1+\textbf{e}_2,-\textbf{e}_2}=\frac{\Omega_{\rm d}}{2}\ee^{-\ii\phi_i}.
\end{split}
\eeq

In comparison to the parametric tunnelling of Eq.~\eqref{eq:parametric_tunneling}, which also leads to a  tunnelling strength with a complex phase (see Eq.~\eqref{eq:tunneling}), we see that the current scheme  leads to a site-dependent  phase as a consequence of the spatial modulation of the optical potential~\eqref{eq:ac_beams}, as depicted in Fig.~\ref{fig:transverse_phonons_syntehtic_ladder}{\bf (b)}. 
This inhomogeneity can be exploited, as depicted in Fig.~\ref{fig:transverse_phonons_syntehtic_ladder} {\bf (c)}, to induce an effective Peierls' phase, such that the  phonons in the synthetic ladder mimic the dynamics of electrons under a magnetic field. In fact, if the local phonon tunnels around the smallest rectangular plaquette $t_{i\textbf{e}_1,\textbf{e}_1}t_{(i+1)\textbf{e}_1,\textbf{e}_2}t_{(i+1)\textbf{e}_1+\textbf{e}_2,-\textbf{e}_1}t_{i\textbf{e}_1+\textbf{e}_2,-\textbf{e}_2}\propto\ee^{\ii\Phi_{\rm AB}}$, it gains a net phase that can no longer be gauged away as in the simple two-mode case of  Eq.~\eqref{eq:tunneling}. In fact, this phase is analogous to the Aharonov-Bohm phase~\cite{PhysRev.115.485} for electrons moving in a plane  under a perpendicular magnetic field
\beq
\label{eq:AB_flux}
\Phi_{\rm AB}=\boldsymbol{k}_{\rm d}\cdot(\boldsymbol{r}^0_{i+1}-\boldsymbol{r}^0_{i})=:2\pi\frac{\Phi_B}{\Phi_0},
\eeq
where $\Phi_B=\int_\square{\rm d}\boldsymbol{S}\cdot\boldsymbol{B}_{\rm bg}$ is the  flux of an effective  magnetic field $\boldsymbol{B}_{\rm bg}$ across the plaquette $\square$, and $\Phi_0=h/e$ is the quantum of flux. As a consequence of this  flux, which can be controlled by tilting the laser wave-vector with respect to the ion chain, one could for instance observe Aharonov-Bohm  destructive interference for $\Phi_{\rm AB}=\pi$, in which a  single phonon cannot tunnel two sites apart along a synthetic plaquette (see Fig.~\ref{fig:transverse_phonons_syntehtic_ladder} {\bf (d)}). Let us note that this Aharonov-Bohm interference occurs at the level of phonons, which have a zero net charge, and thus differs from the interference of charged ions tunnelling between two different crystalline configurations, neatly observed in experiments with a real magnetic field~\cite{Noguchi2014}.

For larger ion crystals, the analogy with a homogeneous magnetic field  is still valid in spite of the existence of dipolar tunnelings, provided that the equilibrium positions of the ions are equally spaced. The equal spacing can be achieved by designing arrays of individual traps with micro-fabricated surface-electrode traps~\cite{https://doi.org/10.48550/arxiv.quant-ph/0501147,PhysRevA.73.032307,PhysRevLett.96.253003,PhysRevA.77.022324,PhysRevLett.100.013001,PhysRevLett.102.233002,Kumph_2011,Welzel2011,Sterling2014,Mielenz2016,Bruzewicz2016,Kumph_2016,PhysRevX.10.031027},  by introducing anharmonic confining potentials in segmented ion traps~\cite{Lin_2009,PhysRevA.95.032341,Johanning2016,PhysRevA.98.032318,PhysRevA.98.032318,Pagano_2019}, or in ring traps~\cite{PhysRevLett.118.053001,PhysRevLett.123.133202}. Let us note that, even if one achieved an homogeneous  spacing, the longer-range nature of the tunnelings imply that the excitations can now enclose larger  plaquettes  potentially changing the interference phenomena. In the pi-flux case, the next-to-nearest neighbour tunnelling leads to plaquettes with zero flux, which may challenge  the existence of the perfect destructive interference of Fig.~\ref{fig:transverse_phonons_syntehtic_ladder} {\bf (d)}. Nevertheless, these larger plaquettes are enclosed at a considerably slower pace, as the tunnelling strengths decay with the cube of the distance~\eqref{eq:tunneling_ions}. We have numerically observed that an almost perfect  Aharonov-Bohm interference can still  occur when considering how a phononic excitation  travels between opposite corners of the synthetic ladder~\cite{msc_thesis_gabriel}.

Let us close this Appendix by highlighting that the effective magnetic field underlying Eq.~\eqref{eq:AB_flux} is not a true dynamical magnetic field, but rather a fixed background field. One can indeed push the analogy to the level of the vector potential using $\boldsymbol{B}_{\rm bg}=\boldsymbol{\nabla}\times\boldsymbol{A}_{\rm bg}$, but the $U(1)$ gauge field $\boldsymbol{A}_{\rm bg}$ would still be a background field, the dynamics of which can only be fixed externally, and has nothing to do with Maxwell electrodynamics. In the main text, we  describe how this scheme of parametric tunnelings can be generalised  to get closer to this situation, and be able to explore  lattice gauge theories.
\begin{figure}[t]
    \centering
     \includegraphics[width=0.9\linewidth]{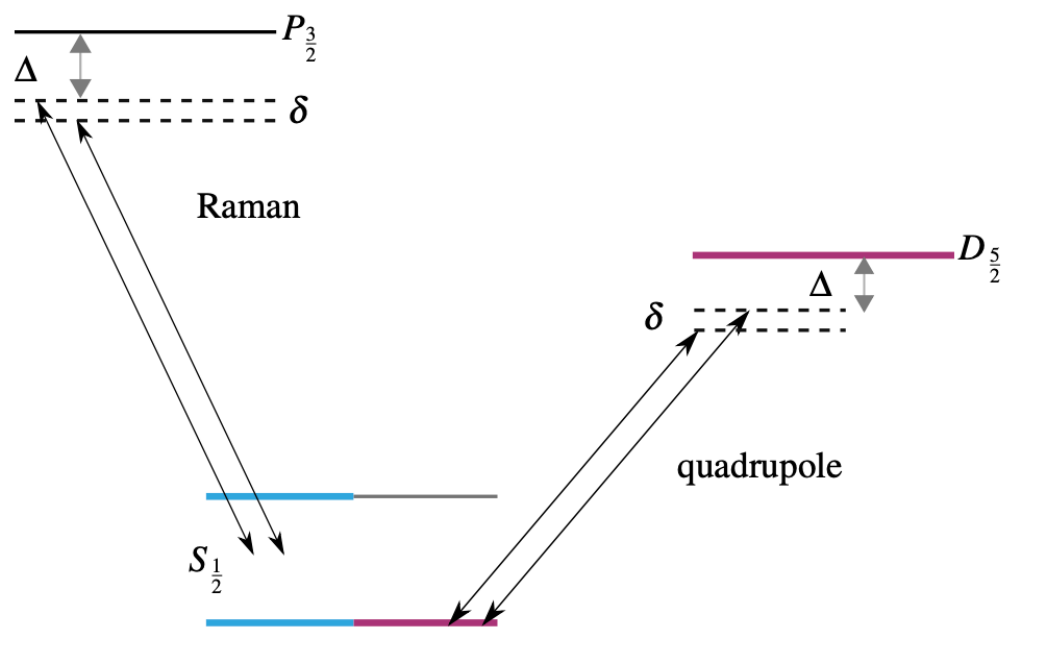}
    \caption{{\bf Beam configurations for  the  light-shift scheme:} We consider the  { ground state} qubit (cyan lines) and the  optical qubit (magenta) in $^{88}{\rm Sr}^+$. The gauge-invariant tunneling for the ground state qubit is created via two far-detuned dipole-allowed Raman transitions of detunings $\Delta$ and $\Delta+\delta$, depicted by black arrows, that virtually couple the qubit states in the $S_{{1}/{2}}$ level to an excited state in the $P_{{3}/{2}}$ level.  Fpr the optical qubit, these  far-detuned  Raman transitions are quadrupole-allowed, and virtually couple   the $S_{{1}/{2}}$ level to  {a metastable}  state in the $D_{{5}/{2}}$ level. When the beat note of the two  tones $\delta$ is on resonance with the difference of two secular trap frequencies, we attain the desired state-dependent parametric tunneling. }\label{fig:LS_style_drive_w_qdp}
\end{figure}

\section{\bf  Dynamical gauge fields and trapped ions}
\label{sec:trapped_ion_dg}

\subsection{Dipole light-shift scheme}\label{app:dp_ls}

In this Appendix, 
we consider a $^{88}\mathrm{Sr}^+$ ion confined in the setup presented in Refs.~\cite{schafer2018a,thirumalai2019high}, an provide a more detailed discussion of the specific experimental parameters, and the errors that arise in the analog scheme  for the $\mathbb{Z}_2$ gauge link  QS. The secular frequency of the axial in-phase mode  can be set to $\omega_z/2\pi = 1.2\,$MHz, while the radial secular frequency is $\omega_x/  2\pi=1.9\,$MHz. The qubit states $\ket{\uparrow_1},\ket{\downarrow_1}$ can be defined by two ground state levels of the $5S_{1/2}$ manifold shown in Fig.~\ref{fig:LS_style_drive_w_qdp}. 
For such a ground state qubit, the light shifts are created by a far-detuned dipole-mediated Raman transition~\cite{gan_2020}. The two beams are assumed to be counter-propagating $\boldsymbol{k}_{{\rm L},1}=-\boldsymbol{k}_{{\rm L},2}=:\boldsymbol{k}$, such that the beat note wave-vector is $\boldsymbol{k}_{\rm d} = 2\boldsymbol{k}$. Moreover, we assume
that the angle between $\boldsymbol{k}_{\rm d}$ and the axial mode ($z$) is $45^\circ$, while the angle with respect to the transverse mode ($x$) is $60^\circ$.The two Raman beams at near $\lambda = 402~\textrm{nm}$ are detuned by $\Delta/2\pi = 10$\,THz from the $S_{1/2}\leftrightarrow P_{3/2}$ transition (see   Fig.~\ref{fig:LS_style_drive_w_qdp}). The Lamb-Dicke factors \eqref{eq:lamb_dicke_regime} of the two motional modes are $\eta_{z} = 2 \times 0.077$ and $\eta_{x} = 2 \times 0.043$. In this system, light shifts of up to $\Omega_{1,2}/2\pi = 1.1$\,MHz can be achieved. In the full Hamiltonian~\eqref{eq:ac_beams_st_dep},
we set the beam detunings to  
$\delta = \omega_x-\omega_z$, resulting in a beat note frequency of $\omega_{\rm d}=\delta$ in Eq.~\eqref{eq:constraints_param_tunneling_ions}. These parameters lead to an estimated coupling of  $\Omega_{1,2}/2\pi = \SI{1.1}{\mega \hertz}$, which  sets the timescale of the targeted dynamics shown in Fig.~\ref{fig:scheme_1_BS_time_scan} of the main text.


\begin{figure}
    \centering\includegraphics[width=1\columnwidth]{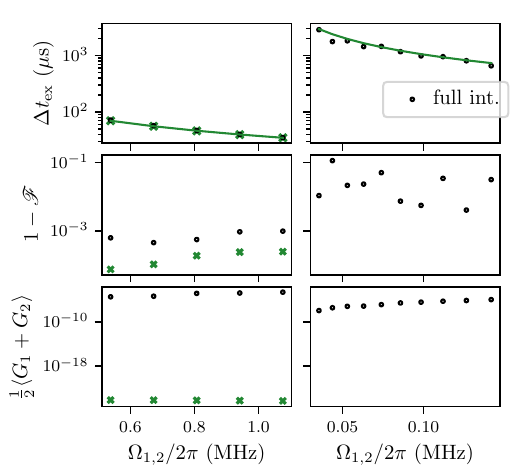}
    \caption{{\bf Light-shift  $\mathbb{Z}_2$ tunneling with Raman couplings:} Numerical simulations of the exchange duration $\Delta t_{\rm ex}$ (upper panel), state infidelity  $1-\mathcal{F}$ (middle panel), and Gauss' symmetry operator $\braket{G_1+G_2}/2$ (lower panel) as a function of the two-photon light shift $\Omega_{1,2}$  for  dipole-allowed couplings (left column) and  quadrupole-allowed couplings (right column). The results are shown for the full Hamiltonian~\eqref{eq:ac_beams_st_dep} (black) and the idealized invariant-gauge tunneling~\eqref{eq:link_ions} (green). The green solid line in the upper panel is the  analytic dependence of the exchange duration on $\Omega_{1,2}$, extracted from Eq.~\eqref{eq:eff_tunneling_light_shift}. For both the full and ideal Hamiltonian the expectation value of the symmetry operator is consistent with $0$, down to $10^{-3}$.}
    \label{fig:raman_ls_vs_omega}
\end{figure}

\begin{figure}
    \centering
    \includegraphics[width=0.75\linewidth]{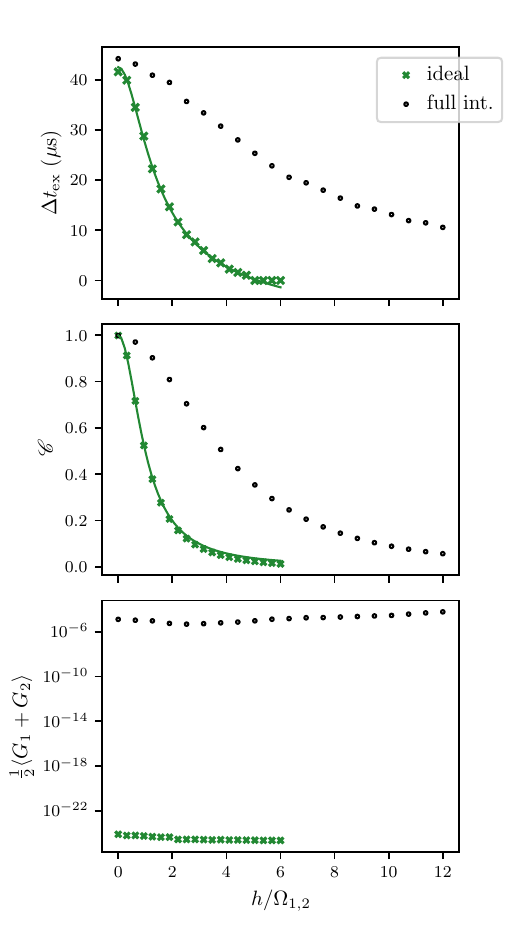}
    \caption{{\bf Light-shift $\mathbb{Z}_2$ tunneling with a non-zero electric field:} we simulate the hopping duration, contrast $\mathcal{C}$ and symmetry operator $\braket{G_1+G_2}/2$ as a function of electric field strength $h$ relative to the two-photon light shift  $\Omega_{1,2}$ of the dipole-allowed transitions. The results are shown for the full Hamiltonian~\eqref{eq:ac_beams_st_dep} (black), including the additional carrier driving~\eqref{eq:second_tone} with a modified resonant condition~\eqref{eq:constraints_carrier_ac_stark}. We also show the results for the ideal gauge-invariant Hamiltonian~\eqref{eq:link_ions_h} (green). The green solid line is the  analytic dependence of the exchange duration and contrast respectively on $\Omega_{1,2}$, extracted from Eqs.~\eqref{eq_eff_rabi_flop} and the effective couplings~\eqref{eq:eff_tunneling_light_shift} 
    and~\eqref{eq:el_field_LS}. For both the full and ideal cases {,} the expectation value of the symmetry operator is consistent with $0$, down to $10^{-3}$.}
    \label{fig:raman_ls_vs_htv}
\end{figure}

In this figure, there is a clear agreement with small deviations from the expected dynamics of the  $\mathbb{Z}_2$ gauge link. To quantify these deviations in more detail, we numerically solve the trapped-ion evolution, starting in $\ket{\textrm{L}}$~\eqref{eq:basis}, where 
we apply the interaction for a duration $t=\Delta t_{\rm ex}$ such that, by the end of it, the overlap squared to the desired state $\ket{\textrm{R}}$ is maximised. If we consider only the idealised tunneling term, Eq.~\eqref{eq:link_ions}, the exchange duration would be given by $\Delta t_{\rm ex}=\pi/2t_{1, {\bf e}_1}$. In the simulation of the more realistic trapped-ion case, there are additional terms neglected in the ideal case that can change the optimal exchange duration. We thus find $\Delta t_{\rm ex}$ by maximising the fidelity $\mathcal{F}(t)=|\langle{\rm R}|{\psi(t)}\rangle|^2$ of achieving the desired state $\ket{\textrm{R}}$.
We also calculate the expectation value of the local symmetry generators~\eqref{eq:generators_link} for $\braket{G_1(t)+G_2(t)}/2$  to check if the effective gauge symmetry is fulfilled. Moreover, when introducing the electric field with magnitude $h$ in the simulations, we plot the maximum contrast $\mathcal{C}$ in the oscillations of  $\overline{s}_x(t)$~\eqref{eq_Rabi flops_link}. Thus, we can evaluate the reduced tunneling probability caused by the energy penalty for stretching/compressing the electric field line as one increases $h>0$.

In Fig.~\ref{fig:raman_ls_vs_omega}, we present simulations of the resulting hopping duration $\Delta t_{\rm ex}$, the fidelity error $1-\mathcal{F}(t)$ and the gauge-invariance operator $\braket{G_1(t)+G_2(t)}/2$ as a function of applied light shift $\Omega_{1,2}$, which can be increased by using higher laser intensities or lower Raman detunings in order to obtain a faster gauge-invariant dynamics~\eqref{eq:eff_tunneling_light_shift}. When comparing the dynamics of the full Hamiltonian to the ideal gauge-invariant tunneling~\eqref{eq:eff_tunneling_light_shift}, we observe excellent agreement with $\Delta t_{\rm ex}=\pi/2t_{1,{\bf e}_1}$. The 
fidelity error and the symmetry operator $\braket{G_1(\Delta t_{\rm ex})+G_2(\Delta t_{\rm ex})}/2$ also remain very low even when including the trapped-ion additional terms such as the off-resonant carrier, which underlies the adequacy of the considered parameters for this specific scheme. We can achieve an effective tunneling  rate of up to $7.1$\,kHz inferred as $1/(4\Delta t_{\rm ex})$. 

To achieve large tunneling rates given the achievable $\Omega_{1,2}$ of current systems, and the low coupling rate of the second-order process $\propto \eta_x \eta_z$~\eqref{eq:eff_tunneling_light_shift}, one needs to push the parameters to a regime that violates the $\delta \gg |\Omega_{1,2}|$ requirement.
This results in off-resonant driving of spurious interactions, most prominently direct off-resonant carrier coupling. As mentioned above we minimise this effect by employing adiabatic amplitude pulse shaping~\cite{Roos_2008} with a rise time of $\SI{10}{ \micro \second}$.

\begin{figure}[t]
    \centering
        \includegraphics[width=0.9\linewidth]{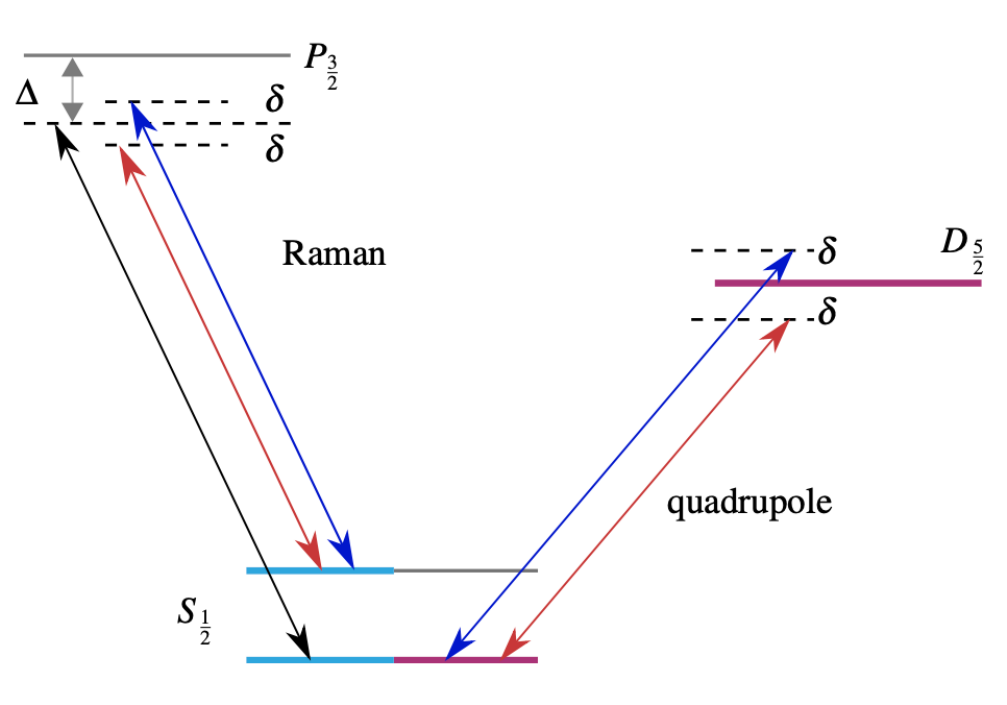}
    \caption{\textbf{ M\o lmer-S\o rensen-type parametric drive:}. For the ground state qubit (cyan lines), the Raman scheme which is now near resonant with the qubit frequency $\omega_0 +\delta$, we need to introduce a third tone, here depicted by a blue arrow . For the optical qubit (magenta lines), which is driven directly via the quadrupole transition, we symmetrically detune the two tones (blue and red) about the qubit resonance by $\pm\delta$.}
    \label{fig:MS_style_drive}
\end{figure}

We now present similar  simulations for the light-shift type scheme as a function of the effective electric-field strength $h$. In Fig.~\ref{fig:raman_ls_vs_htv}, we present the exchange duration $\Delta t_{\rm ex}$ defined at maximum state fidelity $\mathcal{F}$, the maximum contrast $\mathcal{C}$ in $\overline{s}_x(t)$, and the expectation value of the gauge-symmetry generators $\braket{G_1+G_2}/2$, all of them as a function of the ratio between the transverse electric field  $h$ and the differential ac-Stark shift amplitude $\Omega_{1,2}$. While the presence of the non-commuting  carrier coupling ($z$ basis) present in the full Hamiltonian reduces the effect of the transverse term ($x$ basis) from that of the ideal case, the gauge invariance is preserved.

\subsection{Quadrupole light-shift scheme}\label{app:qdp_ls}

\begin{figure}[t]
    \centering
    \includegraphics[width=1\linewidth]{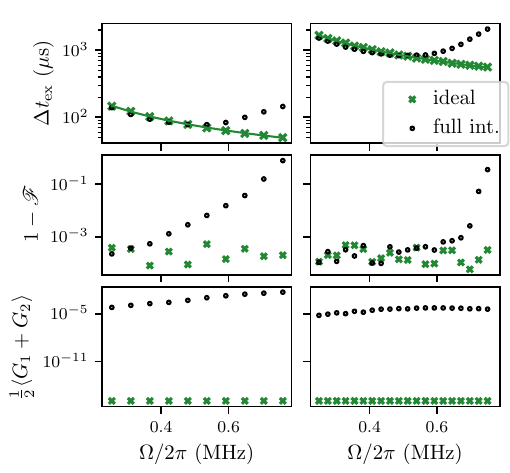}
    \caption{{\bf M\o lmer-S\o rensen-type $\mathbb{Z}_2$ tunneling:}   Simulation of the exchange duration, fidelity error, and symmetry operator $\braket{G_1+G_2}/2$ as a function of the two-photon Rabi frequency $\Omega$ for  dipole-allowed Raman transitions (left column) and the single-photon Rabi frequency $\Omega$ for  quadrupole-allowed transitions (right column). The results are shown for the full Hamiltonian~\eqref{eq:int_term} (black) and the idealized invariant-gauge tunneling~\eqref{eq:MS_second_order_only} (green). The green solid line in the upper panel  is the  analytic dependence of the exchange duration on $\Omega$, extracted from Eq.~\eqref{eq:eff_tunneling_MS}. For both the full and ideal Hamiltonian the expectation value of the symmetry operator is consistent with $0$, as desired, down to $10^{-3}$.}
    \label{fig:merged_ms_vs_omega}
\end{figure}

In this Appendix, we discuss the implementation of the analog scheme  based on optical qubits. For  
 a $^{88}\mathrm{Sr}^+$ ion with the same parameters as in the previous Appendix, the optical qubit is formed by the ground state $5S_{1/2} , m_j = -1/2$ and the metastable state $4D_{5/2} , m_j = -1/2$ (see Fig.~\ref{fig:LS_style_drive_w_qdp}). To estimate realistic numbers for the  {light-shift} scheme on the optical qubit, we consider a narrow-linewidth 674$\,$nm laser system \cite{thirumalai2019high}. The Lamb-Dicke factor changes due to the different wavelength of the two laser beams, leading to $\eta_{z} = 2  \times 0.05$ and $\eta_{x} = 2 \times 0.024$.
We employ two 674-$\mathrm{nm}$ beams detuned with respect to the qubit resonance by $\delta = \omega_x-\omega_z$, where  $\Delta = 2\pi \cdot \SI{3.56}{\mega\hertz}$. The detuning $\Delta$ is much smaller than for the dipole-allowed Raman transitions on the left of Fig.~\ref{fig:LS_style_drive_w_qdp}, as the $D_{5/2}$ level is metastable and its lifetime is much longer than the timescale of interest.
In  {Fig.~\ref{fig:merged_ms_vs_omega}}, we present the results of our numerical simulations for the same quantities as above, but as a function of the Rabi-frequency $\Omega$ for the optical qubit. In this case, we can achieve tunneling coupling rates of up to $0.17$\,kHz inferred as $1/(4\Delta t_{\rm ex})$, which are much slower than the dipole-Raman scheme. Likewise, the state infidelity is larger, showing that the realization with optical qubits will be more challenging.

\subsection{M\o lmer-S\o rensen-type scheme}
\label{app:MS_scheme}
\begin{figure}
    \centering
    \includegraphics[width=1\linewidth]{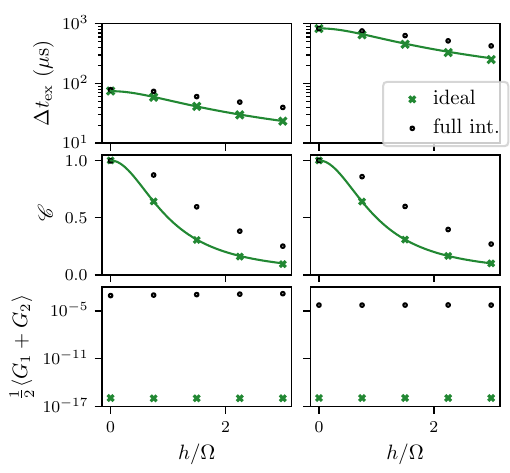}
    \caption{{\bf M\o lmer-S\o rensen $\mathbb{Z}_2$ tunneling with a non-zero electric field:} with Raman couplings (left column) and quadrupole couplings (right column). Simulating the exchange duration, contrast, and symmetry operator $\braket{G_1+G_2}/2$ as a function of electric field strength $h$ relative to the coupling strength $\Omega$. The results are shown for the full Hamiltonian~\eqref{eq:int_term} (black), including the additional carrier driven by a simple shift of the M{\o}lmer-S{\o}rensen detuning~\eqref{eq:el_field_MS}.   We also show the results for the ideal gauge-invariant Hamiltonian~\eqref{eq:ms-z2}   (green). The green solid line in the upper panel  is the expected analytic dependence of the exchange duration on $\Omega$, extracted from Eqs.~\eqref{eq_eff_rabi_flop} and the effective couplings~\eqref{eq:eff_tunneling_MS} and~\eqref{eq:el_field_MS}. For both the full and ideal Hamiltonian the expectation value of the symmetry operator is consistent with $0$, as desired, down to $10^{-3}$.}
    \label{fig:merged_ms_vs_htv}
\end{figure}

A challenge with this scheme when applied to the quadrupole transition is the large resulting light shift 
$\Delta E_{\rm ac}$ on the qubit transition. This spurious term is $1/(\eta_x\eta_z)$ larger than the sought-after tunneling rate. This issue can be circumvented by either combining the tunneling interaction with a spin-echo~\cite{gan_2020} or by tracking the qubit frequency shift in software and feed-forward the acquired phase. The first approach is no longer compatible when a transverse electric-field term is added (see Eq.~\eqref{eq:tunneling_gauge}). The second approach is in principle possible, but relies on the precise calibration of the Stark shift, as the beams used to generate the transverse electric-field term  must be tuned accordingly (see Eq.~\eqref{eq:constraints_carrier_ac_stark}). This is a challenging task, as the shift needs to be calibrated to a precision that goes well beyond that of the effective tunneling  rate. Moreover, pulse shaping makes the calibration more difficult, as the instantaneous light shift changes over the pulse duration. 
In practice, this is difficult as the light-shift amplitude  is $1/\eta_x\eta_z$ larger than the tunneling interaction, and would need to be calibrated to a precision exceeding the tunneling rate. Hence, no simulations are included for the light-shift scheme utilising the quadrupole coupling.

\begin{figure*}[ht]
    \centering
    \includegraphics[width=1\linewidth]{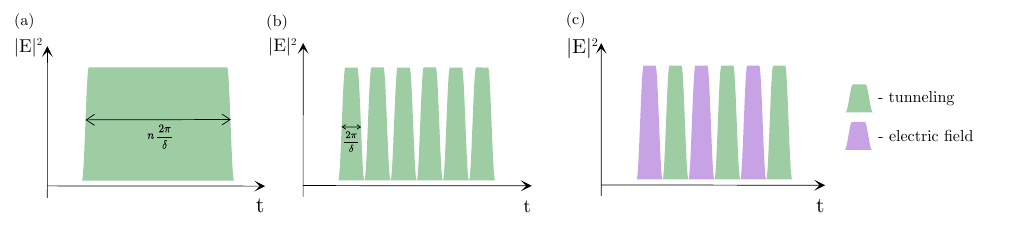}
    \caption{\textbf{Amplitude-shaped pulses for the pulsed scheme:} (a) Shaped pulse that could be used for implementing just the tunneling term. (b) Trotterized pulse sequence for implementing the tunneling term. (c) Trotterized pulse sequence for implementing the full $\mathbb{Z}_2$-link Hamiltonian. }
    \label{fig:scheme_2_pulses}
\end{figure*}

\begin{figure}[t]
	\centering
	\includegraphics[width=0.75\columnwidth]{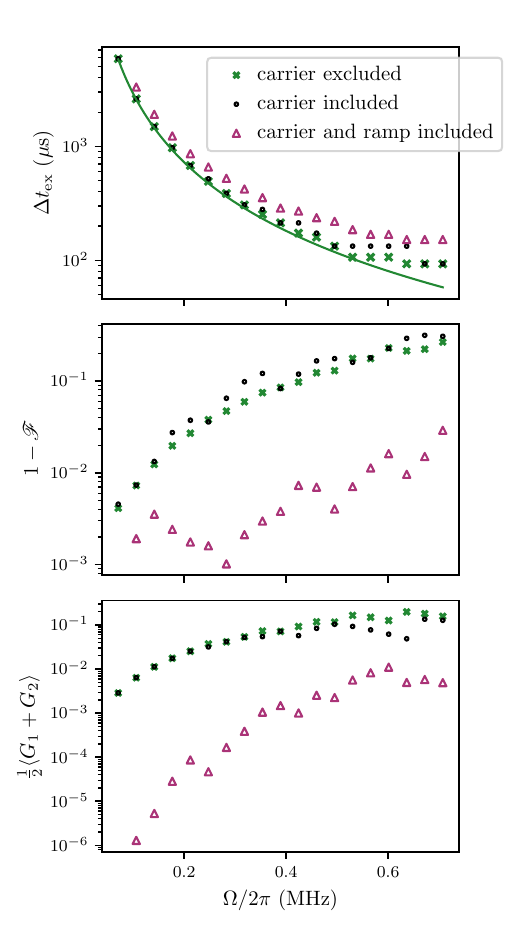}
	\caption{ {\bf Pulsed-scheme $\mathbb{Z}_2$ tunneling with orthogonal forces:} We vary the strength of the tones in the bichromatic field $\Omega$, and evaluate the exchange duration, the infidelity in obtaining the desired state $\ket{R}$, and the expectation value of the local symmetry generators. We do this for three cases: excluding the carrier from the interaction and thus considering Eq.~\eqref{eqn:scheme_2_H_sdf} (green crosses); looking at the full interaction and thus including the spurious carrier terms (black circles); and, finally, considering the full interaction while slowly ramping the pulses on and off (magenta triangles). The green solid line in the upper panel  is the expected analytic dependence of the exchange duration on $\Omega$, extracted from Eq.~\eqref{eq:eff_tunneling_trotter}. For the Hamiltonian including the spurious carrier and adiabatic ramp the expectation value of the symmetry operator is consistent with $0$, as desired, down to $10^{-2}$.}
	\label{fig:scheme_2_tunneling_vs_omega}
\end{figure}

\begin{figure}[t]
	\centering
	\includegraphics[width=0.75\columnwidth]{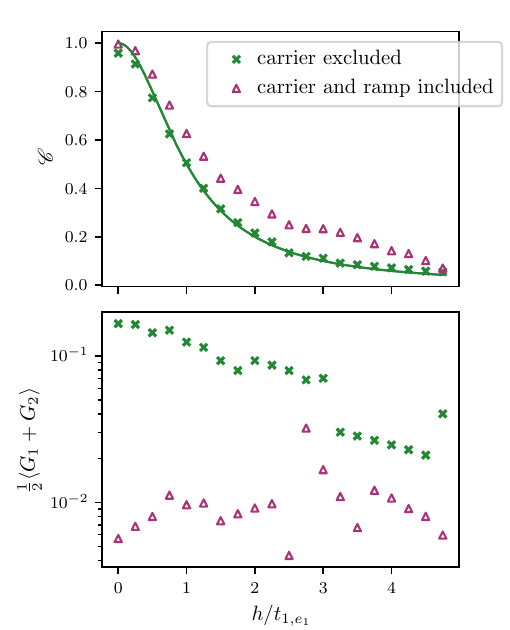}
	\caption{\textbf{ Pulsed-scheme $\mathbb{Z}_2$ tunneling with orthogonal forces in the presence of a non-zero electric field}: We simulate the $\mathbb{Z}_2$ dynamics using two orthogonal spin-dependent forces for different magnitudes of $h$. The value used for $t_{1,e_1}$ was calculated as $1/(4\Delta t_{\rm ex})$ at $h = 0$. We obtain dynamics similar to Fig.~\ref{fig:rabi_flopping_dynamics}~{\bf (b)} and infer the maximum contrast in $\overline{s}_x(t)$ and the expectation value of the local symmetry at the point of maximum contrast. The markers are numerical simulations, while the continuous lines are analytical predictions for different values of $h$ Eq.~(\ref{eq_Rabi flops_link}). For the Hamiltonian including the spurious carrier and adiabatic ramp the expectation value of the symmetry operator is consistent with $0$, as desired, down to $10^{-2}$.}   
	\label{fig:scheme_2_BS_and_H_scan}
\end{figure}

In this Appendix, we present a detailed analysis of possible errors in the MS parametric tunneling, considering  realistic trapped-ion parameters for both the ground state and optical qubit schemes of Fig.~\ref{fig:MS_style_drive}.
Once again, we numerically integrate the full dynamics for \eqref{eq:int_term} and compare it to the idealized tunneling term \eqref{eq:MS_second_order_only}. For the Raman scheme for the ground state qubit, the bichromatic field can be achieved by having one  Raman beams at $\omega_L + \omega_0$ and the other counter-propagating Raman beam consisting of two tones at $\omega_L \pm \delta$, i.e. $\boldsymbol{k}_{\rm d} = 2\boldsymbol{k}$.
For the optical qubit, on the other hand, one can obtain a similar coupling by addressing it with a single beam consisting of two tones at $\omega_0 \pm \delta$, i.e. $\boldsymbol{k}_{\rm d} = \boldsymbol{k}$.
The Rabi frequency $\Omega$ of the blue- $(+\delta)$ and red-detuned   $(-\delta)$ tone must be the same.  We assume the same experimental parameters as in the previous section.
For implementing this on the optical qubit we use the quadrupole  transition at $\lambda=674$~nm. The Lamb-Dicke parameters are $\eta_z = 0.05$ and $\eta_x = 0.024$, and we consider Rabi frequencies of up to $\Omega/2\pi = \SI{1.1}{\mega\hertz}$. As previously, we truncate the phonon occupations to $n_{\rm max}=7$ phonons. We obtain a maximum effective tunneling coupling strength $\SI{3.22}{\kilo \hertz}= 1/(4\Delta t_{\rm ex})$ for the Raman scheme, and a slower one  $\SI{0.30}{\kilo \hertz}= 1/(4\Delta t_{\rm ex})$ for the quadrupole scheme (see Fig.~\ref{fig:merged_ms_vs_omega}).

We note that a similar principle has been recently used to generate single-mode squeezing in reference~\cite{Katz_2022}, which can then be used for the quantum simulation of spin models with multi-spin interactions~\cite{PhysRevA.79.060303,Andrade_2022,https://doi.org/10.48550/arxiv.2207.10550}. We add adiabatic pulse shaping to the simulation as it enables smooth transitioning into the interaction picture and suppresses off-resonant (non-commuting) carrier excitations. This effectively reduces the strength of the tunneling but, importantly, it retains the state dependence. The non-commuting  carrier  sets a limit on the achievable interaction magnitude;, which is reflected by the global minimum in the duration (first row, Fig.~\ref{fig:merged_ms_vs_omega}).

Let us now present the error analysis for the MS scheme when considering an additional electric-field term as discussed around Eq. of the main text.  The resulting exchange duration $\Delta t_{\rm ex}$, the maximum contrast $\mathcal{C}$ in $\overline{s}_x(t)$ and gauge-symmetry generators $\braket{G_1+G_2}/2$, found through numerical simulations of the Hamiltonian \eqref{eq:ms-z2} and are compared to the ideal gauge tunneling \eqref{eq:tunneling_gauge} (see Fig.~\ref{fig:merged_ms_vs_htv}). As before, the presence of the off-resonant carrier  reduces the effective magnitude of the transverse term, but  gauge invariance is preserved.

\subsection{Orthogonal-force pulsed scheme}
\label{app:orthogonal_forces}

In this appendix, we consider the same experimental apparatus as in previous ones, and investigate the implementation of the gauge-invariant tunneling using the two orthogonal state-dependent forces.One way to implement the two orthogonal state-dependent forces in a laser system is by having two sets of M{\o}lmer-S{\o}rensen-style bichromatic fields, see Sec.~\ref{sec:MS_impl}.~One bichromatic field is symmetrically detuned from the carrier by $\delta_1 = \pm(\omega_z + \delta)$ and the other by $\delta_2 = \pm(\omega_x + \delta)$. As described in Sec.~\ref{sec:MS_impl}, the bichromatic fields can either couple levels in the ground state via a Raman transition, or two levels of an optical qubit via a quadrupole transition. Moreover, for having two orthogonal state-dependent forces, we set the phase between the two bichromatic beams such that
\begin{equation}\label{eqn:bich_phases}
\left(\frac{\phi_+ + \phi_-}{2}\right)_2 = \left(\frac{\phi_+ + \phi_-}{2}\right)_1 + \frac{\pi}{2},
\end{equation}
where $\phi_+$, $\phi_-$ are the phases of the blue and red detuned tones in each of the bichromatic fields, $1$ and $2$. Applying these four tones gives rise to the two orthogonal state-dependent forces~\eqref{eqn:scheme_2_H_sdf} that are needed for engineering the tunneling term~\eqref{eqn:scheme_2_H_eff}. We consider the optical qubit and two motional modes with $\eta_z=0.05$, $\omega_z/2\pi = 1.2$ MHz, and $\eta_x=0.024$, $\omega_x/2\pi = 1.9\,$MHz, respectively. We choose the detuning of the bichromatic fields from the respective vibrational mode to be $\delta/2\pi = 75\,$kHz, and set $\Omega/2\pi=0.75/\sqrt{2}~\textrm{MHz}$ for each of the four tones. This should allow us to reach an effective tunneling coupling rate of up to $1.3$\,kHz, inferred as $1/(4\Delta t_{\rm ex})$.

Let us note that the bichromatic scheme also leads to spurious carrier terms that drive off-resonant qubit rotations around axes that are orthogonal to each of the corresponding state-dependent forces. As mentioned above, a technique to mitigate the effect of the carrier term is pulse shaping~\cite{Roos_2008}, which is mainly needed at large coupling strengths, i.e., large $\Omega$ in Eq.~\eqref{eqn:scheme_2_H_sdf}. We describe the amplitude shaping of the pulses and further discuss its effect in our numerical simulations below.

The shortest time step that we can use in order to close the loops in phase space is $2\pi/\delta$, which was used for the simulations above. Hence, the Trotter error is given as $ht_{1,e_1}(2\pi/\delta)^2$
which must be negligible. In the simulation results presented above, for the highest value of $h$, the Trotter error was $\approx 8\times10^{-4}$. Higher values of $\delta$ reduce the $O([\eta\Omega/\delta]^3)$ error, and enable finer time steps in the time scans, as we always want to measure at instants where the loops in phase space that arise from first-order contributions in the Magnus expansion are closed, and the leading effect is the state-dependent tunneling. However, a higher $\delta$ also reduces the effective tunneling rate~\eqref{eqn:scheme_2_H_eff}. This translates into shorter pulses which, when becoming comparable to the ramp length, do not give the effective desired dynamics anymore. 
Comparing the analog and pulsed schemes in terms of observing the matter-gauge field dynamics, the state preparation and measurement stages are the same. The difference is how the effective Hamiltonian in Eq.~\eqref{eq:tunneling_gauge} is experimentally implemented and that is by using a pulse sequence as the one shown in Fig.~\ref{fig:scheme_2_pulses}~{\bf (c)}. This was simulated as a series of pulses that are ramped on and off with 3.6$\,\mu$s ramp durations and with the FWHM duration of $2\pi/\delta$, as shown in Fig~\ref{fig:scheme_2_pulses}~{\bf (b)}. The same dynamics can be achieved by having a single pulse with the FWHM duration equal to an integer multiple of $2\pi/\delta$ (see Fig~\ref{fig:scheme_2_pulses}~{\bf (a)}). By inspecting the figure more closely, we see that there are small deviations with respect to the idealized  Hamiltonian~\eqref{eqn:scheme_2_H_eff} that deserve a more detailed analysis.

In the top panel of Fig.~\ref{fig:scheme_2_tunneling_vs_omega}, we vary $\Omega$ in each one of the tones and evaluate how this affects the exchange duration $\Delta t_{\rm ex}$, calculated by maximising the state fidelity as in the previous section. We conduct simulations for three cases. Initially, we simulate the interaction in Eq.~\eqref{eqn:scheme_2_H_sdf} that only contains the two orthogonal state-dependent forces. We then introduce the off-resonant carrier terms, which would arise from the Lamb-Dicke expansion of Eq.~\eqref{eq:int_term} as the leading off-resonant perturbations to the state-dependent forces~\eqref{eqn:scheme_2_H_sdf}. Finally, we also consider the amplitude pulse shaping. When the carrier terms are excluded, and the pulse is applied for a duration that is an integer multiple of $2\pi/\delta$, the inferred tunneling duration follows closely the theory from Eq.~\eqref{eqn:scheme_2_H_eff}, which is represented by a green solid line according to $\Delta t_{\rm ex}=\pi/2t_{1,{\bf e}_1}$ with the effective tunneling of Eq.~\eqref{eq:eff_tunneling_trotter}. As $\Omega$ is increased, the error term $O([\eta\Omega/\delta]^3)$ becomes significant, and small deviations start to appear. As above, we use the  state fidelity $\mathcal{F}$ (the overlap to the desired state) and $\braket{G_1+G_2}/2$ (the expectation value of the symmetry generators) in order to evaluate the quality of the effective Hamiltonian with respect to Eq.~\eqref{eq:tunneling_gauge}. Introducing the carrier terms does not change the exchange duration $\Delta t_{\rm ex}$, $\mathcal{F}$, or $\braket{G_1+G_2}/2$ significantly. However, amplitude shaping the pulses substantially improves the quality of tunneling as it suppresses the coupling to higher order terms $O([\eta\Omega/\delta]^3)$. Introducing the amplitude-shaping ramp effectively decreases the area of the pulses, which translates into slightly slower exchange durations (top panel), but also leads to smaller errors (middle and bottom panels).

Once the viability of trapped-ion pulsed Scheme for the quantum simulation of the $\mathbb{Z}_2$ gauge-invariant tunneling has been demonstrated, we can consider the errors that would stem from adding the electric field term.
In Fig~\ref{fig:scheme_2_BS_and_H_scan}, we present our numerical results, evaluating the decrease in contrast of the Rabi oscillations between the $\ket{\rm L}$ and $\ket{\rm R}$, and $\braket{G_1+G_2}/2$, as one increases the electric field $h$. When the carrier terms are excluded and we do not use the amplitude-shaped pulses, the decrease, in contrast, follows Eq.~\eqref{eq_Rabi flops_link}, which is represented as a green solid line in the top panel. We find that, in the realistic experimental situation with  carrier terms  present, and  using amplitude-shaped pulses, we need slightly higher values of $h$ to achieve the same contrast. This is mainly due to the effect of the ramp.

\section{\bf  Wannier-Stark solution and MPS benchmark}
\label{sec:WS_solution}
In this Appendix, we discuss in detail the numerical benchmarks of MPS methods using exact  solutions for the confinement dynamics in the one-~\eqref{eq:gauge_chain1} and two-boson~\eqref{eq:2_part_recurssion} sectors.

\subsection{One-boson sector}
Considering that the tunneling strength is homogeneous $t^{\phantom{\dagger}}_{i,\textbf{e}_1}={\Omega_{\rm d}}/{4}$, $\forall i$, and that the initial state contains a single boson~\eqref{eq:gauge_chain1}, the effective Wannier-Stark ladder  can be solved exactly in the thermodynamic limit $N\to\infty$. Shifting the zero energy to the center of the chain, the problem can be mapped onto the dynamics of a single planar rotor~\cite{Hartmann_2004}. Let us discuss some of the details. A particle  in a circle can be described in the basis $\{\ket{\varphi}\}$ determined by its angle $\varphi\in[0,2\pi)$. In this basis, the angular momentum $J_z=-\ii\partial_\varphi$ is a Hermitian operator, and it readily follows that its spectrum is a    countable infinite set $\sigma(J_z)=\mathbb{Z}$. Moreover, one can  {introduce} the unitary ladder operators $J_{\pm}=\ee^{\pm \ii\varphi}$, and find a representation of the $O(2)$ rotor algebra  $[J_z,J_{\pm}]=\pm J_{\pm}$, and $[J_{+},J_-]=0$, which differs from the more standard $SU(2)$ algebra of spin operators. Using the physical states~\eqref{eq:phys_subspace_basis}    of    the $\mathbb{Z}_2$ gauge theory in the thermodynamic limit, one can readily find a specific representation  of the rotor algebra 
\beq
J_z= \sum_{i\in\mathbb{Z}} i\ket{\sim\!\!\sim\!\!\!\!\!\!\bullet_i}\!\!\bra{\sim\!\!\sim\!\!\!\!\!\!\bullet_i},\hspace{2ex}J_{\pm}= \sum_{i\in\mathbb{Z}} \ket{\sim\!\!\sim\!\!\!\!\!\!\bullet_{i\pm 1}}\!\!\bra{\sim\!\!\sim\!\!\!\!\!\!\bullet_i},
\eeq
where we see that the position of the $\mathbb{Z}_2$ charge with the attached electric-field line maps onto the angular momentum of the rotor, whereas the tunneling to the right (left) map onto the rotor ladder operators $J_+(J_-)$.  

\begin{figure}[t]
	\centering
	\includegraphics[width=0.85\columnwidth]{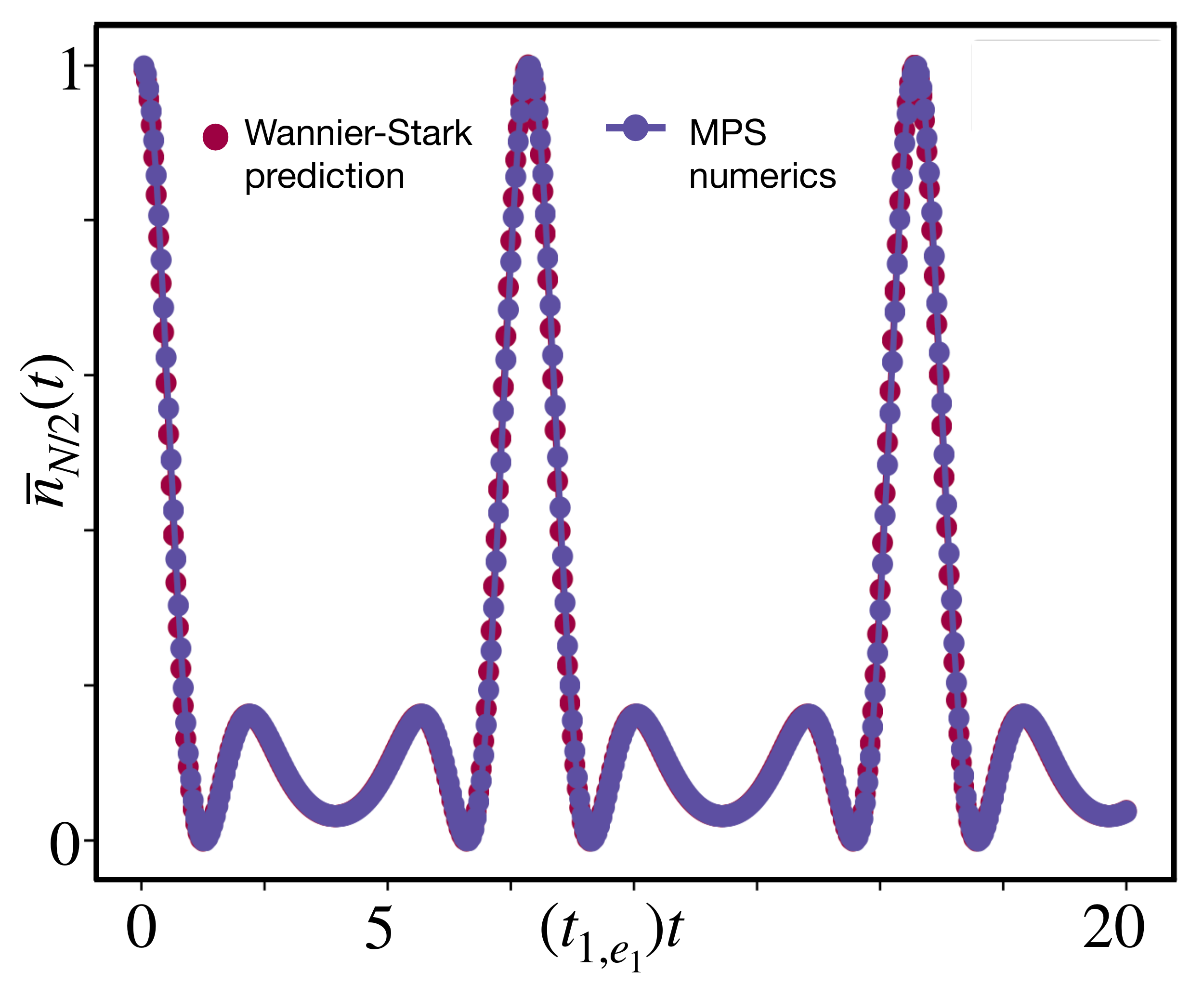}
	\caption{{\bf Wannier-Stark localisation of a single boson: } We compare the analytical prediction for $\overline {n}_{N/2}(t)=\langle a_{N/2}^\dagger a_{N/2}^{\phantom{\dagger}}(t)\rangle$ in Eq.~\eqref{eq_wannier_stark_dynamics} to the numerical results based on Matrix Product states with bond dimension $\chi=100$ for a chain with $N=16$ lattice sites. We set the transverse electric field to $h=0.4t_{1,{\bf e}_1}$ and we use the time step $\delta t=0.05/t_{1,{\bf e}_1}$ . The analytical formula~\eqref{eq_wannier_stark_dynamics} is  based on the mapping of the $\mathbb{Z}_2$ gauge theory for a single boson~\eqref{eq:gauge_chain1}  to the Wannier-Stark ladder  in the thermodynamic limit of an infinitely-long   chain.}
	\label{fig:ws_localistion_comparison}
\end{figure}

It is then straightforward to derive the Heisenberg equations for the mean position and  standard deviation of the $\mathbb{Z}_2$ charged boson. For instance, considering that the boson is initially in the middle of the chain $\ket{\Psi_{\textrm{phys}}(0)}=\ket{\sim\!\!\sim\!\!\!\!\!\!\bullet_{N/2}}$, we find that the mean is
$\langle J_z(t)\rangle=N/2$, while the standard deviation oscillates
\beq
\sigma(t)=(\langle J_z^2(t)\rangle-\langle J_z(t)\rangle^2)^{1/2}=(\sqrt{2}t_{i,\textbf{e}_1}/h)\sin(ht).
\eeq
We thus see that the average position of the boson attached to the electric-field line  remains constant. However, this is not the signature of the aforementioned Wannier-Stark localisation yet. Indeed, setting  $h=0$ yields the same result, as an initially localised  particle in a tight-binding model with the same amplitude of tunneling to the left and right can only disperse around the initial position, but its average position remains static. In this limit, the above expression of the standard deviation leads to a ballistic dispersion $\sigma(t)=(\sqrt{2}t_{1,\textbf{e}_1})t$, which differs clearly from the breathing-type oscillations that appear as soon as $h\neq 0$. Hence, it is the change in the dispersion which provides a signature of the Wannier-Stark localisation of the $\mathbb{Z}_2$ charge, which can only disperse within a localised region by periodically  stretching and compressing the attached  electric-field string. 

To provide a more complete description of this localisation,  we note that
the thermodynamic problem has an exact solution in terms of the so-called Wannier-Stark eigenstates
\beq
\label{eq:WS_eigenstates}
\ket{\epsilon_m}=\sum_{i\in\mathbb{Z}}(-1)^{i-m}J_{i-m}(\gamma)\ket{\sim\!\!\sim\!\!\!\!\!\!\bullet_i}, \hspace{2ex} \gamma=\frac{t_{i,\textbf{e}_1}}{h}
\eeq
where
 $J_n(x)$ is the first-class Bessel function of integer order $n$, and the corresponding energies $\epsilon_m=m(2h)$ define the aforementioned Wannier-Stark ladder for $m\in\mathbb{Z}$.
This solution can be derived by going to momentum space and using the Hansen-Bessel integral representation~\cite{doi:10.1080/01418639608240331,Hartmann_2004} or, more  directly, by looking into the discrete difference equation for the amplitudes of the eigenstates in the physical basis
\beq
\label{eq:sing_part_recurssion}
\ket{\epsilon_m}=\sum_{i\in\mathbb{Z}} c_i^{\phantom{*}}\ket{\sim\!\!\sim\!\!\!\!\!\!\bullet_i},\hspace{2ex} t^{\phantom{*}}_{i-1,\textbf{e}_1}c_{i-1}^{\phantom{*}}+t_{i,\textbf{e}_1}^*c_{i+1}^{\phantom{*}}+2hi\,c_i^{\phantom{*}}=\epsilon_m c_i^{\phantom{*}}.
\eeq
This equation can be rewritten in terms of the recurrence relation of Bessel functions~\cite{Dominguez-Adame_2010}, such that one can identify  $c_i=(-1)^{i-m}J_{i-m}(\gamma)$, and check for the consistency of  normalisation $\sum_{i\in\mathbb{Z}}|c_i|^2=\sum_{i\in\mathbb{Z}}J_{i-m}^2(\gamma)=1$. In light of the asymptotic scaling of the Bessel functions, which vanish rapidly for $|i-m|\gg\gamma$, one can see that the eigenstates~\eqref{eq:WS_eigenstates} are not delocalised over the whole lattice as occurs for $h\to 0$ but, instead, concentrated around the $m$-th site, which is a more direct manifestation of the so-called Wannier-Stark localisation that parallels the definition of Anderson localisation in disordered  systems~\cite{PhysRev.109.1492}.

 With these eigenstates, one can construct the full unitary propagator of the problem. Using the Neumann-Graff addition formula of Bessel functions~\cite{bessel_book},  the probability to find the  boson with the attached electric-field line $r$ sites apart is
$
p_r(t)=|\bra{\sim\!\!\sim\!\!\!\!\!\!\bullet_{{N}/{2}+ r}}\Psi_{\rm {phys}}(t)\rangle|^2=J^2_{ r}\big(2\gamma\sin(ht)\big)=p_{-r}(t).
$
In comparison to the Rabi oscillations of the single-link case in Fig.~\ref{fig:rabi_flopping_dynamics}, where the boson and the gauge field oscillate in phase according to the observables of Eq.~\eqref{eq_Rabi flops_link}, we now have  correlated Wannier-Stark oscillations in both the number of bosons  
\beq
\label{eq_wannier_stark_dynamics}
\overline {n}_{i}(t)=\langle a_i^\dagger a_i^{\phantom{\dagger}}\!\!(t)\rangle=p_i(t)=J^2_{i-N/2}\big(2\gamma\sin(ht)\big),
\eeq
and the position of the electric-field line attached to the boson, which can be inferred from the two-point correlation function
\beq
\label{eq:domain_wall_correlations}
\langle \sigma^x_{i-1,{\bf e}_1}\!\!\sigma^x_{{i},{\bf e}_1}(t)\rangle=1-2J^2_{i-N/2}\big(2\gamma\sin(ht)\big).\\
\eeq
In Fig.~\ref{fig:ws_localistion_comparison} , we present a quantitative comparison of the analytical prediction for the boson number operator at the center of the chain $\overline {n}_{N/2}(t)$ in Eq.~\eqref{eq_wannier_stark_dynamics} with the numerical results based on MPS  discussed in the main text. The agreement is remarkable, which serves to benchmark the validity of our approach.

\subsection{Two-boson sector}

In the two-boson sector,  by introducing the center-of-mass $x_{\rm cm}=\half(i+j)$, and relative coordinate $r=j-i\geq0$,   the problem~\eqref{eq:2_part_recurssion} reduces to the  Wannier-Stark ladder for a single particle in  a one-dimensional chain. Noting once more that the tunneling strengths are homogeneous, one finds 
\beq
\label{eq:recurrence_2_partciles}
 c_{i,j}^{\phantom{*}}=\ee^{\ii P x_{\rm cm}}c(r),\hspace{2ex} t_{P}c(r-1)+t_{P}^*c(r+1)+2hr\,c (r)=\epsilon_{m,P} c(r),
 \eeq
 where we have introduced the conserved total momentum $P=(p_i+p_j)$, the momentum-independent  Wannier-Stark ladder energies $\epsilon_{m,P}=m(2h)$, and the dressed tunneling strength 
 \beq
 t_P=2t_{1,\textbf{e}_1}\cos\left({P}/{2}\right).
 \eeq

\begin{figure}[t]
	\centering
	\includegraphics[width=0.85\columnwidth]{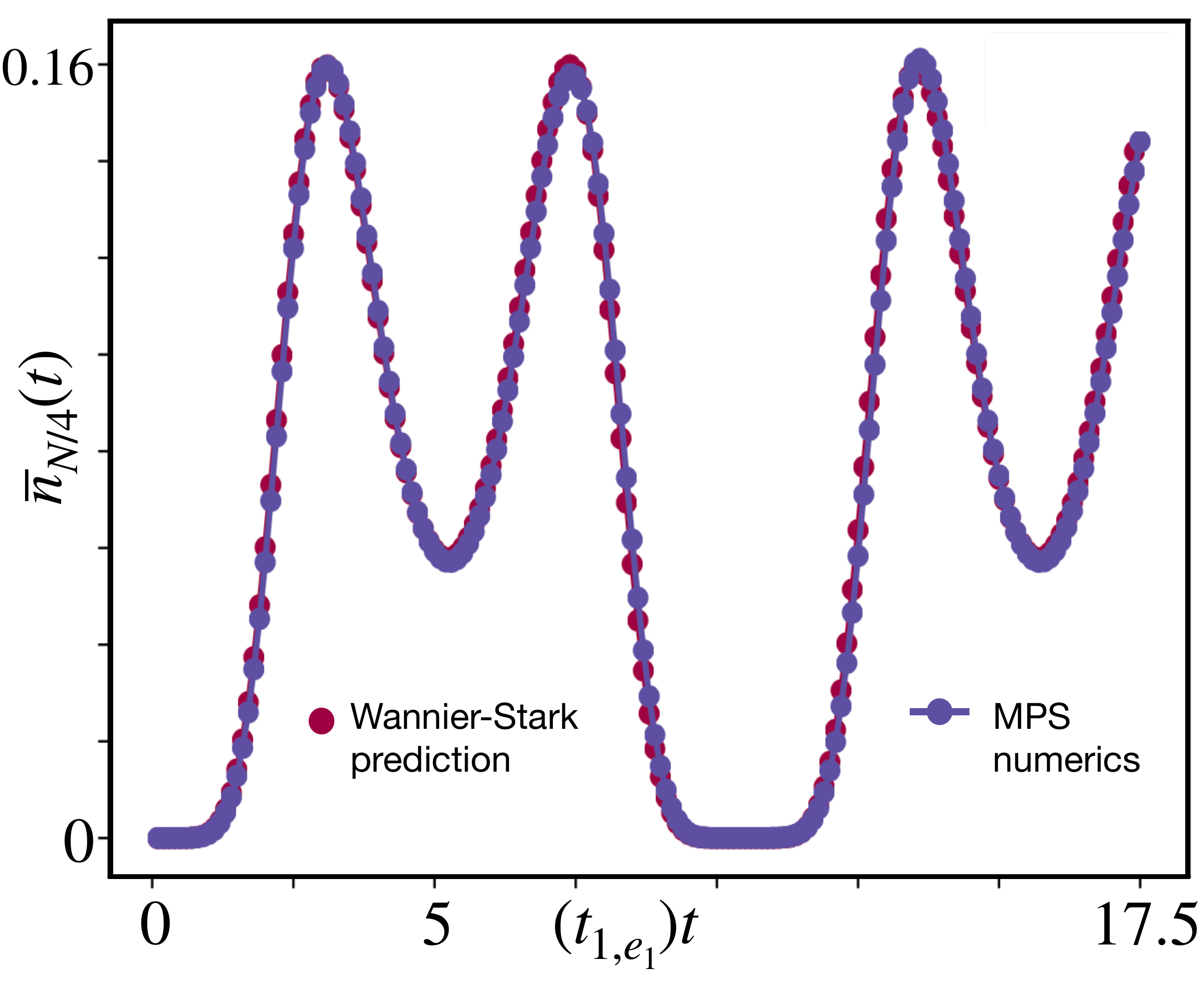}
	\caption{{\bf Wannier-Stark confinement for two  bosons: } We compare the analytical prediction for $\overline {n}_{N/4}(t)=\langle a_{N/4}^\dagger a_{N/4}^{\phantom{\dagger}}(t)\rangle$ in Eq.~\eqref{eq_wannier_stark_dynamics_2_bosons} to the numerical results based on Matrix Product states with bond dimension $\chi=100$ for a chain with $N=32$ sites. We set the transverse electric field to $h=0.3t_{1,{\bf e}_1}$ and we use the time step $\delta t=0.05/t_{1,{\bf e}_1}$. The analytical formula~\eqref{eq_wannier_stark_dynamics_2_bosons} is  based on the mapping of the $\mathbb{Z}_2$ gauge theory for a boson pair~\eqref{eq:gauge_chain1}  to the Wannier-Stark ladder  for the particle of reduced mass  in the thermodynamic limit of a   chain.}
	\label{fig:ws_confinement_comparison}
\end{figure}

 In contrast to Bose-Hubbard-type models with finite range interactions, which can lead to both  scattering and bound states for a pair of bosons~\cite{PhysRevA.81.042102}, the above recurrence equation~\eqref{eq:recurrence_2_partciles} describes a relative particle that tries to tunnel against a linear potential with a dressed tunneling strength that depends on the center-of-mass momentum. Once again, by   taking the thermodynamic limit, the recurrence equation corresponds to that of a Wannier-Stark ladder~\eqref{eq:sing_part_recurssion} for the relative particle.  In this way, we obtain the following solutions 
 \beq
 \label{eq:eigsnetates_two_body_problem}
 \ket{\epsilon_{m,P}}=\sum_{i,j}{}^{\!\!'}\sum_{m\in\mathbb{Z}}\ee^{\ii P(i+j)/2}(-1)^{j-i+m}J_{j-i-m}(\gamma_P)\ket{_i\bullet\!\!\!\!\sim\!\!\sim\!\!\!\!\!\!\bullet_j}
\eeq
where $\gamma_P=2t_{1,\textbf{e}_1}\!\cos(P/2)/h$. As occurred for the single-boson sector, where any of the $m$-th eigenstates~\eqref{eq:WS_eigenstates} is localised around the $m$-th site, the two-particle solutions~\eqref{eq:eigsnetates_two_body_problem} only consist of bound states regarding their relative distance instead of scattering states. The original bosons are  thus confined  in pairs, forming wavefunctions  of energy $\epsilon_{m,P}$, which decay exponentially fast as their relative distance $r$ increases $|r|>m$. These solutions are a toy analog  of  mesons in higher-dimensional non-Abelian gauge theories. The original particles, which carry a net $\mathbb{Z}_2$ charge, cannot be observed as individual excitations, just like quarks in quantum chromodynamics. They become instead confined in  pairs of zero net charge, which are associated to  a specific quantised bound energy depending on their respective confinement. In the present case, these meson-like particles  can freely move as a whole, i.e. non-zero center-of-mass momentum.

Let us now consider an initial state  in which the bosons are symmetrically positioned about the center of the chain with  relative distance $r_0$, namely $\ket{\Psi_{\rm phys}(0)}=\ket{_{(N-r_0)/2}\bullet\!\!\!\!\sim\!\!\sim\!\!\!\!\!\!\bullet_{(N+r_0)/2}}$. This state has a vanishing total momentum $P=0$, such that the center of mass will remain localised at the center of the chain while the two particles disperse and interfere. According to our previous discussion, the number of bosons~\eqref{eq_wannier_stark_dynamics} should now evolve according to
\beq
\label{eq_wannier_stark_dynamics_2_bosons}
\overline{n}_{i}(t)=\left|J_{i-N/2-r_0/2}\big(2\gamma\sin(ht)\big)+J_{i-N/2+r_0/2}\big(2\gamma\sin(ht)\big)\right|^2.
\eeq
In Fig.~\ref{fig:ws_confinement_comparison}, we present a quantitative comparison of this analytical expression for $\overline{n}_{N/2}(t)$ with the TDVP numerical results. As found in the single-particle sector of Fig.~\ref{fig:ws_localistion_comparison}, the agreement of the numerical TDVP results with the analytical prediction in terms of the sum of Bessel functions is remarkable, and serves to benchmark the validity of our approach, which will be extended to situations beyond  analytical  solutions below.

\end{appendix}

\bibliographystyle{apsrev4-1}
\bibliography{bibliography}

\end{document}